%% file: ms.tex
\documentclass[12pt,preprint]{aastex}

\newcommand {\nh}{N$_{2}$H$^{+}$\ }
\newcommand {\nhns}{N$_{2}$H$^{+}$}     
\newcommand {\co}{C$^{18}$O\ }
\newcommand {\cons}{C$^{18}$O}          
\newcommand{\twelco}{$^{12}$CO\ }
\newcommand {\thirco}{$^{13}$CO\ }	
\newcommand {\thircons}{$^{13}$CO}
\defcitealias{Kirk06}{KJD06}
\defcitealias{Kirk07}{KJT07}
\defcitealias{CB06}{CB06}
\defcitealias{BCW09}{BCW09}
\defcitealias{BCDW09}{BCDW09}

\newcommand{\hk}{}

\begin{document}
\title{The Interplay of Turbulence \& Magnetic Fields in Star-Forming Regions:
Simulations and Observations}

\author{Helen Kirk\altaffilmark{1,2}, Doug Johnstone\altaffilmark{1,2}, 
	and Shantanu Basu\altaffilmark{3}}
\altaffiltext{1}{Department of Physics \& Astronomy, University of Victoria, 
        Victoria, BC, V8P 1A1, Canada; hkirk@uvastro.phys.uvic.ca}
\altaffiltext{2}{National Research Council of Canada, Herzberg Institute of 
        Astrophysics, 5071 West Saanich Road, Victoria, BC, V9E 2E7, 
        Canada}
\altaffiltext{3}{Department of Physics and Astronomy, The University of 
	Western Ontario, 1151 Richmond Street, London, ON, N6A 3K7, Canada}

\begin{abstract}
We analyze a suite of thin sheet magnetohydrodynamical
simulations based on the formulation of Basu, Ciolek, Dapp \& Wurster.
These simulations allow us to examine the observational consequences
to a star-forming region of varying the input level of turbulence
(between thermal and a Mach number of 4) and the initial 
magnetic field strength corresponding to a range of mass to flux
ratios between subcritical ($\mu_{0}=0.5$) and supercritical
($\mu_{0} = 10$).  The input turbulence is allowed to decay over
the duration of the simulation.  We
compare the measured observable quantities with those found 
from surveying the Perseus molecular cloud.
We find that only the most turbulent of simulations
(high Mach number and weak magnetic field) have
sufficient large-scale velocity dispersion (at $\sim$1~pc) 
to match that observed
across extinction regions in Perseus.  Generally, the simulated 
core ($\sim$0.02~pc) and line
of sight velocity dispersions provide a decent match to observations.
The motion between the simulated core and its local 
environment, however, is far too large in simulations with high 
large-scale velocity dispersion.

\end{abstract}


\section{INTRODUCTION}
\label{s_intro}

The study of star formation is no longer in its infancy. Both observational 
and theoretical advances have taken the field beyond simple descriptions and 
heuristic models toward a more critical examination of the
physical processes involved in assembling a star within a molecular cloud. 
Indeed, perhaps the biggest advance in star formation studies has been the 
ability to connect the formation of a star with the physical properties 
of the molecular cloud.  Theorists and simulators have thus 
begun to model the larger scale properties of the molecular cloud to
attack the stellar Initial Mass Function (IMF)
and the star formation efficiency (SFE) \citep[see review by][]{Bonnell07}. 
At the same time, observers have 
begun to investigate statistically meaningful samples of prestellar cores 
and protostars, often within a single star-forming region 
\citep[see reviews by][]{Difran07,Wardthomp07}. 

Most stars are born in groups and associations \citep{Lada03}, 
thus the traditional pursuit of how an isolated star
forms requires either an interest in a small subset of all stars or a 
hope that the underlying physics
of star formation is not dominated by conditions external to the local 
prenatal core. These investigations have nevertheless 
proved to be extremely useful, yielding a simple 
yet powerful five stage model (prestellar/Class0-III)
for the collapse and ignition of an isolated stellar-mass core
\citep{Lada87,Andre93}. In this manner, the timescale for
collapse, the evolution of the accretion luminosity, and the formation 
of disks and jets has been
examined both theoretically and observationally 
\citep{Terebey84,Henriksen97,Shu94}. 
It is hoped that if the initial physical conditions 
in the core can be characterized - mass, temperature, turbulent energy, 
angular momentum, magnetic 
field strength and geometry - the stellar (plus disk) properties can be 
inferred (at least probabilistically).

Once the simplifying assumptions of isolated collapse are removed, physical 
intuition becomes more difficult.  Theoretical calculations become 
complex, and simulations take precedence in attempting
to discern how a molecular cloud fragments into stars. The large 
discrepancy between the Jeans mass in the cloud, typically tens to 
hundreds of Solar masses, and the mass of either the total cloud
(typically hundreds of thousands of Solar masses) or the average star 
formed (typically around half a Solar mass) provides incentive for 
studies which take star formation to be part of the evolution
of the cloud itself. In this scenario, there may never be an equilibrium 
core stage and the final distribution of stars may be determined by 
the flow of mass, energy, and magnetic fields within
the evolving cloud. A large compendium of work on this topic is now 
available \citep[e.g.,][]{MacLow04,Elmegreen07}. 

Despite the complexity inherent in following the star formation process from 
the molecular cloud through dense core formation to the collapse to 
individual stars, the final outcome is quantifiable.
Most simulations thus start with a particular set of initial conditions, 
evolve the cloud, and compare the outcome with observable quantities, 
primarily the IMF and SFE.  The initial conditions usually attempt
to recreate the observed physical conditions within the bulk cloud, 
concentrating on the additional support
beyond thermal motion that keeps the bulk cloud from collapsing and 
fragmenting on a dynamical time. Thus supersonic turbulence and/or 
magnetic fields are utilized to provide partial support,
at least initially, to the cloud.  As the turbulence decays and the 
neutrals slip past the ions, gravity
becomes dominant and the cloud collapses. Occasionally a suite of 
simulations with a range of initial conditions are computed and thus 
the variety in outcomes can be examined 
\citep[e.g.,][hereafter \citetalias{BCW09} and 
\citetalias{BCDW09} respectively]{BCW09,BCDW09}. 

Recent observations are providing new and more stringent conditions on 
the simulations. It is now possible to observe the dense gas and dust 
within an entire molecular cloud, revealing the manner in which 
structure forms - location, mass function, etc 
(e.g., Onishi et al. 1998; Johnstone, Di Francesco \& Kirk 2004; 
Hatchell et al. 2005; Kirk, Johnstone \& Di Francesco 2006, hereafter KJD06;
Di Francesco et al. 2007; Ward-Thompson et al. 2007).
As well, the kinematics of 
these dense cores and their surrounding envelopes can be detailed
(e.g., Benson \& Myers 1989; Jijina et al.~1999; Walsh et al.~2004, 2007; 
Kirk, Johnstone \& Tafalla 2007, hereafter KJT07).
It is thus time to consider
how to utilize these additional constraints on star formation within 
the simulations.

In this paper we consider how to appropriately `observe' a molecular cloud 
simulation in order
to compare the model output with the properties of real star-forming clouds. 
In \S\ref{s_sim} we describe the details of the numerical simulations.  
In \S\ref{s_obs} we introduce the observational data set used in this analysis. 
\S\ref{s_obssim} discusses the manner in which the simulations are 
`observed' while \S\ref{s_results} discusses the results and compares
with the observations.  Finally, we relate the results back to the
larger theoretical framework in \S\ref{s_disc} and conclude in \S\ref{s_conc}.   
In Appendix~\ref{s_care} we furthermore show 
that the trends found in the preceding sections are robust.

\section{SIMULATIONS}
\label{s_sim}
\subsection{General Overview of Simulations}

In order to analyze the interplay between the competing processes 
responsible for converting molecular cloud material into stars, along 
with the ensuing physical attributes, it is necessary to run 
(magneto)hydrodynamic simulations \citep[see review by][]{MacLow04}. 
The purpose of the present paper is to test which simulations show 
similarities with the observations of real molecular clouds and
which observable measures provide the best discrimination
between simulations with different initial conditions. We thus 
begin with a brief overview of the evolutionary implications of both 
turbulence and magnetic fields within molecular clouds, in order to 
set the stage for the results which follow. 

In almost all simulations, the initial conditions begin with much more 
mass than can be supported under a simple thermal Jeans calculation. 
From the start of the calculation, the region is therefore inherently 
unstable to collapse unless additional processes are at play.
Global gravitational collapse is typically delayed via the inclusion
of initial supersonic turbulent motions.
In keeping with both observations \citep[e.g.,][]{Larson81} and 
theoretical considerations, the largest length scales dominate the 
input turbulence. The simulated cloud therefore initially forms 
sheets and filaments with box crossing lengths, and it is within these 
over-dense regions that collapse takes place.  Given the dynamic and 
fast evolution of these clouds, the applied initial conditions are 
extremely important and the explicit results should be taken with 
caution. The simulations provide instead an ideal laboratory for 
considering the time, length, and mass scales, and the
efficiency of turbulent induced star formation. 

An alternate mechanism which can significantly delay the global
collapse of the simulated cloud is the 
inclusion of strong magnetic fields.  The low density and moderate 
ionization fraction within molecular clouds leads to effective field 
freezing and thus magnetic fields act as an additional buoyant force 
opposing gravity.  In sheet-like geometries, sufficiently strong 
magnetic fields are capable of overcoming gravity entirely and 
preventing collapse for an ambipolar diffusion time during which the 
neutral particles diffuse past the load-bearing, 
magnetic-field-coupled ions, eventually accumulating 
approximately a Jeans mass of 
material which is then capable of gravitational collapse. 

While it is possible that either of these approaches alone may be 
responsible for delaying global cloud collapse, it is more likely that 
they are both present in which case the interplay between them is not 
at all straightforward.  
In addition to turbulence included in the initial conditions
of a simulation, significant non-thermal
motions can be generated through the gravitational collapse of
non-uniformly distributed material (e.g., Burkert \& Hartmann 2004).
While magnetic fields oppose gravity by applying additional pressure 
support, supersonic turbulence both opposes gravity 
through enhanced kinetic energy in the gas and aids gravity through 
strongly compressive shocks.  When both turbulence and magnetic fields
are present, the turbulent shocks can be 
suppressed by magnetic fields. Even small enhancements in the density 
within these compressed regions, however, can significantly lower the 
ambipolar diffusion time.  Meanwhile, shock compressions and 
rarefractions dissipate energy from the turbulent waves themselves, 
producing a competition between decaying turbulence and ambipolar 
diffusion. 

Parameter studies following the evolution of molecular clouds with 
both magnetic fields and turbulence are greatly needed. The 
complicated numerical equations that must be solved in such cases 
make such simulations computationally expensive, however.  Few 
large parameter studies have thus been carried out.  Here, we utilize an 
extensive study \citepalias{BCDW09}
which followed the evolution of a sheet-like 
cloud in which significant simplifications are possible. 
The resolution of the simulations (128 by 128 cells) is modest,
however, this is sufficient
for our purposes -- the simulations resolve the critical
thermal length with a minimum of sixteen pixels and are stopped 
at an early enough stage that grid-scale
fragmentation has not occurred.  The analysis of BCDW09 further
demonstrates the lack of impact that higher resolution has
on the evolution of these simulations. {\hk As well, in Appendix~B,
we include our own analysis of higher resolution simulations
(512 by 512 cells), which demonstrate that the coarser resolution
simulations are adequate.}
We also note that 
while the geometry is simplified, and thus the turbulence and 
magnetic fields are idealized and unlike those expected in true three 
dimensional clouds, we are not attempting to prove any particular set 
of parameters as `true'. Our goal is less ambitious: to determine the 
effect of the input parameters on observables.

We note that there are other aspects of the input parameters that 
still remain to be properly investigated. In particular, whether 
turbulent energy does indeed decay or is driven \citep[see the recent
analysis of simulated dense core observables by][]{Offner08}. 
Other simulations 
have addressed this issue in order to determine the scale on which 
driving would need to occur to prevent collapse of cores
\citep[e.g.,][]{Klessen05}. 
This remains, however, an open question to be considered with respect to the 
competition between magnetic fields and turbulence.

\subsection{Details of the Simulations}
\citet[][hereafter CB06]{CB06} and BCW09 introduce and 
discuss in detail the setup and equations
governing the evolution of the simulation with linear perturbations
\citep[see also][]{BC04}.  The setup of simulations with turbulent
intial conditions are discussed in BCDW09.  Here, a few key details
are highlighted.  Note that the scalings applied to the simulation
in order to convert from scale-free to observable quantities are
given in Section 4.1.
The simulations we analyze have a thin sheet geometry and are
supported in the vertical direction by thermal pressure working
against gravity and magnetic pinching.  In the lateral direction,
the evolution is governed by the interplay of thermal pressure,
gravity, magnetic pinching and magnetic tension;
see CB06 for a discussion on the physical justification
and a detailed description of this geometry.  
The simulations
start with uniform column density and a constant vertical
magnetic field.  Velocity perturbations are added in Fourier 
space to wavenumbers $k$ corresponding to wavelengths up to the 
simulation length in both the x and y direction.  Velocity amplitudes 
are chosen from a Gaussian distribution scaled to the power spectrum 
$v_k^2 \propto k^{-4}$ (i.e., most of the power is on the largest scales).
The turbulent energy in these simulations is not driven but is allowed
to decay.  Unlike some purely turbulent simulations, gravity 
is always ``on''; there is no initial period where turbulence 
is driven without gravity.

The material in the simulation is bounded in the direction perpendicular
to the sheet by external pressure and gravity, as well as smaller
contributions from magnetic pressure and magnetic tension.  
The relative importance of the former two quantities is expressed by 
the dimensionless parameter
$\widetilde{P}_{\rm{ext}}$, the initial ratio of the external pressure to the 
self-gravitational pressure 
of the sheet in the vertical direction.  We analyze simulations
where the external pressure is minimal and the sheets are gravitationally
dominated ($\widetilde{P}_{\rm{ext}} = 0.1$), not unreasonable for
the denser regions within molecular clouds.  The other extreme, 
where the sheet is dominated by external pressure, would be
appropriate in a situation where star formation is triggered by a 
supernova shock wave, for example.  BCW09 analyzed a suite of simulations
with both weak and strong external pressures 
($\widetilde{P}_{\rm{ext}} = 0.1$ and 10) 
and found that high external pressure environments lead to the 
formation of much
smaller cores that attain high column density significantly
faster than their low external pressure counterparts.

Considering only thermal support, the simulations begin with
gravitationally unstable conditions.  
In the absence of external pressure, the simulations have
a length of eight times the critical thermal length, or
four times the lengthscale of maximum growth (see CB06 for further
discussion), thus approximately sixteen objects would be 
expected to form.
Additional support can be derived
from both the input magnetic fields and turbulent velocity.
We analyze simulations ranging from subcritical mass to magnetic
flux ratios
where there is no gravitational collapse possible without
ambipolar diffusion ($\mu_{0} = 0.5$ and 0.8), to critical ($\mu_0 = 1$), 
and supercritical ($\mu_0 = 2$ and 10)\footnote{We only analyze 
one simulation with $\mu_0=10$}.  
The turbulent velocity field input varies from having essentially 
no turbulent velocity (thermal Mach number of 0) to
a Mach number of 4 in integer steps. 
Note that the simulations with higher Mach numbers, particularly
Mach 4, showed significant qualitative differences from observations
as discussed in Section~4.4, hence simulations with even higher
Mach numbers were not run.

\begin{figure}[htbp]
\begin{centering}
\begin{tabular}{cc}
\includegraphics[width=8.5cm]{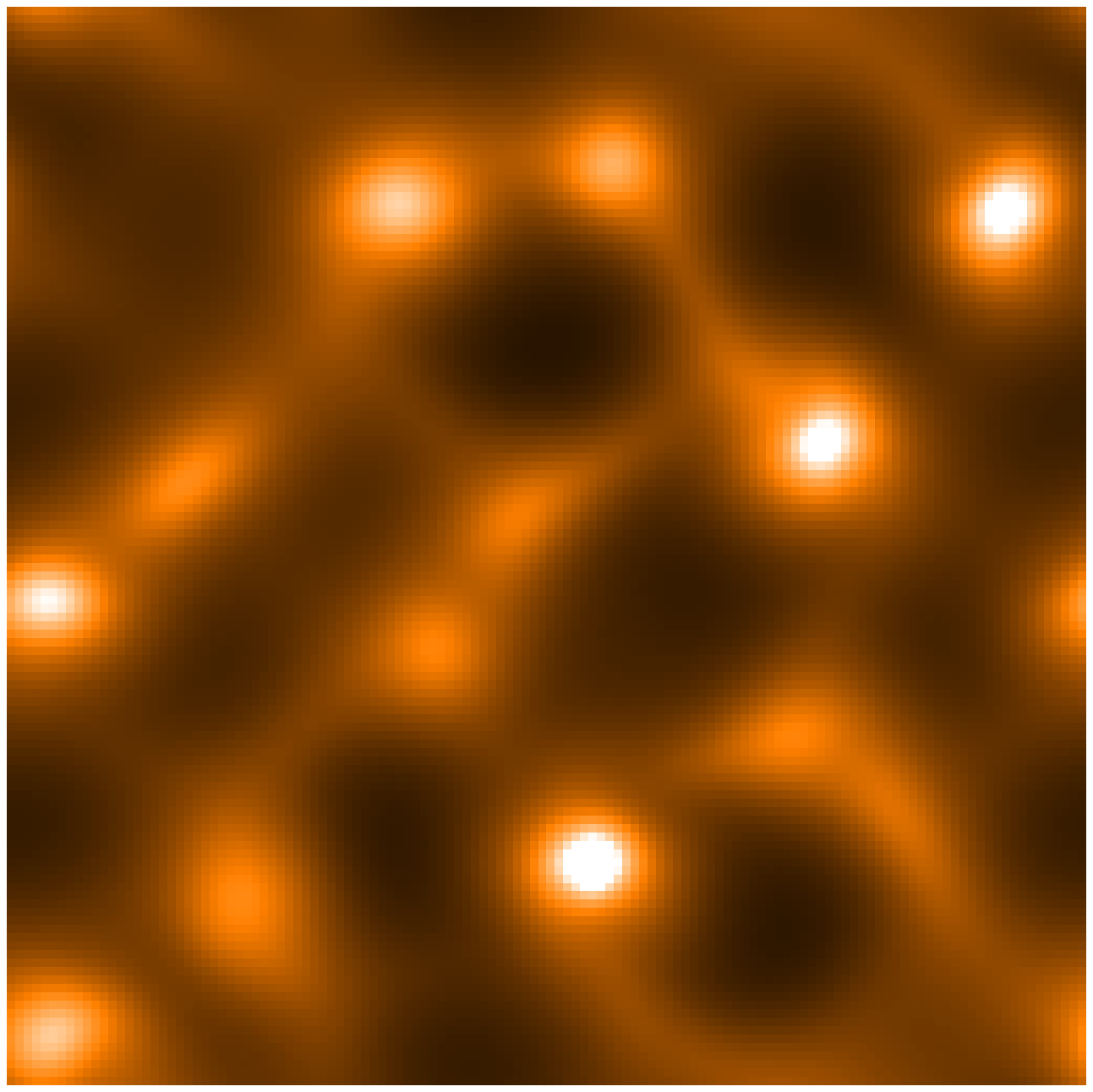} &
\includegraphics[width=8.5cm]{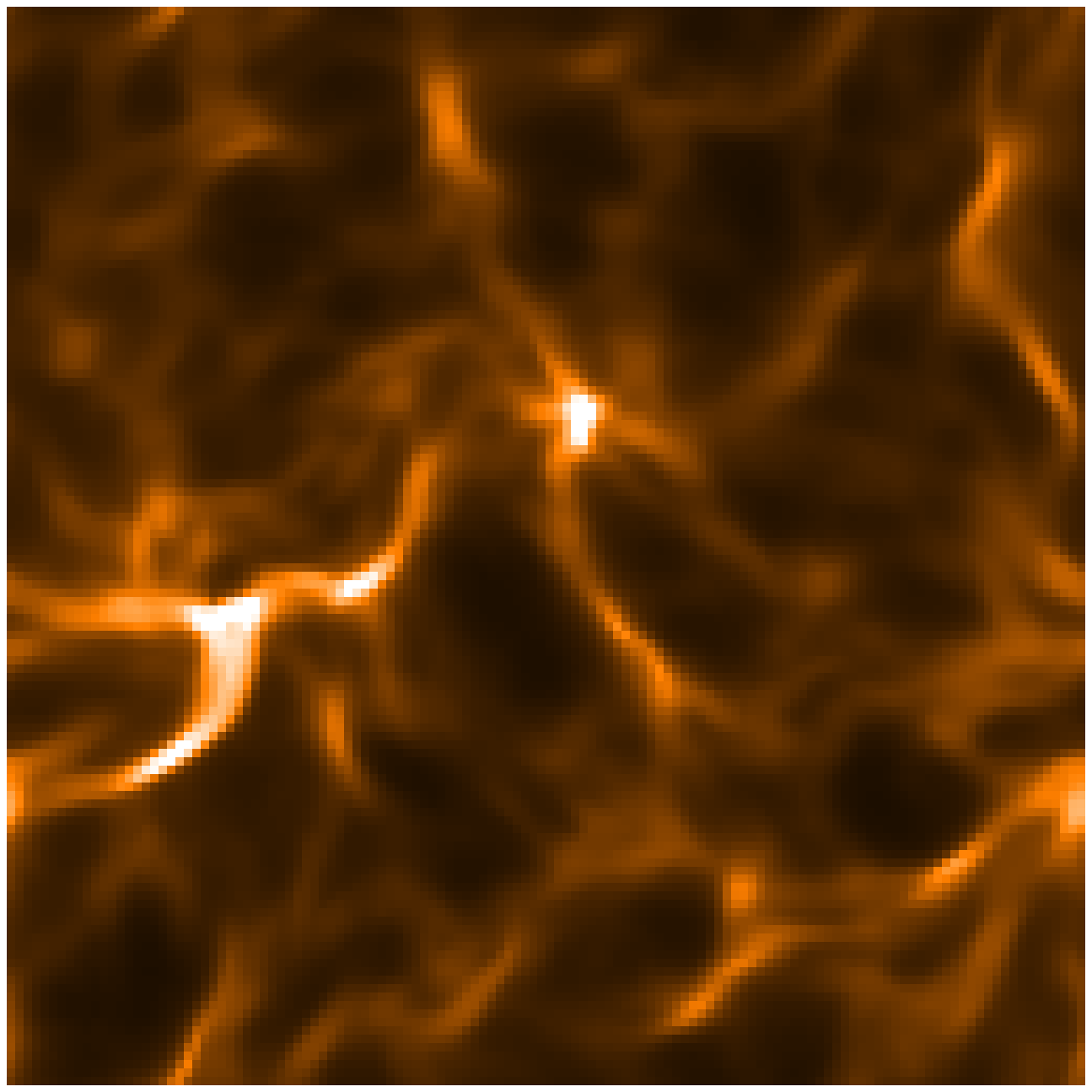} \\
\includegraphics[width=8.5cm]{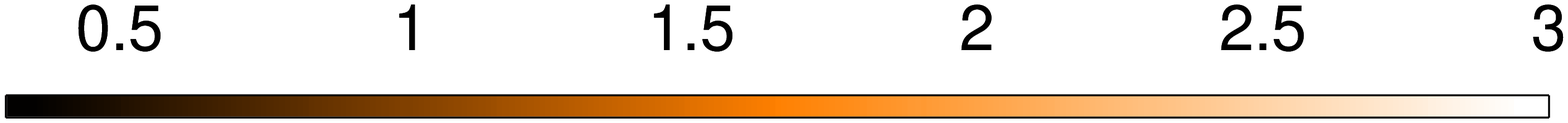} &
\includegraphics[width=8.5cm]{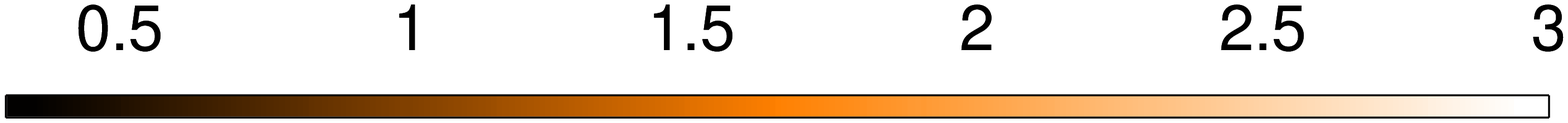} \\
\end{tabular}
\end{centering}
  \caption{The final column density in two simulations -- with 
	$\mu_{0} = 0.5, M = 0$ (left) and $\mu_{0} = 1.0, M = 3$ (right).  The 
	scale, in units of the initial column density, is shown in 
	the bar.  Under the scalings we applied, a value of 1 corresponds
	to 3.5$\times 10^{21}$~cm$^{-2}$.}
\label{fig_sims}
\end{figure}

The simulations are evolved until a column density of
ten times the initial uniform value is reached.  
Figure~\ref{fig_sims} shows two sample column 
density maps from the suite of simulations we analyzed.  In simulations 
where the pressure from the magnetic field exceeds the pressure from
the input turbulent motions, the magnetic field has a 
strong influence on the dynamics of the simulation.  In particular, 
turbulent compressions are moderated by the restoring force of the 
magnetic field, hence density enhancements are smaller and tend to 
rebound rather than collapse due to gravity.  When the magnetic 
fields are strong, the simulations thus require longer periods of time to
reach a column density sufficient for the simulation to stop.
The input turbulence is therefore able to
decay to a much greater extent in these simulations than 
in turbulence-dominated ones.  This causes the large-scale velocity
dispersions to be smaller in the strongly magnetic simulations
than their weakly magnetic counterparts, as seen in the analysis
below.  We postpone a full analysis of how the turbulent velocity
field changes over the run of the simulation in these cases for 
a future paper.


\section{OBSERVATIONS}
\label{s_obs}

\subsection{Summary of Previous Results}
\label{s_obs_sum}
Previously, we conducted a large-scale pointed survey of cores and their
dynamics in the Perseus molecular cloud \citep{Kirk07}.  
The Perseus 
molecular cloud is an intermediate mass star forming region located 
at a distance of
$\sim$250~pc \citep[e.g.,][]{Cernis93} that displays more active and clustered
star formation than seen in the low mass Taurus star forming region, but much
less than seen in the high mass Orion star forming region.  Our survey was
thus ideal for probing the conditions of a more clustered mode of 
star formation without suffering from the complexity present in a region
such as Orion.  
The dense core candidates were selected from cores
observed in the submillimetre continuum using SCUBA \citepalias{Kirk06}
and supplemented with visually-selected regions of higher extinction
in the Palomar plates (and large-scale 2MASS extinction peaks).
In our survey, we made single-point observations of 
dense core candidates simultaneously in \nhns(1-0) and \cons(2-1), 
with beamsizes of 25\arcsec\ and 11\arcsec\ respectively, using the 
IRAM 30~m telescope in Spain.  \nh is a good tracer of dense gas,
with a critical density of $\sim10^{5}$~cm$^{-3}$ \citep{Tafalla02}, 
while \co traces lower density gas \citep[critical density of
$\sim10^{3}$~cm$^{-3}$][]{Ungerechts97} and freezes out of the
gas phase onto dust grains at $\sim10^{5}$~cm$^{-3}$ \citep{Tafalla02}, 
hence the
survey obtained information simultanenously about the dense core
and the surrounding `envelope' of material.  
We differentiated between protostars and starless cores on the
basis of Spitzer data \citep{Jorgensen07}.  Supplemental information
is also available from the COMPLETE Survey \citep{Ridge06} --
large scale column density measures from stellar reddening observed
with 2MASS (Alves \& Lombardi in prep) using the NICER technique
\citep{Lombardi01} as well as
\twelco and \thirco FCRAO observations across the cloud \citep{Ridge06}.

The survey of \citetalias{Kirk07}, 
based on submillimetre dust emission and
optical / near-IR dust extinction - identified cores
provides a useful set of dynamical
constraints 
that simulations of star formation should match; we
highlight several which are relevant to this work. 

1) {\it Cores have little internal turbulence.}  Starless cores have velocity
dispersions which are dominated by thermal motions; protostars tend to
have slightly larger (but still thermally-dominated)
velocity dispersions.  This result has
been found by many other surveys including \citet{Benson89} and
\citet{Jijina99}.

2) {\it Material surrounding cores has a larger velocity dispersion.}  The 
less-dense material (traced by \co in our survey) surrounding the dense 
cores tends to display
much larger non-thermal motions, also consistent with previous 
observations \citep[e.g.,][]{Benson89,Jijina99}.

3) {\it The relative velocity between the core and 
surrounding material
is small.}  The vast majority of dense cores (nearly 90\%) have 
less than thermal motions with respect to their surrounding material.
Previous observations \citep{Walsh04,Walsh07} concur with this result.

\subsection{Additional Observations}
\label{s_obs_obs}
In order to compare our previous results with the analysis of 
simulation data below, we must fold in a few additional pieces of
information about larger scales.  First, we note that the dense cores
in Perseus are clustered within larger-scale regions, as seen in
the 2MASS extinction map.  We defined these regions in \citetalias{Kirk06} 
(referred to there as `extinction super cores' and here as
`extinction regions' for clarity); here we adopt the size and mass
estimates of these regions from \citetalias{Kirk06} 
(typically 10\arcmin\ and several hundred solar masses; 
see Table~\ref{ext_regs}).  
These regions have
roughly comparable sizes and masses to that found in the simulated
box as discussed in the following section.

Additionally, we use the \thirco data from COMPLETE \citep{Ridge06} 
to gather
information on the dynamics of each extinction region.  While the
spatial resolution of this data cube is lower (46\arcsec) than our IRAM \co 
data, it has the advantage of uniformly covering each
extinction region rather than sampling a set of sparse and 
biased locations.  We calculate the velocity dispersion along
each line of sight by fitting a single Gaussian to each \thirco spectrum.
Additionally, we calculate the total velocity dispersion seen in each
extinction region by summing all of the spectra within the extinction
region and fitting the resultant spectrum with a single Gaussian.
The \thirco data cube has a smaller areal coverage than the extinction
map thus we only consider the eight of eleven extinction regions
that have $> 80\%$ coverage in the \thirco data cube.  
For each extinction region, Table~\ref{ext_regs} gives the
region number, the mass, the radius, the velocity dispersion 
required to prevent gravitational
collapse (in Gaussian sigma units), the observed velocity dispersion 
in \thirco (in Gaussian sigma units), the percentage
of the region observed in \thircons, and a descriptive location. 
For those extinction regions which have at least 80\%
coverage in the \thirco data, Table \ref{obs_effic} gives the
observed \thirco velocity dispersion relative to the sound speed
(assuming a temperature of 15~K),
the number of dense cores identified in the SCUBA data of
\citetalias{Kirk06}, the number of dense cores observed in the IRAM
data of \citetalias{Kirk07}, and the estimated percentage of
the region's mass found in each set of dense cores 
(or core formation efficiency; the calculation
of the number and mass of dense cores in each region is discussed in
more detail in \S5.4).

\section{`OBSERVING' THE SIMULATIONS}
\label{s_obssim}

\subsection{Scalings}
\label{s_obssim_sca}
The simulations are scale-free, with units in terms of the mean
density and sound speed.   To convert to observable quantities, 
we adopt the nominal pressure scaling of \citet{CB06}, a density
scaling roughly three times their nominal density,
plus the temperature and distance we previously adopted for
our observations of the Perseus molecular cloud in \citetalias{Kirk06} and
\citetalias{Kirk07},
i.e., :
\begin{eqnarray}
	n_{n,0} = 10^{4}~{\rm cm}^{-3} \\
	T = 15~{\rm K} \\
	D = 250~{\rm pc}
\end{eqnarray}

Recall that we analyze simulations where the sheet is gravitationally
dominated with minimal external pressure, i.e,
\begin{equation}
	\widetilde{P}_{\rm{ext}} = 0.1
\end{equation}

Using these scalings, the box length of the simulated region is 1.5~pc 
corresponding to an
angular size of 21\arcmin\ at the distance of the Perseus molecular
cloud.  This is comparable to the sizes of the
largest structures identified in the extinction maps of \citetalias{Kirk06}
(see Table~\ref{ext_regs}).
Each pixel in the simulation is 0.012~pc, corresponding to an 
angular size of 9.8\arcsec, somewhat smaller than the beamsize
of our SCUBA and IRAM observations.  The mean initial column density
is 3.5 $\times 10^{21}$~cm$^{-2}$, corresponding to a value of 1 in the
scale-free simulations.  If we adopt a dust opacity of 
$\kappa_{850} = 0.02$~cm$^{2}$~g$^{-1}$ and a dust temperature 
of 15~K (and assume a dust-to-gas ratio
of 1:100) for our SCUBA observations, then 1~Jy~bm$^{-1}$ corresponds 
to $N_{H} = 6.37 \times 10^{22}$~cm$^{-2}$ and the mean column density in the
simulation corresponds to 55~mJy~bm$^{-1}$~\footnote{Note,
however, that since the SCUBA maps are created from chopped observations,
this constant component would not be seen in the data.}.  

The simulations are stopped when the maximum {\it column} density in the
simulation reaches approximately a factor of 10 times the initial
value, or $\sim 3.5\times10^{22}$~cm$^{-2}$ (corresponding to a density of
$\sim 10^{6}$~cm$^{-3}$).  In simulations with strong turbulence and 
weak magnetic fields, this maximum is reached quickly, 
while in simulations with strong magnetic fields,
more time is required to reach the maximum.  
The time each simulation runs is given in Table~\ref{tab_time}
in units of $t_{0}$ :
\begin{equation}
t_{0} \equiv \frac{c_{s}}{2\pi G \sigma_{n,0}}  
\end{equation}
where c$_{s}$ is the sound speed, G the gravitational
constant, and $\sigma_{n,0}$ the initial column density in the
simulation (see CB06).
The unit of $t_{0}$ is related to the sound crossing time of the sheet -- 
when thermal pressure is the dominant term in determining the 
thickness of the sheet, we can write 
\begin{equation}
t_{0} = \frac{Z_{0}}{2c_{s}}
\end{equation}
(using eqn 13 of CB06), where Z$_{0}$ is the initial half-thickness of
the sheet.  Thus $t_{0}$ is approximately one quarter the sound
crossing time of the entire thickness of the sheet.  In their
linear analysis, CB06 found that away from the critical mass to magnetic 
flux ratio
($\mu_{0} = 1$), the lengthscale for maximum growth of instabilities
was 2$\lambda_{T}$ where $\lambda_{T}$ is the critical thermal length
scale given by
\begin{equation}
 \lambda_{T} = \pi 
	\Big(\frac{1 + 3\widetilde{P}_{\rm{ext}}}{1+\widetilde{P}_{\rm{ext}}}\Big) Z_{0} 
\end{equation}
(CB06 eqn 40), or $\lambda_{T} \simeq \pi Z_{0}$ for a low external
pressure environment.  Thus the sound crossing time across the lengthscale
of maximum growth is $4\pi t_{0}$.  Table~\ref{tab_time} shows that
in the non-turbulent simulations, the run-time for the weakest
magnetic field cases is roughly equal to this timescale.  In the
highly turbulent simulations, the run-time can reach a few tenths of
$t_{0}$, much shorter than even the turbulent crossing timescale of
the linear lengthscale of maximum growth.  This implies that in these
highly turbulent simulations, the resultant peaks are due to 
compressions directly from the input initial conditions, 
as there has not been sufficient time for gravitational information 
to have propagated across the relevant lengthscales.
Note that the entire length of the simulated box is
16$\pi c_s t_0$ or 8 $\pi Z_0$, corresponding to four times the
lengthscale of maximal growth for a low pressure environment with
a mass to magnetic flux ratio away from critical.  All of the
turbulent simulations are therefore stopped well before the thermal
crossing time of the box, $16 \pi t_{0}$.

\begin{figure}[htb]
\begin{centering}
\begin{tabular}{cc}
\includegraphics[width=8.5cm]{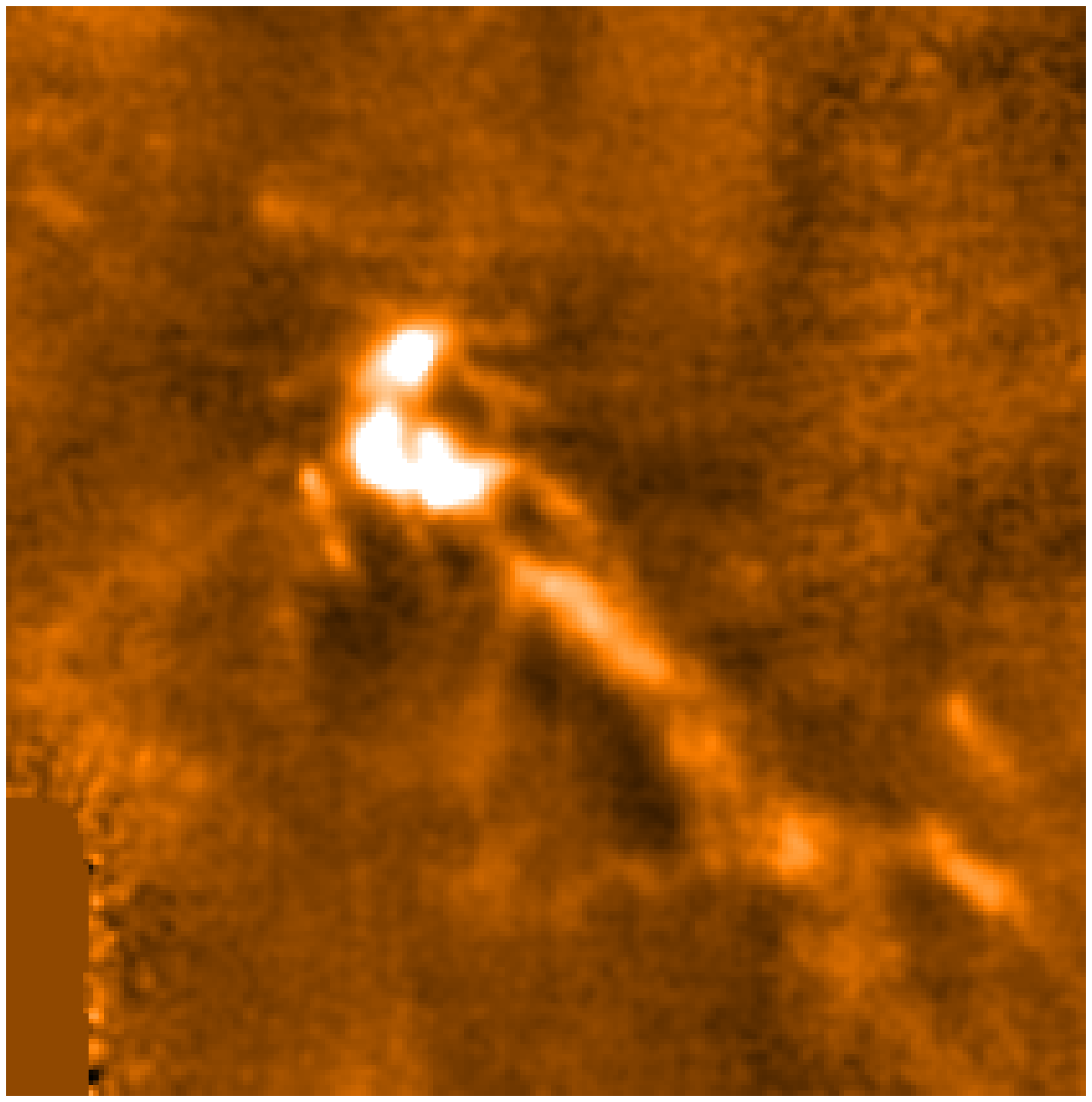} &
\includegraphics[width=8.5cm]{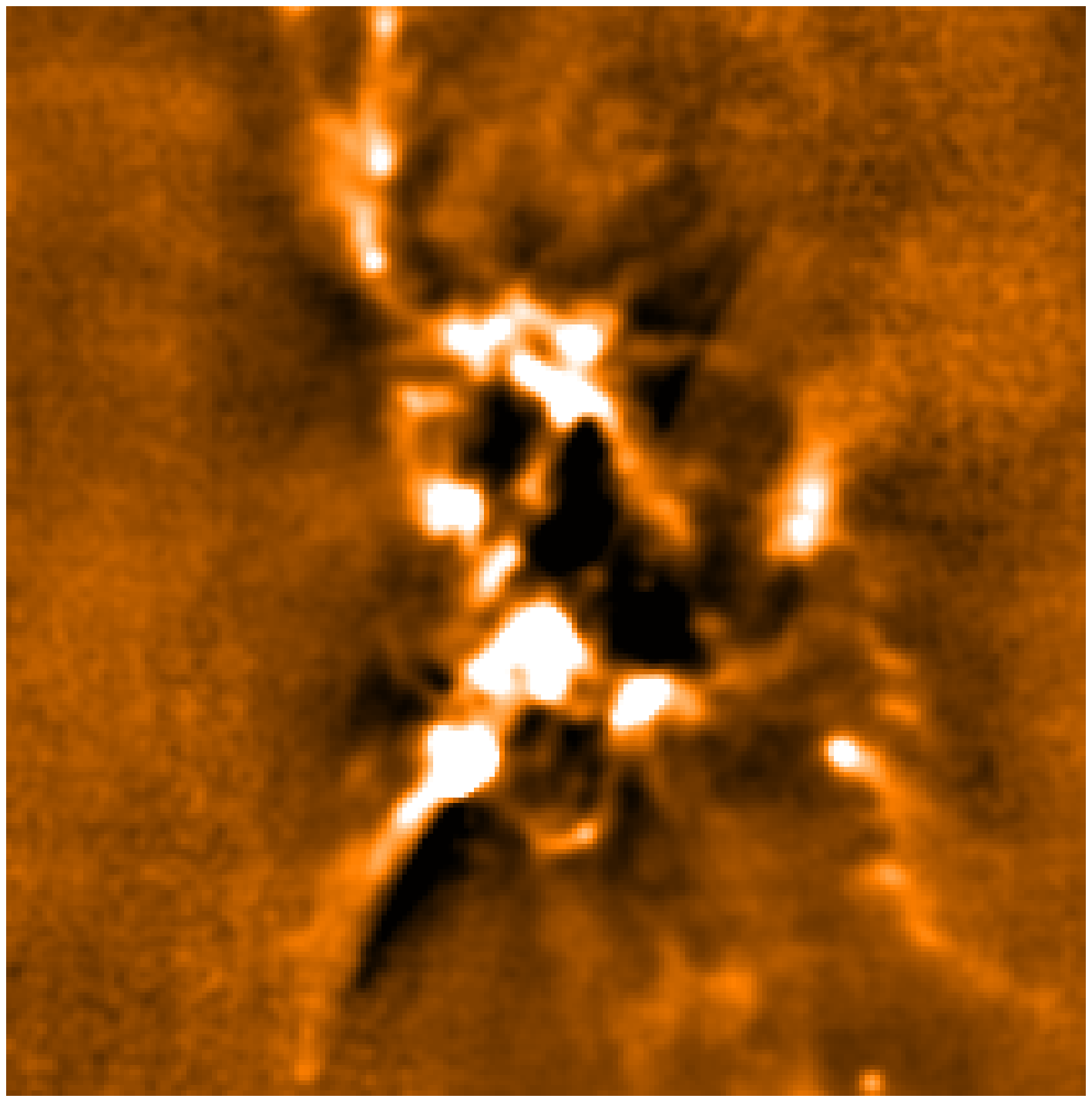}\\
\includegraphics[width=8.5cm]{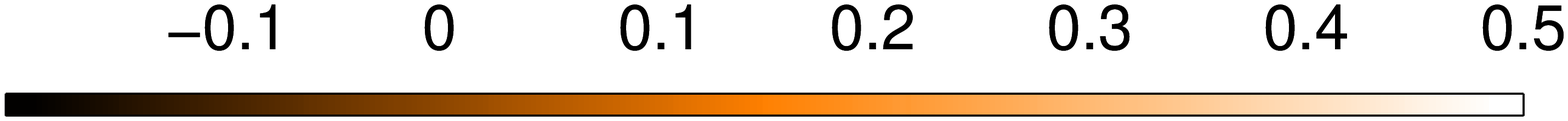}&
\includegraphics[width=8.5cm]{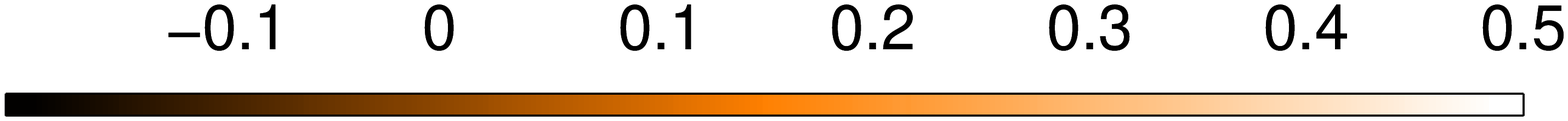}
\end{tabular}
\end{centering}
  \caption{SCUBA submillimetre maps of two well-known star forming
	regions -- B1 and NGC1333.  Each image spans 21\arcmin\ ($\sim$1.5~pc)
	and the scale bar is in units of Jy~bm$^{-1}$.}
\label{fig_obs_vs_sim}
\end{figure}

\begin{figure}[h]
\begin{centering}
\begin{tabular}{cc}
\includegraphics[width=8.5cm]{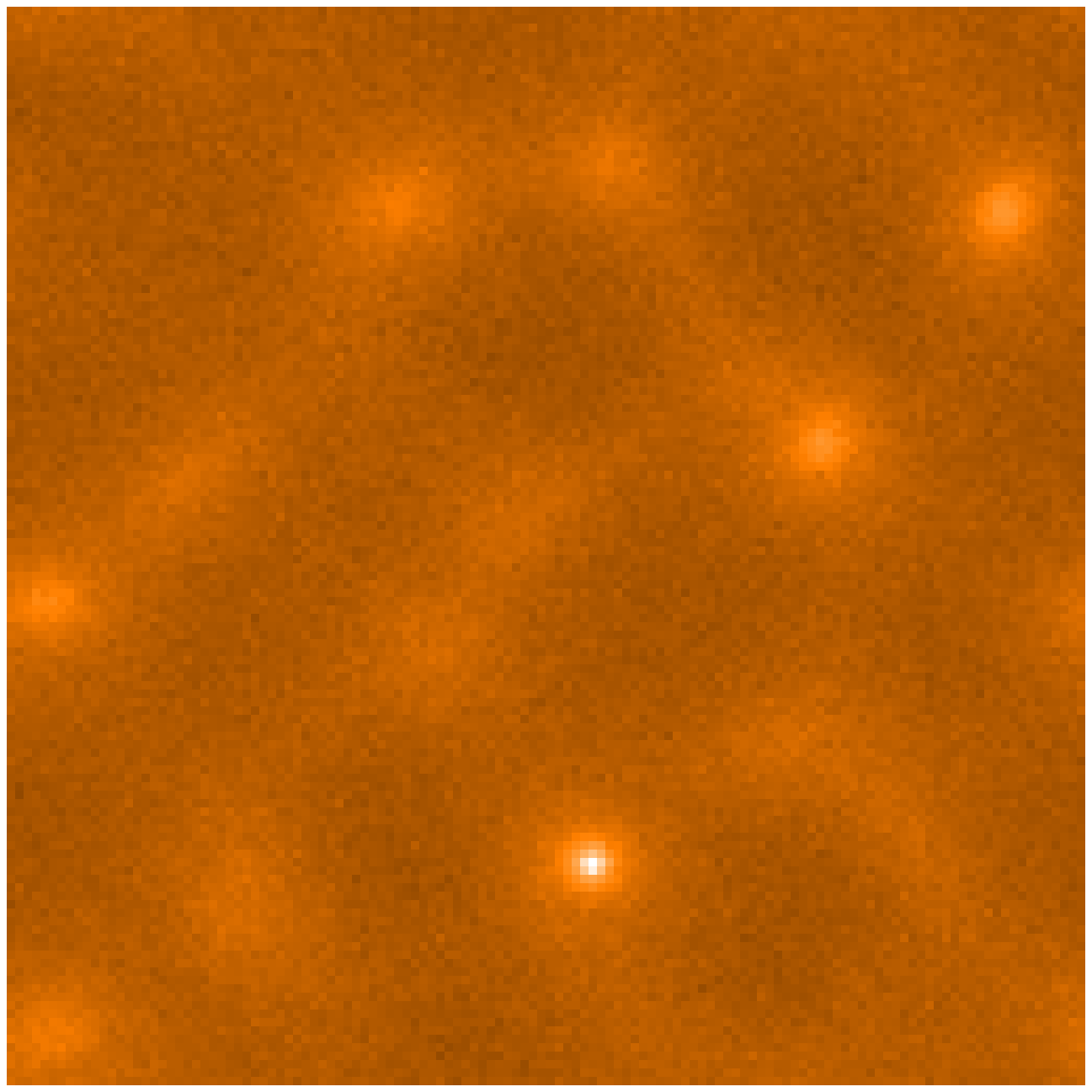} &
\includegraphics[width=8.5cm]{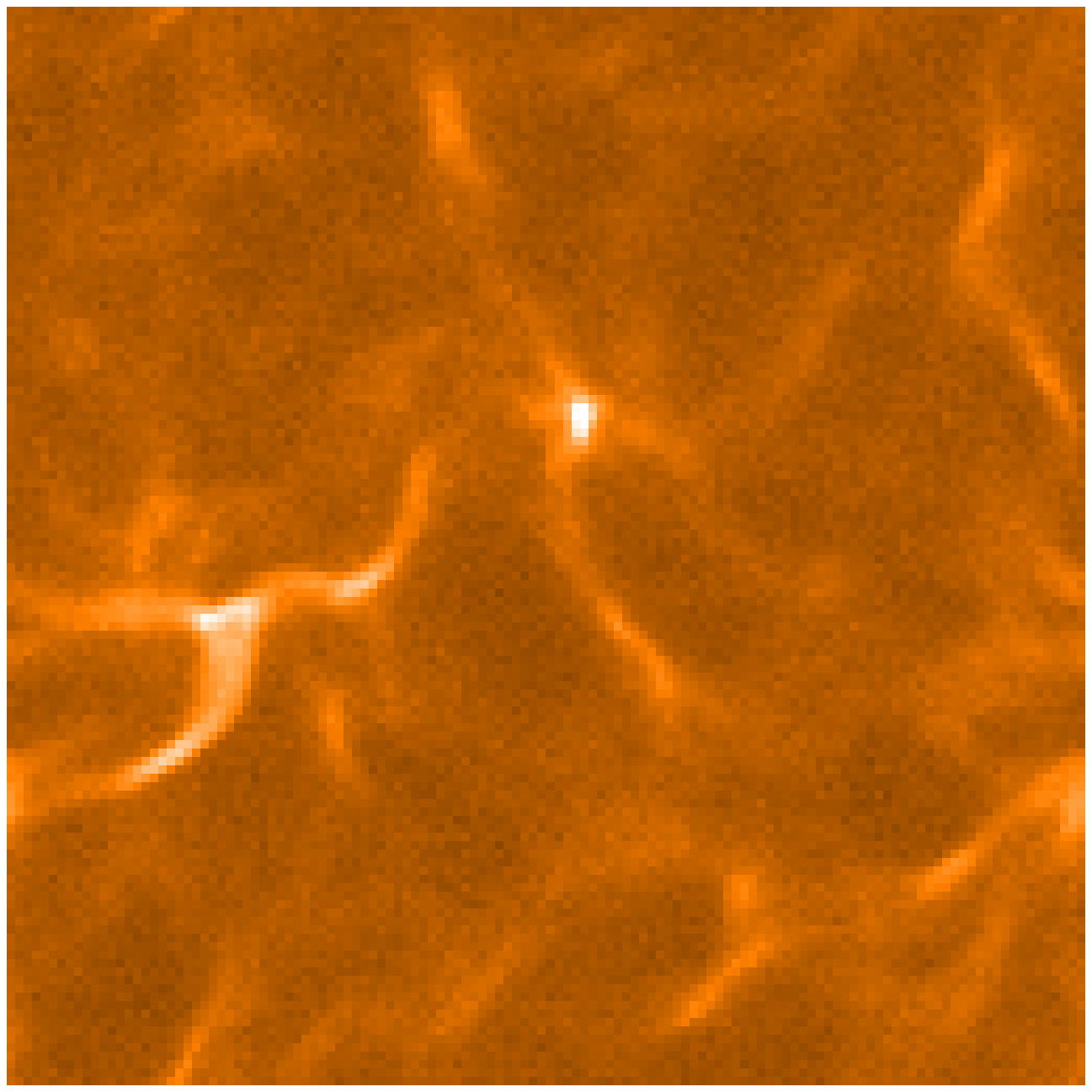} \\
\includegraphics[width=8.5cm]{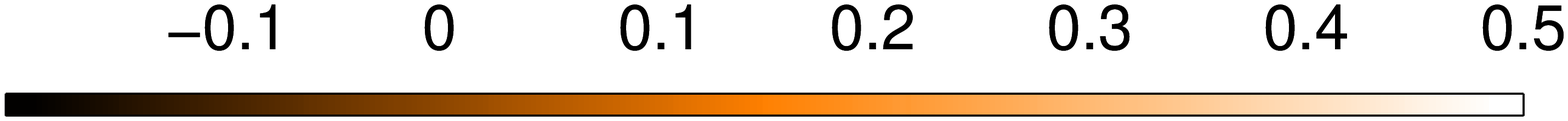} &
\includegraphics[width=8.5cm]{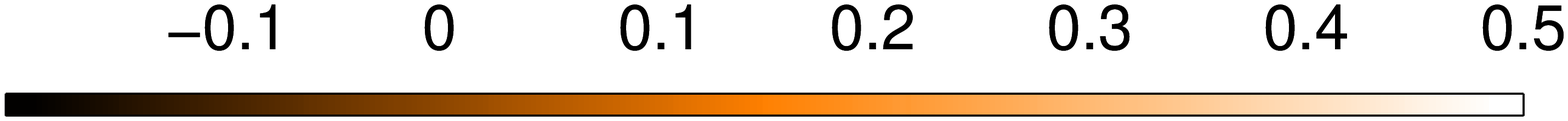}
\end{tabular}
\end{centering}
  \caption{The same simulations shown in Figure 1, with the column
	density converted to equivalent SCUBA flux.  White noise has
	been added at approximately the same level as the noise in
	the SCUBA observations.  The linear size and
	scale bar are the same as in Figure 2.}
\label{fig_obs_vs_sim2}
\end{figure}

Figure~\ref{fig_obs_vs_sim} shows several examples of well-known
star forming regions in Perseus observed in the submillimetre
while Figure~\ref{fig_obs_vs_sim2} shows the same example 
simulations as shown in Figure~\ref{fig_sims} but scaled to the 
same flux range and size as the observations.
White noise has been added to the simulations at approximately
the same level as is present in the observations.

\subsection{Identification of Simulated Cores}
\label{s_obssim_id}
\begin{figure}[htb]
\plottwo{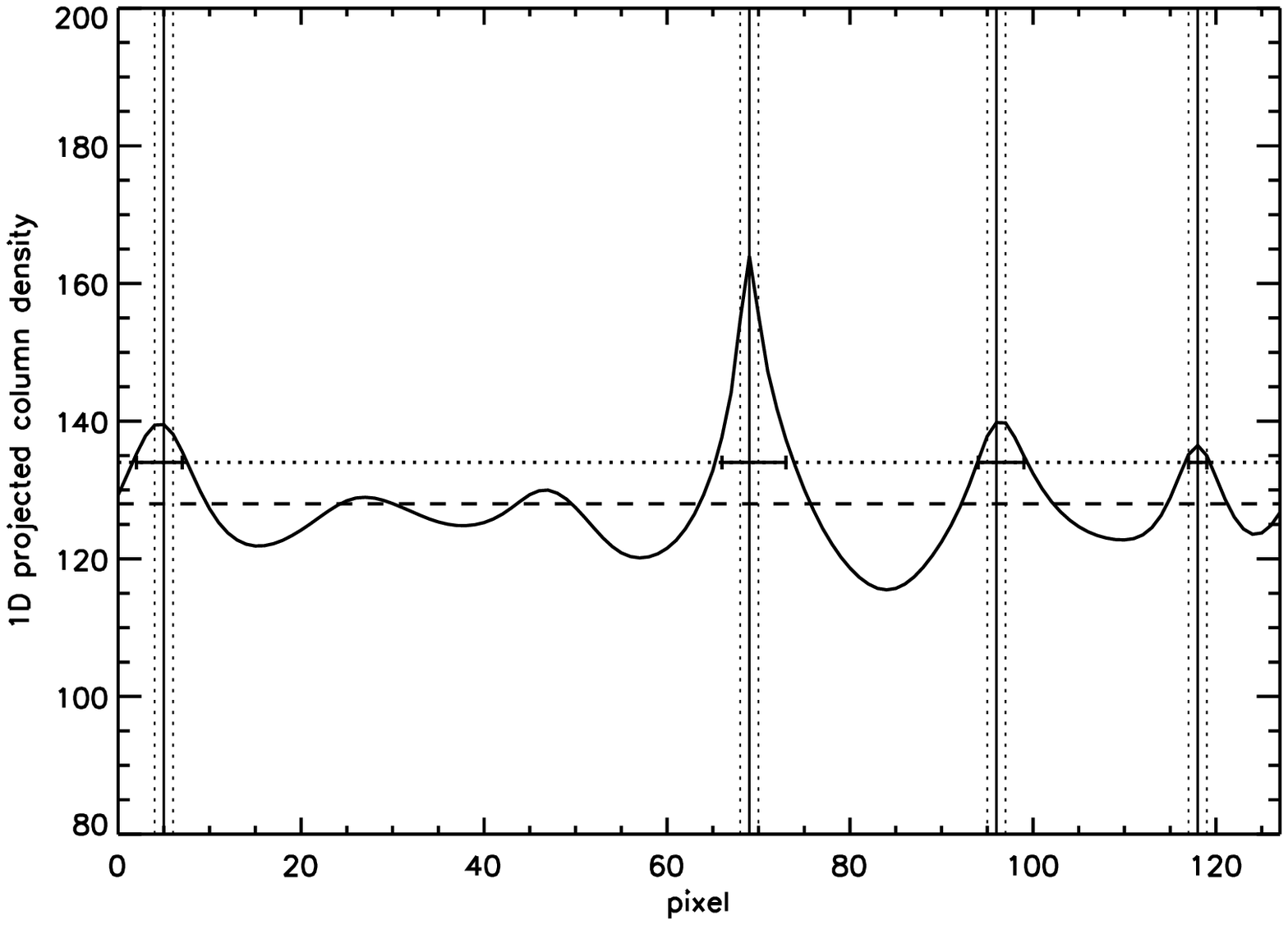}
	{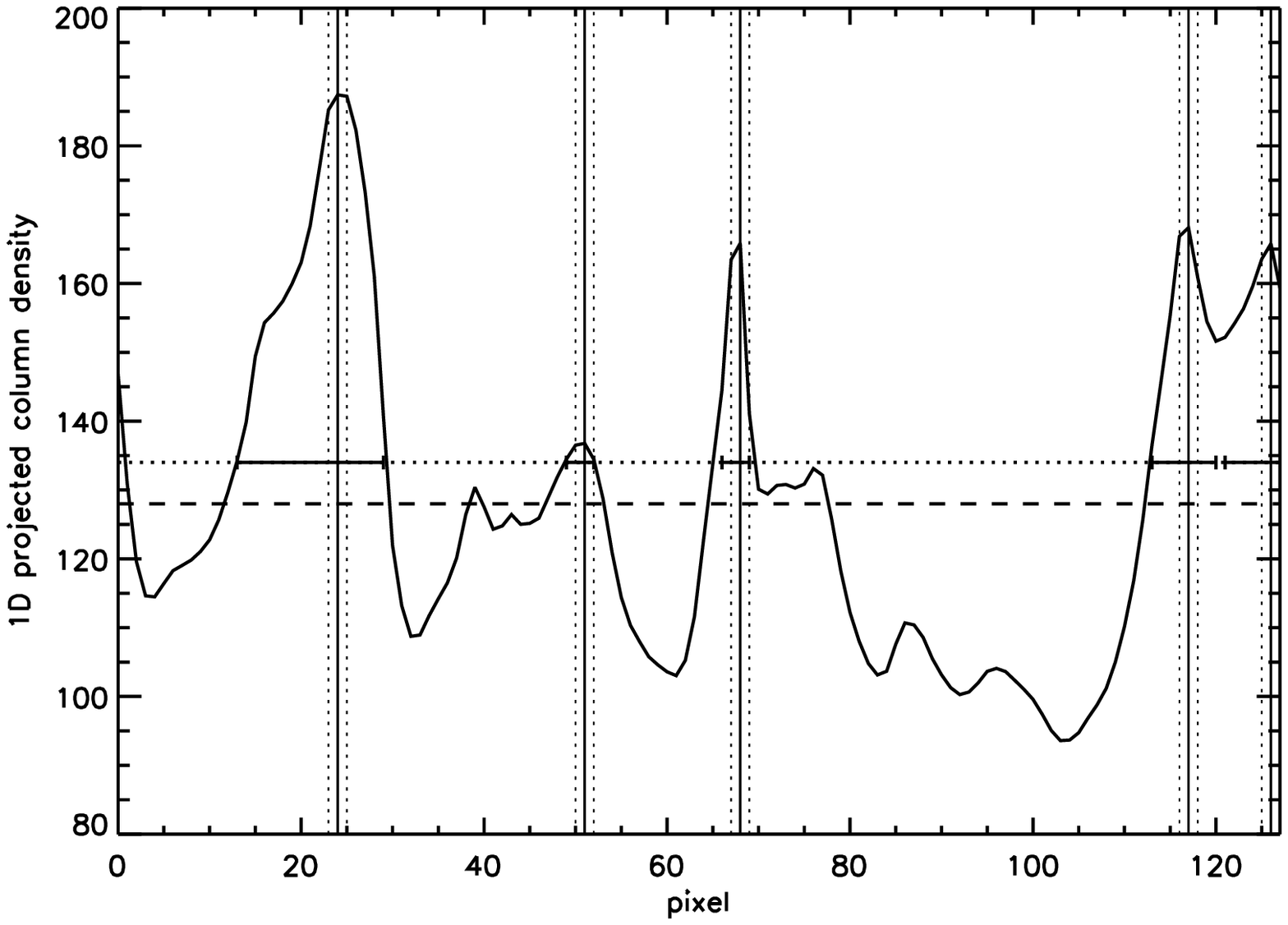}
  \caption{Projected 1D column density distribution (in the x direction) for the
	simulations shown in Figure~\ref{fig_sims} 
	($\mu_{0} = 0.5$, $M = 0$ on the left and $\mu_{0} = 1.0$, $M = 3$
	on the right).  Using our
	assumed scalings, a 1D column density value of 100 corresponds
	to $3.5\times10^{23}$~cm$^{-1}$.
	Cores are identified as peaks above
	a specified threshold and separated by at least three pixels
	corresponding to roughly 29\arcsec or one IRAM \nh beam.  
	The dashed horizontal line indicates  
        the mean projected column density while the dotted horizontal line 
	indicates  
        the threshold for core identification.  The solid horizontal
	lines indicate the extent of each core identified while the
	solid and dotted vertical lines indicate the core centres and
	width respectively.}
  \label{1dpeaks}
\end{figure}

Direct comparison can be made between the column density structure
found in the simulations and observations, since both are two-dimensional 
projections.
This is not, however, the case for comparing dynamics.  Since the
velocity is only calculated in the two dimensions of the simulation,
the simulations must be considered in a one dimensional projection
along x or y in order to obtain the motion towards or away from
an observer.  We first project the column density map in the x and
y directions to identify density peaks.  Since
the simulated sheet is thin, we do not account for changes in
the projected column density due to variations in the thickness
of the sheet.  Note that the
locations of peaks in the one dimensional projections tend to
have good correspondence with overdensities in the original
two dimensional column density map, although there can be differences
where two structures lie along the same line of sight (particularly
for filamentary structures seen in the highly turbulent simulations).
Figure~\ref{1dpeaks} shows examples of the projected 1D column
density.

To identify cores, we find all the peaks in the projected column
density which lie above a specified threshold and are separated by
at least one IRAM \nh beam.  We set the minimum peak threshold
to a value corresponding to the typical column density threshold
we would expect in two dimensions -- assuming a typical core spans
little more than an IRAM \nh beam (3 pixels), and has a minimum
column density threshold of 3 times the mean (which, 
for particles uniformly distributed along the column corresponds to
a density of 9 $\times 10^{4}$~cm$^{-3}$, roughly the critical
density of \nhns), then the minimum
1D column density is $(N-3)\times 1\sigma_{0} + 3\times3\sigma_{0}$,
where $\sigma_{0}$ is the mean column density (1) and $N$ is the number 
of pixels in the simulation in 1D (128).
This core identification is designed to mimic our observational
method as best as is possible (\nh targets in KJT07 were based on dust column
density measurements, primarily from SCUBA data).  Note that the results
in our later analysis show little variation with any changes in the
peak identification parameters (see Appendix~\ref{s_care} for
further details).

\subsection{Calculation of Simulated Spectra}
\label{s_obssim_cal}
\begin{figure}[htb]
\plottwo{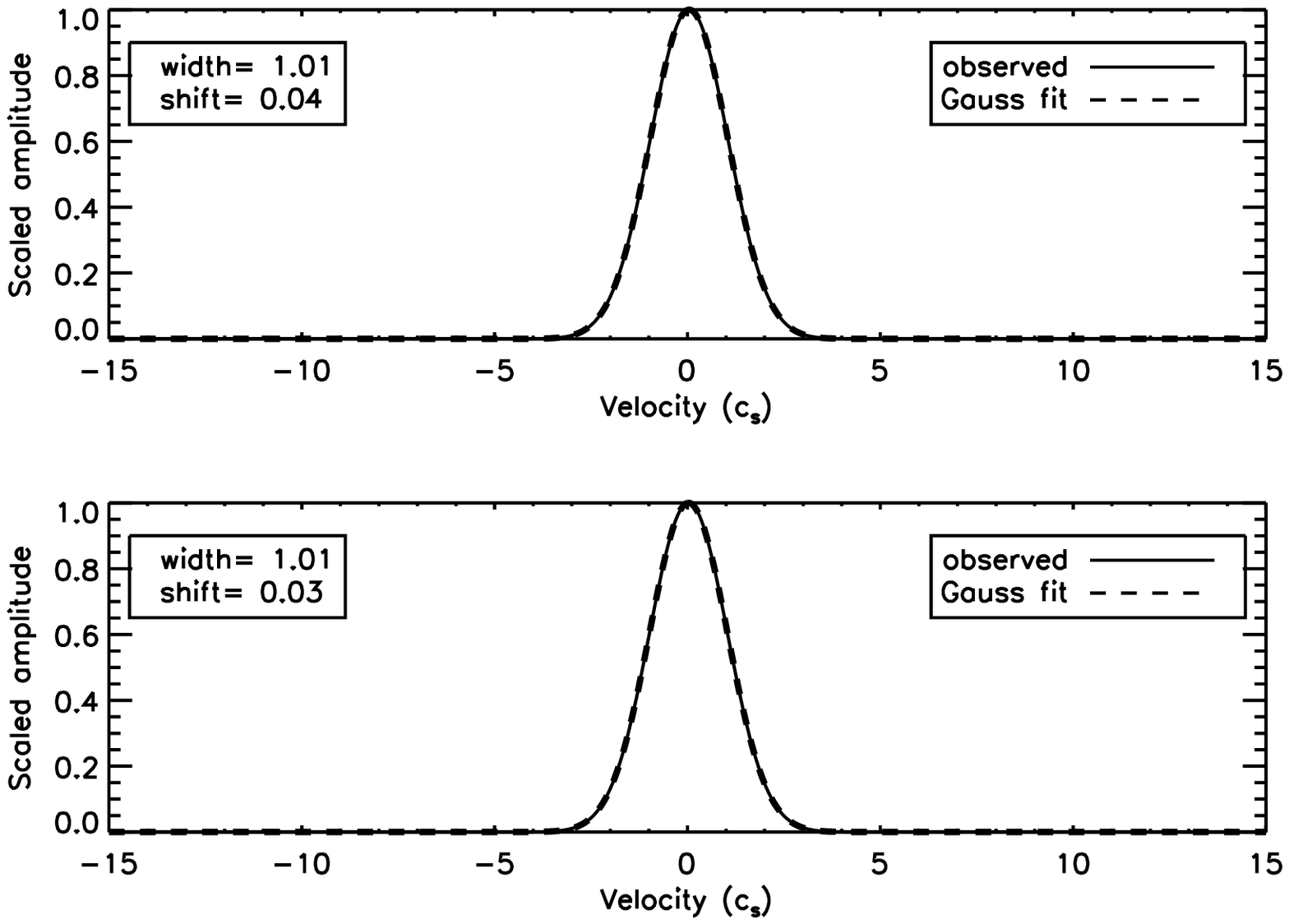}
	{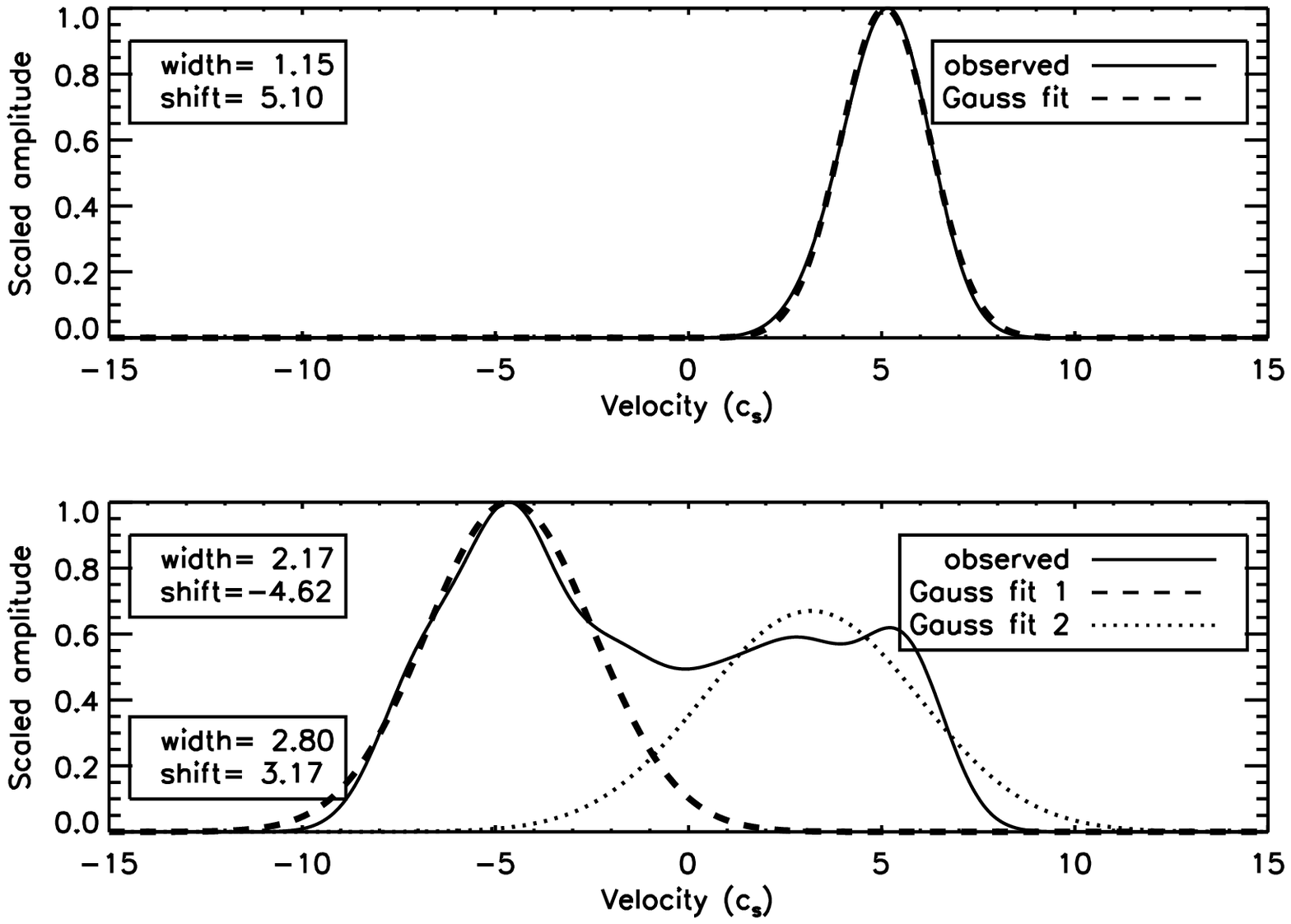}
  \caption{Two examples of spectra calculated for the core and LOS LDG material.
	The solid lines show the calculated (`observed') spectra 
	while the dashed and dotted lines show the model (Gaussian) 
	spectral fit.  The top panels show the core spectra while the 
	bottom panels show the LOS LDG material spectra.
	The left figure shows an example from a non-turbulent simulation
	with a subcritical mass to magnetic flux ratio 
	($\mu_{0} = 0.5$, $M = 0$) where the spectra
	are well-described by a single Gaussian.  The right figure 
	shows an example from a highly turbulent simulation with a
	critical mass to flux ratio ($\mu_{0} = 1$, $M = 3$) 
	where the LOS LDG spectrum is less well-fit by even a two Gaussians.}  
  \label{spectra}
\end{figure}

As described in \S2, our IRAM survey consisted of pointed observations
of dense cores in two molecules -- \nhns, which traces the dense
gas, and \co which only traces the less dense gas.  For
every peak identified in the simulations, we can
calculate spectra corresponding to both of these tracers.
For each projected 1D peak, we consider 
the material
along the corresponding line of sight and split it into the dense
material (cells with values above 3 times the mean
column density), corresponding to the material which \nh
traces, and less dense material (cells with values
less than 3 times the mean column density), corresponding to
the material which \co traces.  
This threshold corresponds to a 
mean density of $9\times10^4$~cm$^{-3}$ or nearly $10^5$~cm$^{-3}$.
We then calculate the spectra by assuming that 
each cell emits a thermally broadened Gaussian centred
on the velocity of that cell and weighted by the amount
of material (column density) present in that cell.
For clarity, we term the two spectra calculated for every 1D column density
peak as the core (high density gas) spectrum, which corresponds to the
\nh dense core spectrum, and the LOS LDG (line of sight low density gas)
spectrum, which corresponds to the \co `envelope' spectrum.
Figure~\ref{spectra}
shows two example sets of core and LOS LDG spectra.

\begin{figure}[htb]
\plotone{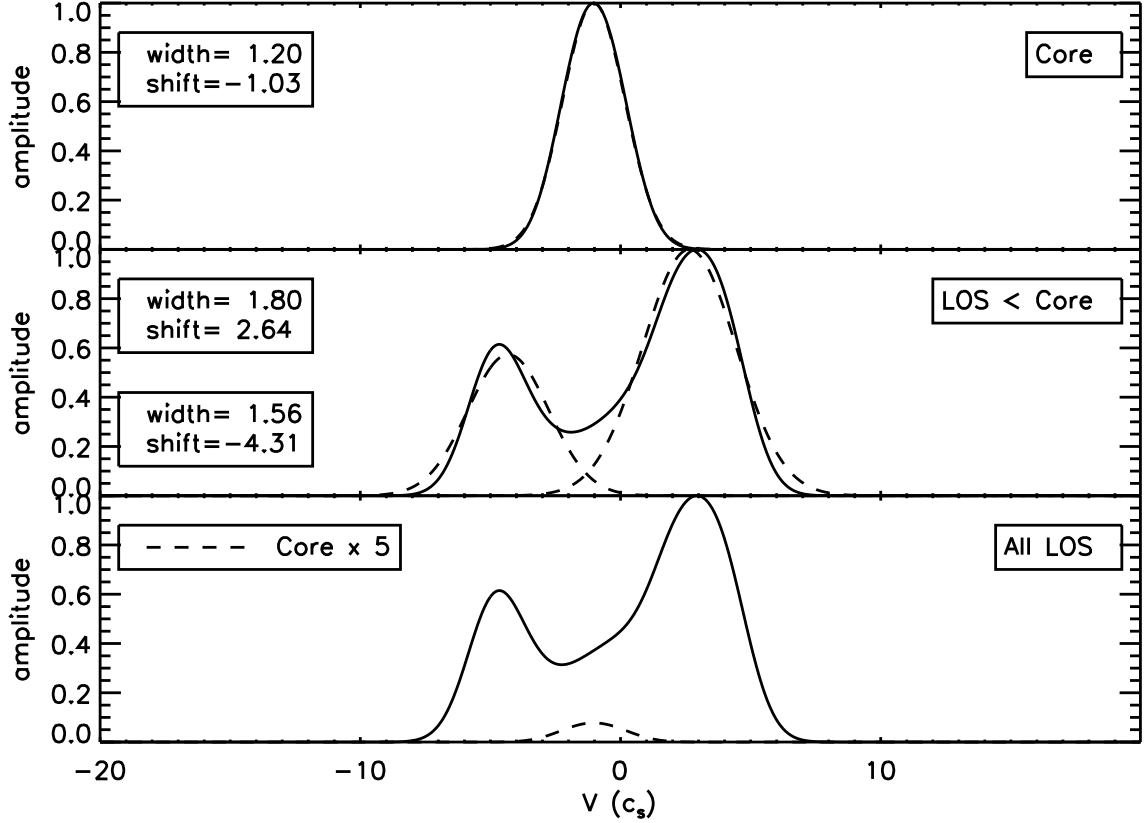}
  \caption{The spectrum calculated for a core identified in a simulation
	with a subcritical mass to magnetic flux ratio ($\mu_{0}$=0.8) and high
	input turbulence ($M$=3).  This core was found to have supersonic
	motion between the core and LOS LDG material.
	The top panel shows the normalized spectrum of the core, with
	the best fit Gaussian shown by the dashed line (the fit
	parameters are given in the legend).  The middle panel
	shows the normalized spectrum for the LOS LDG material around the core,
	with the two Gaussian fits indicated by the 
	dashed lines.  The lowest panel shows the normalized spectrum that
	would have been calculated for the LOS LDG material had all
	of the material along the LOS been used (rather than
	excluding that above the core detection threshold), as well as
	the core spectrum multiplied by a factor of five for
	visibility.  There is little difference between this spectrum 
	and the one used in our analysis (middle panel), indicating that 
	our analysis is little effected by the inclusion of the core material
	in the LOS LDG spectrum (as also discussed in \S\ref{s_obssim_cal}).}
  \label{appendix_1}
\end{figure}

As an aside, we note that the LOS LDG spectra are little affected
by the exclusion of material above a column density
threshold, rather than all material along the line of sight.
The physical basis for this is that the majority
of the mass is found in the lower density material, so the behaviour
of the high density material has very little effect on a cumulative
spectrum.  
 This is demonstrated in Figure~\ref{appendix_1} which shows
spectra of the core (top panel), the LOS LDG material
used in our regular calculations (middle panel), and all material
along the LOS (bottom panel).  This is one of many cases
we found where the simulated core spectral peak lies
roughly in the minimum between two spectral peaks in the 
associated LOS LDG material.
Clearly, the addition of the extra material in determining
the LOS spectrum (bottom panel) has very little effect on the LOS LDG
spectrum.  This is important, as it demonstrates that our choice of
column density threshold for the LOS LDG material has little effect
on the measure of core to LOS LDG velocity difference (see also
Appendix~\ref{s_care}).

In our IRAM survey, the beamsize of the \nh observations is more
than twice as large as the \co observations (25\arcsec\ 
and 11\arcsec\ respectively), 
as mentioned in \S\ref{s_obs}.  In our calculation of the
simulation `observables', we similarly consider cells within
the appropriate beamsize (3 pixels for the core, 1 pixel 
for the LOS LDG) for their contribution
to each spectrum.  An odd number of pixels is necessary to allow an
equal sampling on either side of the core's peak.

A few peaks identified in the 1D projection have no cells within the
beam above the (column) density threshold set for the core material.
This occurs where two lower density structures, e.g. filaments,
lie along the same line of sight, usually in the
high turbulence simulations where the structure is less regular.
We use data from these points only in our analysis of the distribution
of LOS LDG velocity dispersions.   This is analagous to our IRAM data, 
for which we had a few more detections in \co than in \nh of candidate
dense cores.

We also create spectra to compare with the \thirco FCRAO observations.
Like \cons, the \thirco observations trace the lower density material, 
however, we
have \thirco observations across the entire Perseus cloud (albeit at
a lower resolution than our \co observations).  These observations allow
us to gain a more
global understanding of the dynamics of lower density material on larger 
scales.  We therefore calculate LDG spectra along every LOS
of the simulations
using a beamsize of 5 pixels to correspond to the 46\arcsec\ beam
of the \thirco data (i.e., 26 = 128/5 spectra for each x and y projection
of a simulation).  These simulated spectra can be thought of as arising 
from a chord across the projected centre of the star-forming region, and thus 
we compare these spectra with \thirco spectra that run across the
approximate centre (in right ascension and declination) of each 
extinction region.
Comparison with all observed \thirco spectra in the extinction region
instead of only the central chord would include a larger fraction of
lines of sight that pierce only the small edge of the 3D
cloud, whereas the simulations trace the mid-plane of the cloud only.

\subsection{`Observed' Properties}
\label{s_obssim_prop}
We fit the resultant core and associated LOS LDG spectra with Gaussians 
-- a single Gaussian
where possible, and two Gaussians where the fit is significantly
improved; this method was also used in our IRAM survey.  
Figure~\ref{spectra} shows some examples of the
spectra generated.  

Simulations with strong turbulence tend to produce spectra which are
more irregular and not well-fit even with a double Gaussian.  In
\citetalias{Kirk07}, we found that \nh dense core spectra were 
well fit by single (and occasionally double) Gaussians, and the 
\co spectra were mostly well fit by single or double 
Gaussians.  The poorer fits in the simulated spectra for the
high input turbulence suggests that these simulations are qualitatively
different than the observed regions.

We used our Gaussian fits of the simulated spectra to identify the relevant 
dynamical properties of each
core and associated LOS LDG material -- linewidth and centroid velocity.
In the following sections, we analyze these results and compare them with
the observational analyses in \citetalias{Kirk07}.  Note that
while a small number of cores are identified in a given
simulation, small number statistics are not a major concern,
i.e., similar properties are measured across multiple simulations
run with the same input Mach number and magnetic field strength\footnote{
We analyze data from three 
additional simluation runs with a $\mu_{0}$ of 0.5 and a Mach number of 2 to 
determine the magnitude of statistical uncertainty in the derived core 
and LOS LDG properties.  Qualitatively, we find that the column density
maps of the additional $\mu_{0} = 0.5$, $M = 2$ simulations reveal
structures similar to the one we analyzed in the bulk of the paper.
We find a similar
result quantitatively -- comparison of the mean value (and where appropriate,
standard deviation) of all of the `observables' analyzed in this
paper shows these are consistent with the mean and standard deviation
of the observables found in the extra three runs of the simulation.
We also analyzed a single simulation run which again has $\mu_{0} = 0.5$,
$M = 2$ but an input turbulent velocity spectrum with $v_k^2\propto k^{-3}$
where $k$ is the wavenumber.  This simulation also
showed observables consistent with the additional runs with $\mu_{0} = 0.5$ and
$M = 2$ and our standard $v_k^2\propto k^{-4}$.
We therefore conclude that our results are not severely affected by
small number statistics and furthermore that the results are somewhat
insensitive to the power spectrum of the input turbulent velocity field.
Further discussion of the latter point can be found in \citetalias{BCDW09}.
}.

\begin{figure}[htb]
\plotone{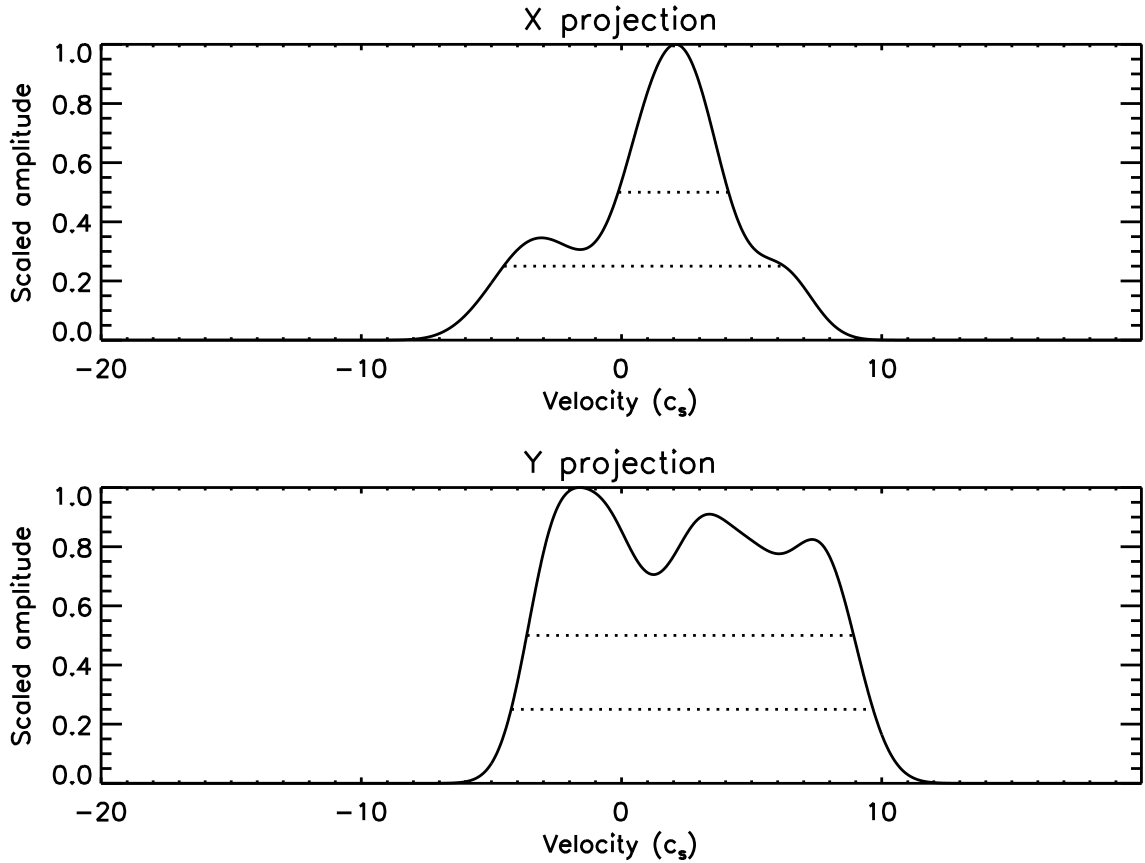}
  \caption{Calculated \thirco spectra for one of the LOS's in the most
	turbulent simulation, with the weakest magnetic field ($\mu$=2)
	and highest input turbulence ($M=4$).  The spectrum's amplitude
	has been normalized to the maximum intensity.  The dotted horizontal
	lines indicate the extent of the spectrum at half- and
	quarter- maximum intensity.  Note how it is necessary to measure the
	width well below the half-maximum mark in order to be
	sensitive to low-lying large scale modes.}
  \label{fig_sample_13co_spectra_sim}
\end{figure}

For the larger-scale LDG LOS simulated spectra corresponding to our
\thirco observations, we found that both the simulated and
observed spectra were less well described by a Gaussian fit.
Since our analysis of these spectra requires only a measurement
of the linewidth, we found that measuring the full width of the 
emission at one quarter of the peak (FWQM)
was effective.  Figure~\ref{fig_sample_13co_spectra_sim}
shows an example of a simulated larger-scale LOS LDG spectrum 
in one of the most turbulent simulations.  As seen in the figure,
we found the width of the emission at half of the peak (FWHM) 
often picked out
narrow emission peaks while missing underlying wide emission,
while the FWQM was sensitive to the wider emission underneath.
Measurement at levels lower than the quarter maximum sometimes presented
a problem for the \thirco observations where these levels became 
too close to the noise level.  Where multiple peaks 
were found to be separated even at the quarter maximum, we only counted
regions above the quarter max for our width measurement.  This
was especially important in our \thirco observations to prevent
bias from large noise spikes.  (Note we also smoothed our
\thirco spectra to five spectral channels, or 0.33~km~s$^{-1}$ 
to reduce errors introduced by noise.)  
For a single Gaussian, the FWQM
can be converted to a Gaussian $\sigma$ through division by
$4\sqrt{\ln{2}}$ or $\sim3.3$.  We refer to the Gaussian $\sigma$
corresponding to the measured FWQM as the ``effective'' velocity
dispersion.  

\begin{figure}[htb]
\plotone{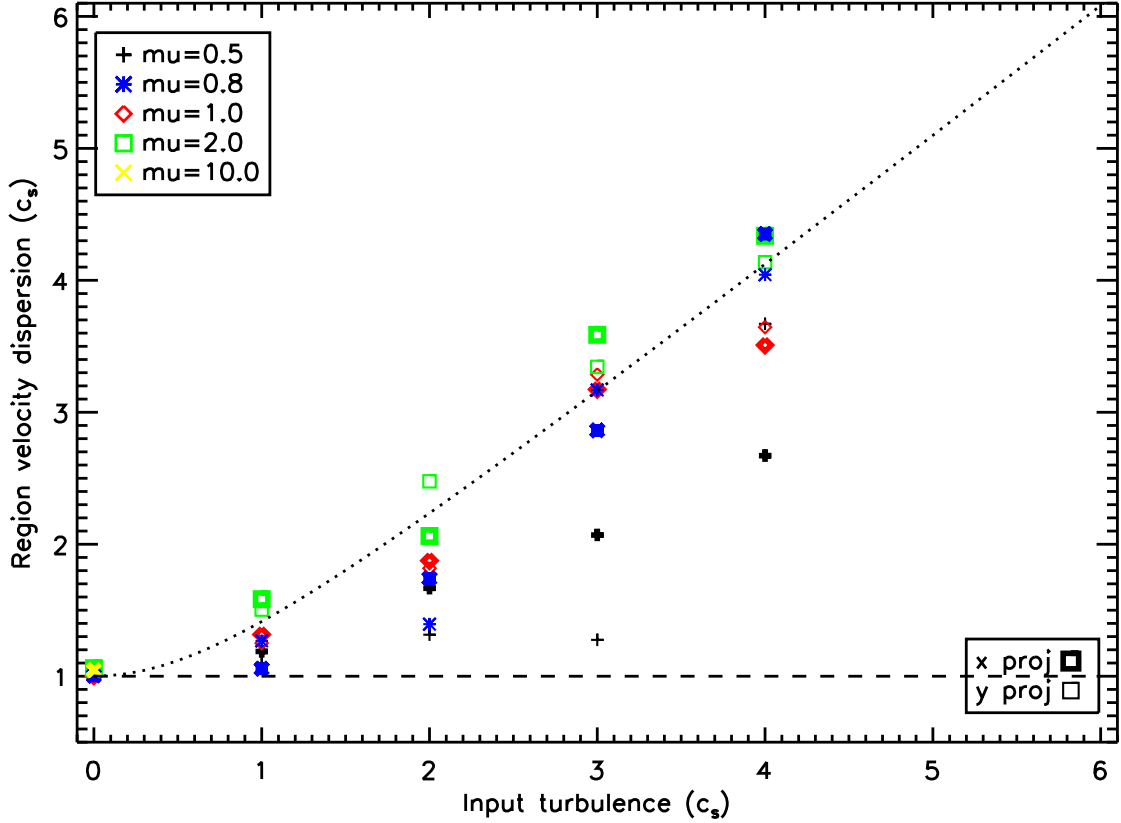}
  \caption{The velocity dispersion of the material in the entire simulated
	region measured at the end of the simulation versus the input 
	level of turbulence.  The dashed line
	indicates a thermal velocity dispersion while the dotted line
	shows the relationship for a velocity dispersion equal to the
	input turbulence.  Squares and plusses show the results for the 
	x and y projections of the simulations respectively.
	The simulations with strong magnetic fields
	show much smaller velocity dispersions than would be expected
	from the original input turbulence.}
\label{fig_box_disp_vs_turb}
\end{figure}

Finally, we created a composite spectrum describing the whole
of each simulation (in both projections), and similarly summed all
of the \thirco spectra within each extinction region.  
To estimate the velocity dispersion of the entire region, we fit these 
spectra with a single Gaussian, which provides reasonable agreement.  
Note that in the
simulations with weak magnetic fields, the velocity dispersion
measured is nearly identical to what would be expected from the
input turbulence level; simulations with stronger magnetic fields
tend to have lower measured dispersions due to damping of the turbulence.
Figure~\ref{fig_box_disp_vs_turb} shows the velocity dispersion 
found across the
simulation as a whole versus the level of input turbulence.
In our observations, we only include 8 of the 11 extinction regions
identified in \citetalias{Kirk06}, as the other 3 have poor
coverage ($< 80\%$) in the \thirco map which prevents proper measurement
of the dynamics.

Table~\ref{sims_dyn} provides the overall statistics for all of the
`observed' dynamic quantities in the simulation -- the mean and standard
deviation of the velocity dispersion of the LDG material
along all lines of sight (corresponding to the \thirco observations),
the velocity dispersion of the LDG material along lines
of sight where dense cores were identified (corresponding to
the \co observations), the velocity dispersion of the 
dense cores (corresponding
to the \nh observations), and the difference in centroid velocity
between the core and LOS LDG material.  Table~\ref{sims_effic} shows
the number of cores found in each simulation and the percentage
of mass found within the cores (core formation efficiency, or CFE). 
{\hk Note that Appendix~B includes the analysis of several higher resolution
simulations and demonstrates that the resolution does not affect
the `observable' dynamic quantities reported here.}

\section{RESULTS}
\label{s_results}

\subsection{Internal Velocity Dispersion}
\label{s_turb}
\subsubsection{Across the Cloud}
\begin{figure}[htb]
\plotone{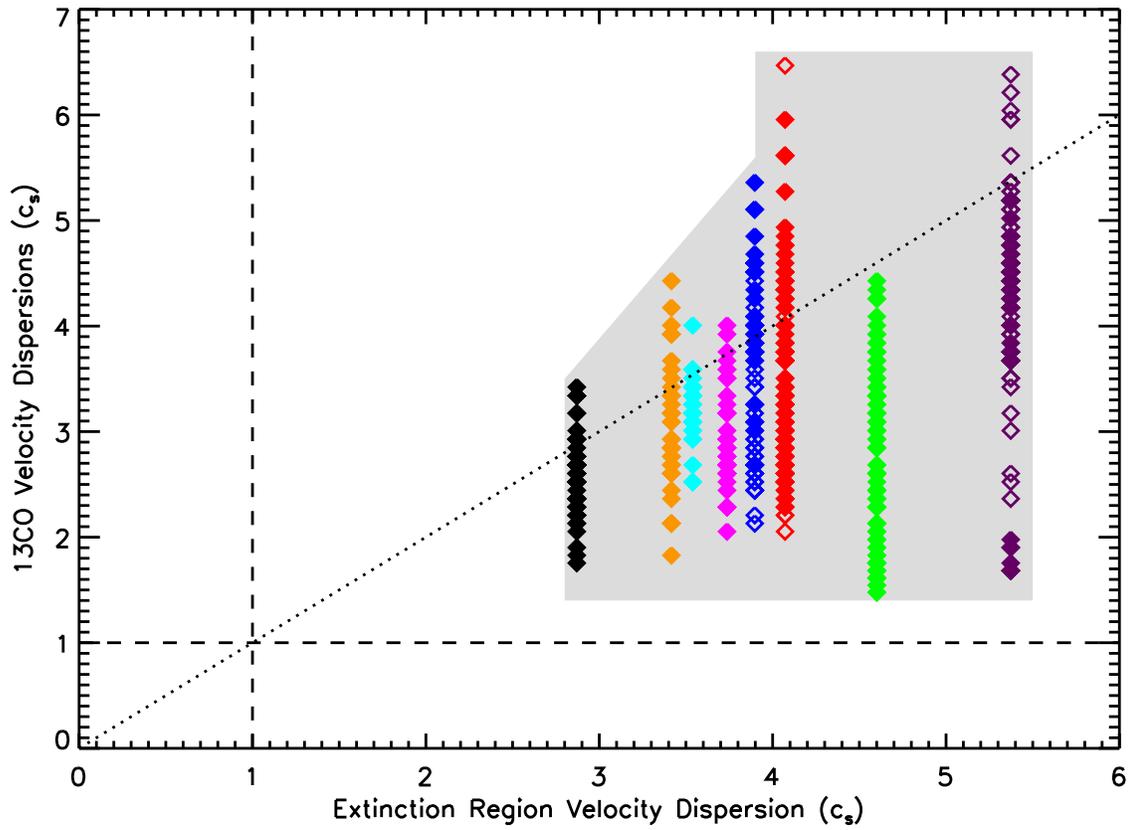}
  \caption{The velocity dispersion of low density material (\thirco)
	measured along the central chord through each extinction region versus
	the velocity dispersion seen across the region as a whole.
	The filled symbols indicate regions where star formation is
	more recent, while the shading indicates the region spanned
	by these observations.}
\label{fig_LOS_allwidths_obs}
\end{figure}

\begin{figure}[htb]
\plotone{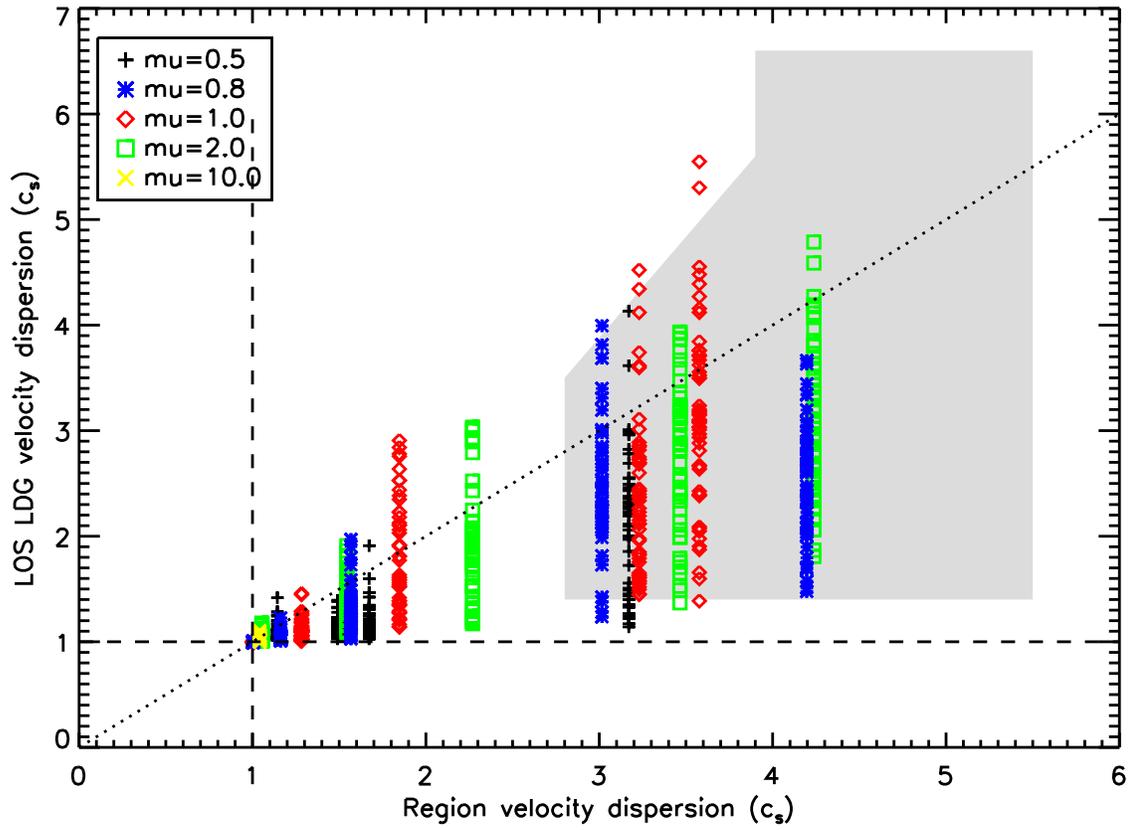}
  \caption{The velocity dispersion of low density material measured
	along all lines of sight in the simulation versus the
	velocity dispersion in the region as a whole.  The
	shading indicates the region spanned by the observations
	(see Figure~\ref{fig_LOS_allwidths_obs}).}
\label{fig_LOS_allwidths_sim}
\end{figure}

We first examine the behaviour of the larger-scale, less dense material
in order to gain perspective on the environments in which the
dense cores form.  
In Figure~\ref{fig_LOS_allwidths_obs}, we show the
effective velocity dispersions observed in \thirco along two central 
chords (in right ascension and declination) through each 
extinction region versus the velocity 
dispersion observed for the entire extinction region.  
(Note that the observational results do not change significantly if 
all lines of sight in the extinction region are used instead.)
In Figure~\ref{fig_LOS_allwidths_sim}, we show the corresponding 
measurements for the simulations -- the effective velocity dispersions 
of LDG along all LOSs compared to the velocity dispersion of the
simulated region as a whole.  (These values are included in 
Table~\ref{sims_dyn}.)  
Both Figures~\ref{fig_LOS_allwidths_obs} and \ref{fig_LOS_allwidths_sim} 
show a range in LDG velocity
dispersions along individual LOSs, ranging from small to larger than
the mean for the cloud as a whole.  
Note that while the range in individual LOS LDG velocity dispersions are 
similar for both the observations and simulations, the extinction 
regions observed in Perseus tend to have higher
velocity dispersions as a whole than within the simulations.  
Only simulations with
the highest input turbulence (Mach number 3 or 4) show regional velocity
dispersions similar to the lower end of the range spanned by our
\thirco observations.  Since feedback
is not included in the simulations, it may be more appropriate
to compare the simulations only with regions in Perseus that
do not appear to have formed previous generations of stars;
the filled symbols in Figure~\ref{fig_LOS_allwidths_obs} indicate the 
extinction regions which appear to have had less recent active star
formation (i.e. excluding NGC1333, IC348, and B1).  These
regions do tend to have slightly lower regional velocity dispersions
than their active star-forming counterparts.

\subsubsection{LDG Along LOSs Associated with Dense Cores}
We next examine the velocity dispersions seen in the LOS LDG
material associated with the dense cores.
As discussed above, the LOS LDG motion could provide hints for
the scale of turbulence in the molecular cloud.
There also may be a difference between lines of 
sight that contain cores and those that do not, if, for example,
cores form at stagnation points in the velocity field.  
Our pointed IRAM \co observations are at higher spatial
resolution than the \thirco map, but still allow LOSs with cores
to be compared with the overall distribution of LOS LDG dispersions.

\begin{figure}[htb]
\plotone{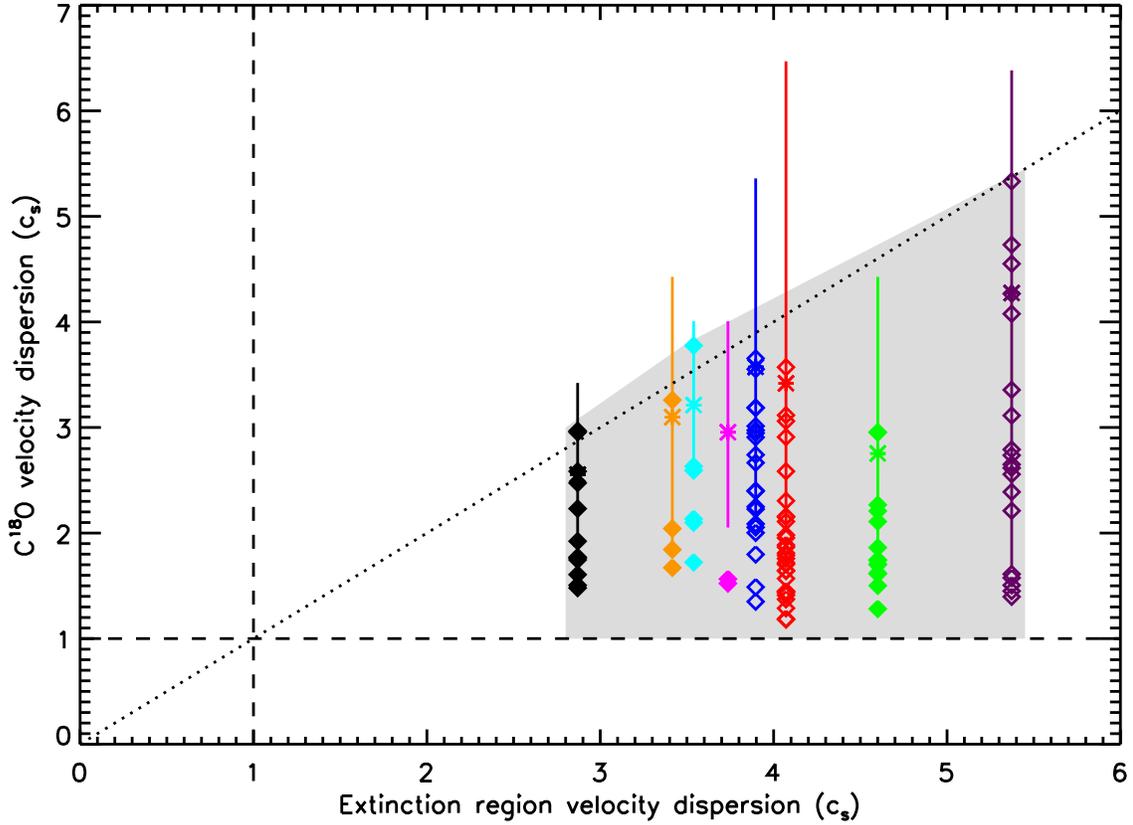}
  \caption{The velocity dispersion of low density material (\cons) 
	versus the velocity disperison measured in the larger
	extinction region (diamonds).  The solid vertical lines 
	indicate the range
	of velocity dispersions through the central chord of each
	extinction region, while the asterisk indicates the mean
	velocity dispersion.  The dotted line indicates a 1-1
	relationship while the dashed lines indicate thermal
	objects.  The filled symbols 
	indicate regions where star formation is more recent.  The
	shaded region represents roughly the area spanned by our
	observations.} 
\label{fig_LOS_widths_obs}
\end{figure}

\begin{figure}[htb]
\plotone{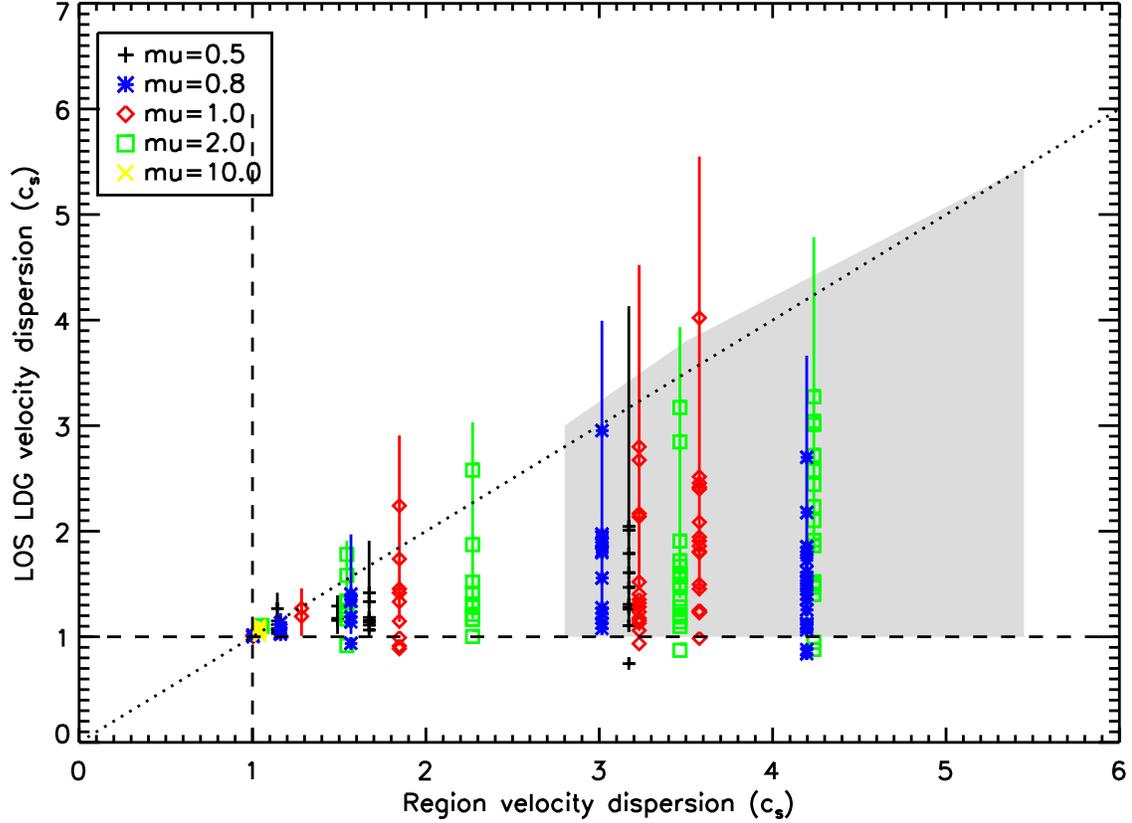}
  \caption{The velocity dispersion of low density material
	in the simulations versus the total velocity dispersion
	observed in the region (plus signs).  The solid vertical lines
	indicate the full range of velocity dispersions that
	are seen along all lines of sight in the region within a `beamsize'
	matching that of the \thirco observations.  The shaded region
	represents roughly the area spanned by our observations
	(see Figure~\ref{fig_LOS_widths_obs}).}
\label{fig_LOS_widths_sim}
\end{figure}

Figures~\ref{fig_LOS_widths_obs} and
\ref{fig_LOS_widths_sim} show the distribution of velocity dispersions
measured for LOS LDG material where cores were detected (points) for 
the observations and simulations (at the higher
resolution of the IRAM survey) versus the velocity dispersion
of the entire region.  Each figure also
shows the distribution of LDG velocity dispersions (mean plus and minus
the standard deviation) for {\it all} line of sight material (at
lower resolution / larger beamsize).  Note that as in the previous section,
the larger-scale LOS LDG linewidths are measured from the FWQM, with only
the LOSs along the central right ascension and declination line of the 
extinction regions
included in the observed plot. 
The dashed lines show the
minimum expected observationally, i.e., the thermal width, while
the dotted line shows a 1:1 relationship between the velocity
dispersion of the larger environment (simulated region or extinction
region) and the LOS LDG material.  The figures illustrate that the
bulk of the LOS LDG material has a similar velocity dispersion to
the larger environment, while the LDG along LOSs that contain cores 
often have much lower velocity dispersions.  Note the spectral
line-fitting technique and resolution could play a role in this 
difference -- the FWQM measure used for the bulk of the LOS LDG
spectra is more sensitive to low levels of extended velocity
dispersion than a single (or double) component Gaussian fit
which may preferentially probe the local core environment.

\subsubsection{Dense Cores}

We next turn our attention to the cores.  Recall that emission from
the dense cores originates from relatively compact regions of overdensity. 
In some turbulent simulations,
dense cores are transient features of overdensity formed by the intersection
of two colliding streams of material \citep{BP03}.  
In such a case, one might expect
to see a large internal velocity dispersion within the core
for some fraction of the population
-- although the centre of the turbulent compression typically has
low velocity dispersion, the core as a whole can encompass a sufficiently
large region so as to reveal a larger velocity dispersion.
{\hk Indeed, in turbulent simulations for which the distribution of dense
core internal velocity dispersions were calculated
\citep{Klessen05,Ayliffe07,Offner08}, a small but non-zero population 
of cores with supersonic internal velocity dispersions is present. 
In each case, the highest internal velocity dispersions in the simulations
were larger than those seen in the observations of \citetalias{Kirk07}. }
On the other hand, cores which are bound and evolving more quiescently 
would be expected to have small internal velocity dispersions.  As
discussed in \S\ref{s_obs}, previous
dense core surveys \citep[e.g.,][]{Benson89,Jijina99} found that dense
cores have small, thermally dominated internal velocity dispersions,
although some turbulent motion may still be present,
as discussed in \S\ref{s_obs}.

\begin{figure}[htb]
\plotone{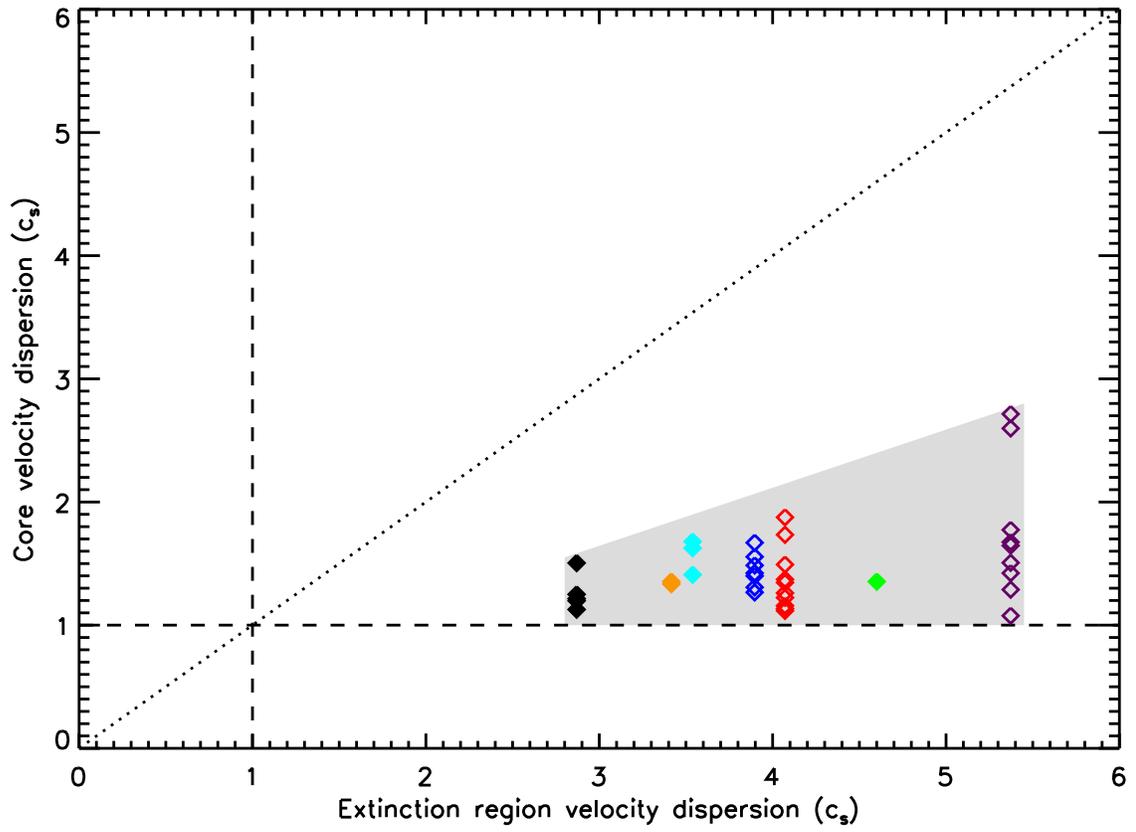}
  \caption{The internal velocity dispersion of the cores versus the
	velocity dispersion of their large-scale environment.  
	The filled symbols indicate regions where star formation
	is more recent.  The shaded region represents roughly
	the area spanned by our observations.}
\label{fig_core_widths_obs}
\end{figure}

\begin{figure}[htb]
\plotone{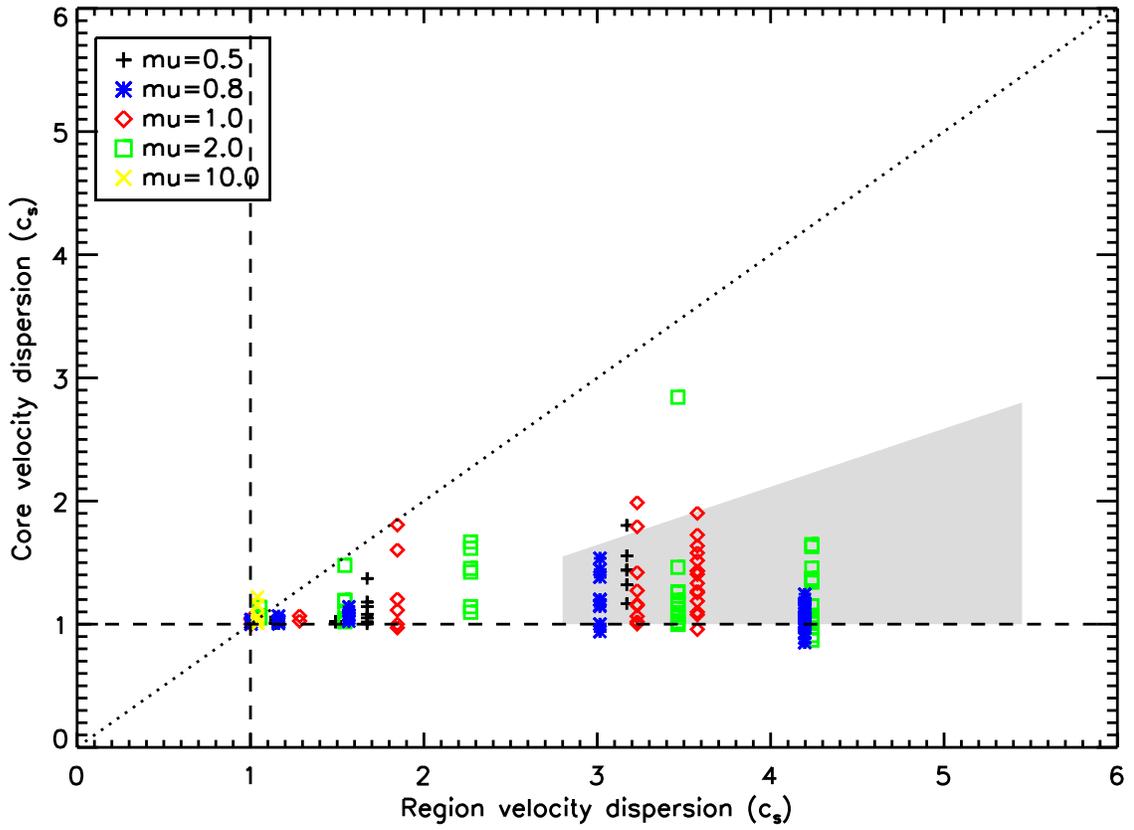}
  \caption{The internal velocity dispersion of the cores versus the
	velocity dispersion of their 
	large-scale environment for the simulations.  The shaded
	region represents roughly the area spanned by our 
	observations (see Figure~\ref{fig_core_widths_obs}).} 
  \label{fig_core_widths_sim}
\end{figure}

Figures~\ref{fig_core_widths_obs} and \ref{fig_core_widths_sim} show 
the internal velocity dispersion of the cores with respect to their larger
environment (the extinction / simulation region) versus
the velocity dispersion of the entire region.  The cores have 
a similar range of velocity dispersions in both the simulations and 
observations, although as noted previously, the observed cores tend
to inhabit large-scale environments with higher turbulent energy.

\subsection{Core-to-LOS Motions}
\label{s_mot}

A second discriminant between the various simulations as well as the
observations is the relative motion
of the core and the surrounding LOS LDG material.  
Observations
indicate that cores do not have significant motions within their
local environment, with most displaying smaller motions than either
the mean velocity dispersion of their local environment or the
local sound speed (Walsh et al. 2004, 2007; KJT07).  
\citet{Walsh04} have argued that in the turbulent framework
of star formation, if cores form via competitive accretion,
one might expect to see significant motion between the core
and its local environment.  Some simulators have since argued
that competitive accretion simulations can display
similarly small velocity dispersions \citep{Ayliffe07}.  It is
unclear, however, whether these simulations are able to simultaneously 
meet all observational constraints \citepalias[e.g., both the core to envelope
motions and core internal velocity dispersion; see][]{Kirk07}.    
In any event,
the motion between the core and its surrounding material is a 
useful measure to examine in the context of turbulent models.

\begin{figure}[htb]
\plotone{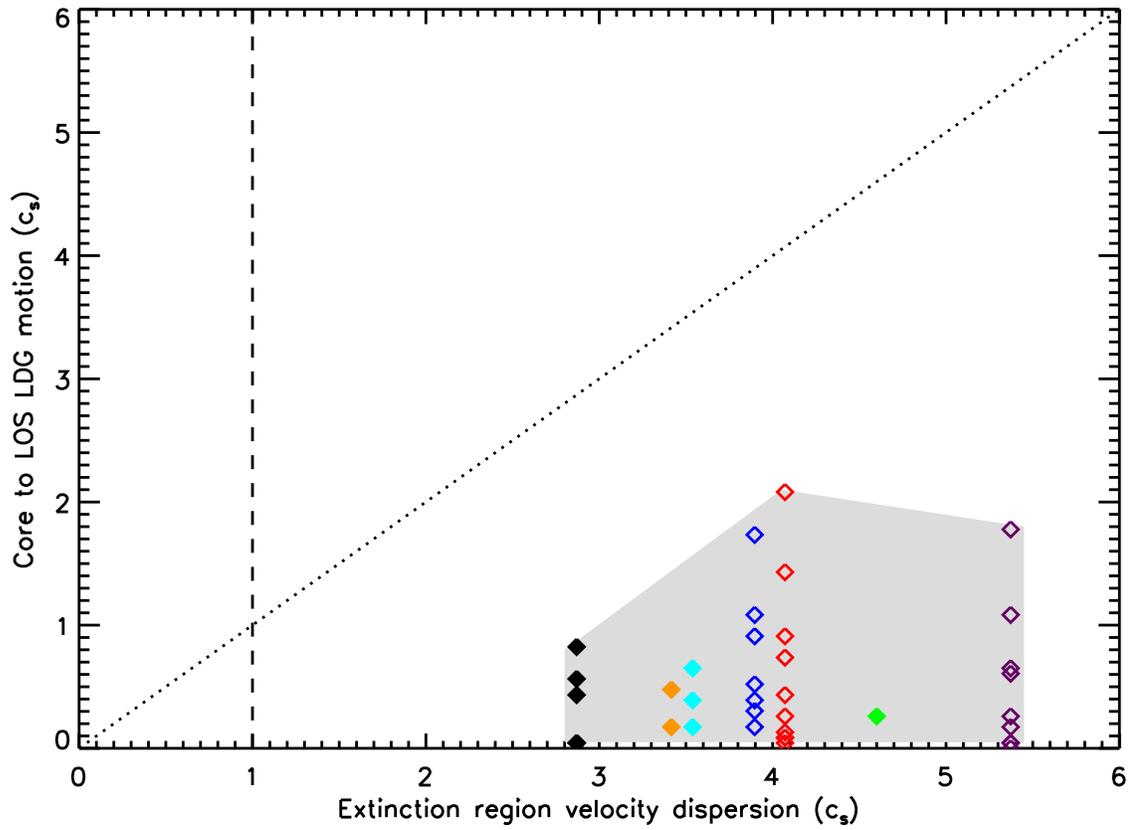}
  \caption{The observed relative motions of the core and its surrounding LOS LDG
	material versus the velocity dispersion 
	of the extinction region.  The filled symbols indicate regions
	where star formation is more recent.  The shaded region 
	represents roughly the area spanned by our observations.}
  \label{fig_core_to_LOS_obs}
\end{figure}

\begin{figure}[htb]
\plotone{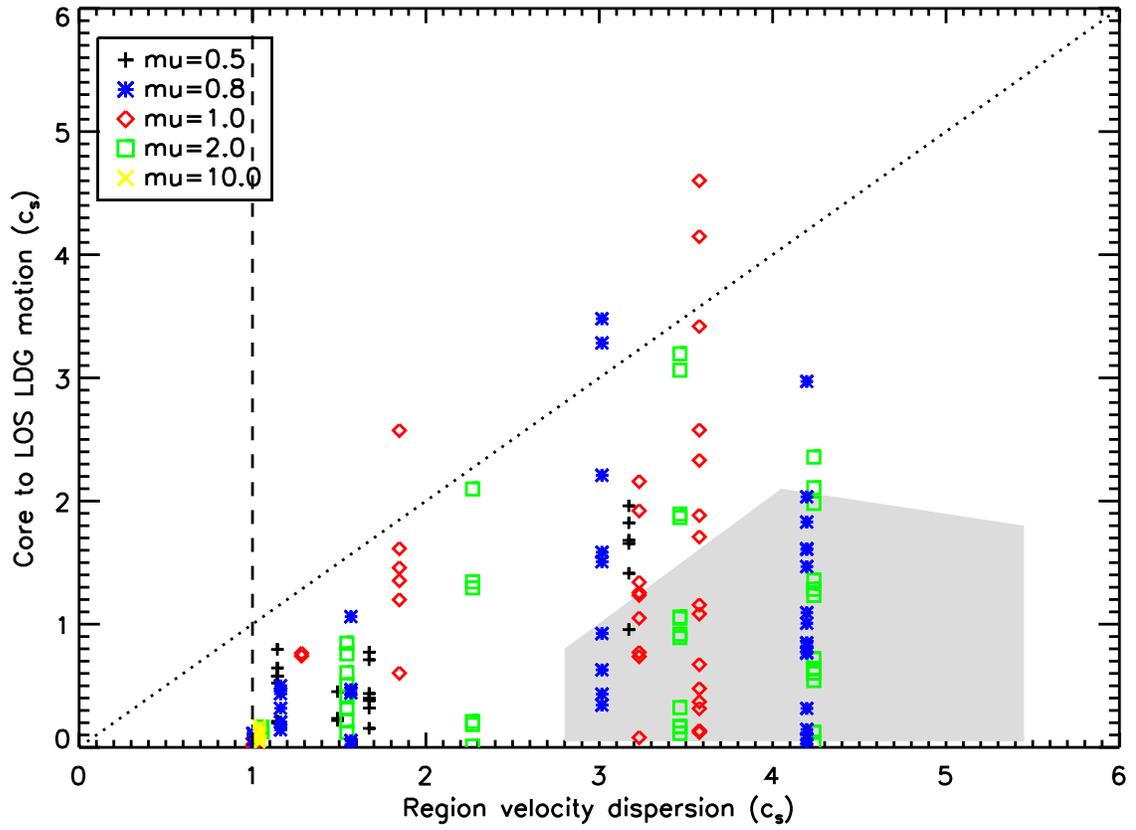}
  \caption{The relative motions of the core and its surrounding LOS LDG material
	in the simulations versus the velocity dispersion of the region.  
	The shaded region represents roughly the area spanned by our 
	observations (see Figure~\ref{fig_core_to_LOS_obs}).}
  \label{fig_core_to_LOS_sim}
\end{figure}

Figures~\ref{fig_core_to_LOS_obs} and \ref{fig_core_to_LOS_sim} show
the relative motions of cores and their LOS LDG material versus
the velocity dispersion of the entire region for 
the Perseus observations and the simulations respectively.   What is
striking about the two figures is that observations display a much
greater connection between the core and LOS LDG material -- the differences
in velocity are quite small despite having more non-thermal motion on 
larger scales.
In the simulations, the core-to-LOS LDG motion becomes large even for 
only moderate large-scale velocity dispersions.

\begin{figure}
\plotone{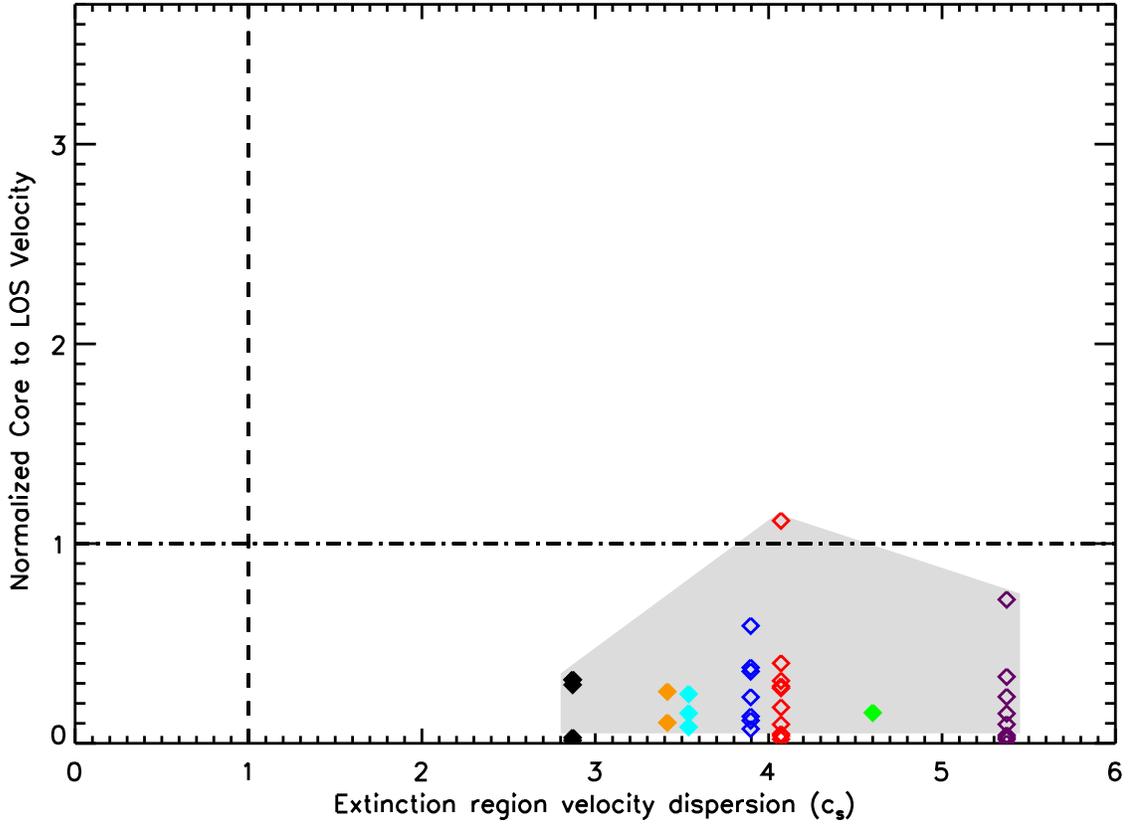}
  \caption{The ratio of observed core to LOS LDG motion and LOS LDG velocity 
	dispersion versus the velocity dispersion of the extinction region.
	The dash-dot line indicates the point at which cores
	move {\hk faster than the velocity dispersion of the LOS LDG} 
	with respect to their local environment
	(see \S5.2).
	The filled symbols indicate regions where star formation is more
	recent.  The shaded region represents roughly the area
	spanned by observations.}
  \label{fig_core_to_LOS2_obs}
\end{figure}

\begin{figure}[htb]
\plotone{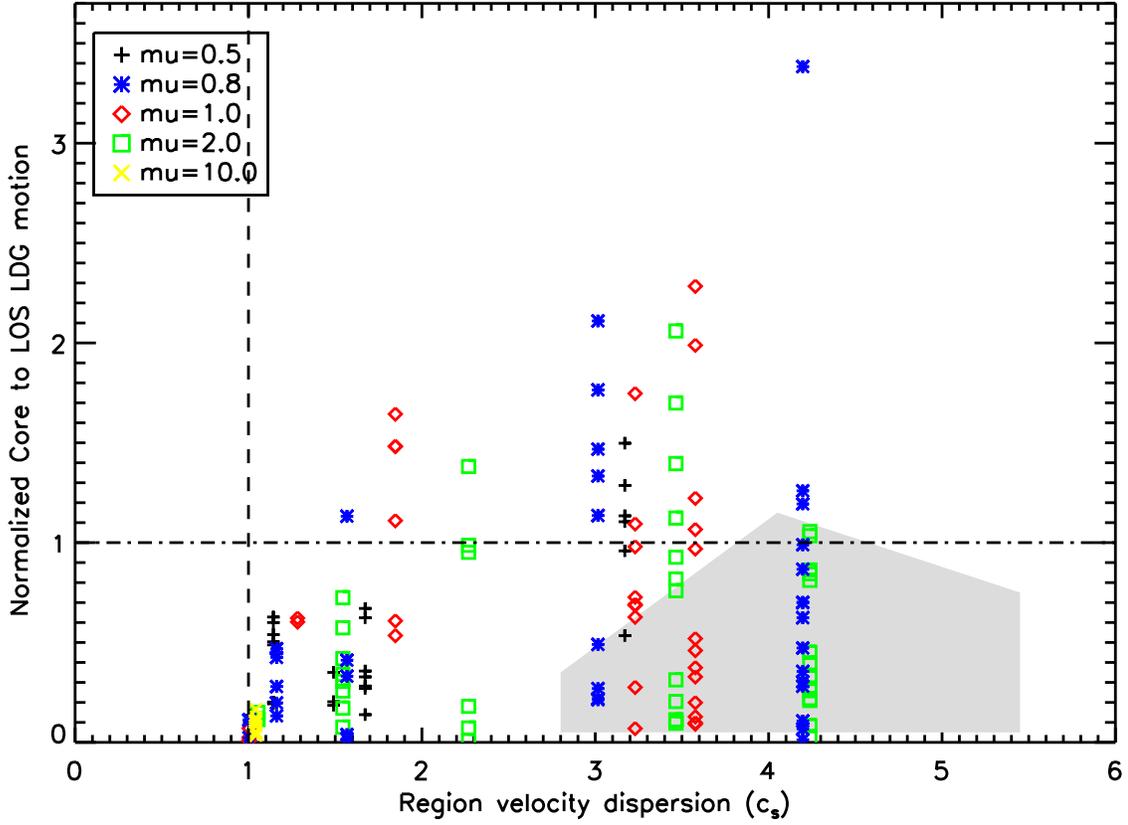}
  \caption{The ratio of core to LOS LDG motion and LOS LDG velocity dispersion 
	in the simulation versus the velocity dispersion of the entire region.
	The dash-dot line indicates the line at which cores
	move {\hk faster than the velocity dispersion of the LOS LDG} 
	with respect to their local environment (see \S5.2).
	The shaded region represents roughly the area spanned by our 
	observations (see Figure~\ref{fig_core_to_LOS2_obs}).}
  \label{fig_core_to_LOS2_sim}
\end{figure}

A second way to measure the motion is to compare the core-to-LOS LDG
motion to the typical motion found in the LOS LDG material, i.e.,
the LOS LDG internal velocity dispersion.  The difference between
the observations and simulations using this measure are even more pronounced.
Observations
show the cores are quite quiescent within their nearby surroundings,
while the simulations show cores {\hk often have motions larger than
the LOS LDG velocity dispersion}.
Figures \ref{fig_core_to_LOS2_obs} and \ref{fig_core_to_LOS2_sim}
show the core-to-LOS LDG motion divided by the LOS LDG velocity dispersion
versus the velocity dispersion of the region as a whole.

In the simulations with high input Mach numbers, 
the LOS LDG spectra often had complex structures and most required
fits with two Gaussians.  In instances of two Gaussian fits, we 
selected the velocity component with the closest centroid
velocity to compare with the core centroid velocity.
This was also the procedure used in our IRAM observations
in \citetalias{Kirk07}, 
although in that work we found fewer spectra that required multiple
Gaussian fits than in the turbulent simulations, where LOS LDG material
spectra frequently displayed more than two peaks.  
We visually checked all core and LOS LDG
spectra and fits to ensure the Gaussian fits to the more
complex structure did not bias the fits to higher velocity
differences.  We found there are a few cores where somewhat
smaller velocity differences would have been obtained had
all of the peaks in complex LOS LDG spectrum been used for the spectral
fitting, however, there were also several instances where the
velocity differences would have been larger had all the peaks been
used in the spectral fitting.  We therefore conclude that the
overall effect on the velocity difference from using only two Gaussian
components is negligible.  Note that for the observations, the
selection of the closest \co velocity component did not bias the
results to smaller velocity differences (see KJT07 for more details).


\subsection{Core to Region Motions}
In order to further investigate the large motions between core and 
associated LOS LDG material in the most turbulent simulations, we
examined the motion between the core and the entire region.  This
provides an indication of whether the cores sit relatively
unperturbed within the larger environment while the low density
LOS material moves at high velocities, or whether both core and 
LOS LDG material move at (differing) high velocities.

\begin{figure}[htb]
\plotone{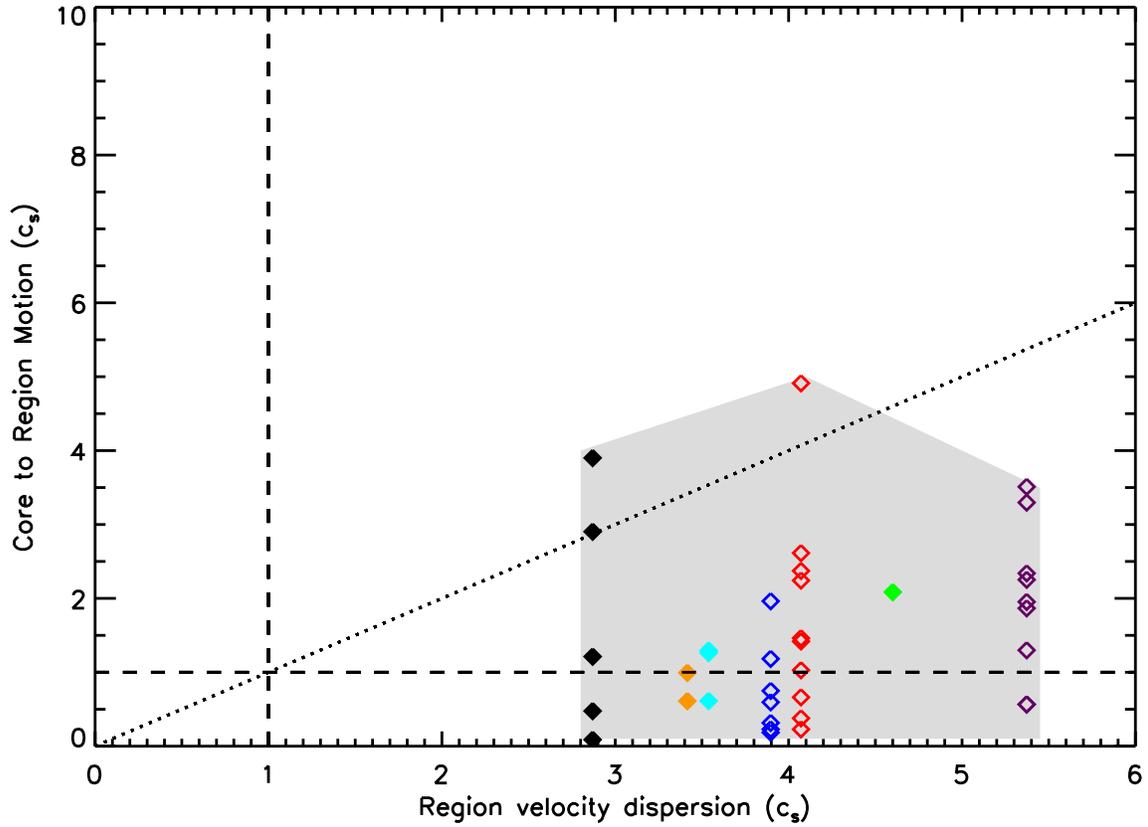}
  \caption{The motion of the observed \nh cores in Perseus within their
	parent extinction regions versus the velocity dispersion of
	the extinction regions they inhabit (as measured in \thirco).  The
	dotted line indicates a 1-1 relation.  The filled symbols indicate
	regions where star formation is more recent.  The shaded region
	represents roughly the area spanned by our observations.}
\label{fig_core_to_box_obs}
\end{figure}

Figure~\ref{fig_core_to_box_obs} shows the absolute difference
in centroid velocity between the observed dense cores and the extinction region
as a whole versus the velocity dispersion of the extinction region.  
While a significant fraction of the cores do
display supersonic motions, nearly all have motions that are 
significantly smaller than the total velocity dispersion observed in the 
region.  The velocity dispersion of each extinction region 
tends to be similar (within a factor of 1.5) to that 
required for the region to be in energy equipartition between
gravitational and internal kinetic energy
(Table~\ref{ext_regs}) \footnote{Note that this is often referred
to being in virial equilibrium although it is not in the strict
sense of the definition.}.  This implies that the dense cores 
observed tend to move with sub-virial velocities.

\begin{figure}[htb]
\plotone{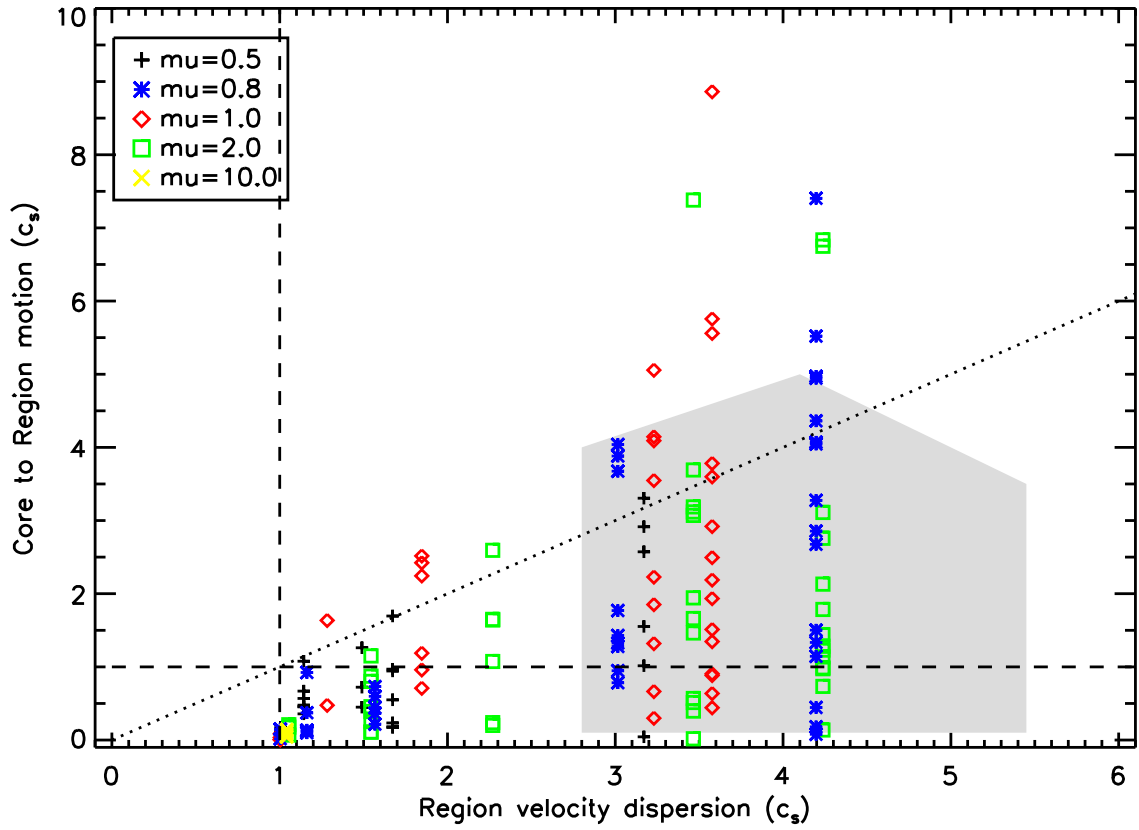}
  \caption{The difference between core and region centroid velocity
	versus the velocity dispersion of the region for the simulations.
	The dotted line indicates a 1-1 relation.  The shaded region
	represents roughly the area spanned by our observations
	(see Figure~\ref{fig_core_to_box_obs}).}
\label{fig_core_to_box_sim}
\end{figure}

Such small velocities are not seen in the simulations, however.  
Figure~\ref{fig_core_to_box_sim} shows
the absolute difference in centroid velocities between
the dense core and the region as a whole versus the velocity
dispersion of the region.  This indicates that the dense cores can 
reach quite large velocities with respect to the simulated region
as a whole (the centroid velocity of the region is
nearly zero as one would expect).  Figure~\ref{fig_core_to_box_sim}
shows that the simulated cores often move much faster than the
sound speed, and sometimes even in excess of twice that of
the velocity dispersion of the region as a whole!
In these simulations, the timescale for the formation of
the first high column density peak is very short -- much less
than a crossing time (see Table~\ref{tab_time}).  The short timeframe
before dense structure formation allows some of the cores to
inherit the kinematics of material at the high end of the
velocity tail of the input turbulence.  (Note that for every high
velocity core, there are several at lower velocties in the same
simulation.)  The short simulation timescale also originates from the high
velocity tail -- high velocities lead to strong turbulent
compressions which quickly compresses the column density 
to ten times the mean, at which point the simulation is stopped.

The observations therefore paint a picture of cores moving slowly with
respect to their turbulent surroundings, whilst in the 
simulations, compressed density enhancements can be formed with a 
much greater velocities than their surroundings.


\subsection{Formation Efficiency}
\label{s_form}
The final quantities that can be compared between the simulations 
and the observations are the number and total mass of the cores
formed within each region.  Star formation is known to be
an inefficient process, with the fraction of mass ending up in
stars ranging from a few percent to a few tens of percent in clustered
environments \citep{Lada03}.  Within the size scale probed by the 
simulation, the fraction of mass in dense cores is expected to be on
the order of 10\%, since this is a more clustered environment than
the cloud as a whole.  \citet{Jorgensen08} found that protostars
make up 17\% of the mass within the clustered regions of Perseus
and 3\% of the mass of the cloud as a whole.  Similarly, 
\citetalias{Kirk06} found that dense cores make up roughly 1\%
of the entire cloud mass in Perseus.

Our observations, unfortunately, do not lend themselves to a straightforward
computation of either the number of dense cores or the mass within
them.  Our IRAM \nh survey, which is the basis for the calculations of 
dense core properties, was based on a target list created using several 
methods.  The majority of targets were based on cores found 
in the SCUBA dust continuum map of the cloud, while a still significant 
fraction was based on visual selection of targets
from Palomar plates.  The SCUBA observations spanned roughly the
entire Perseus molecular cloud (to A$_{V} \sim 5$) so the list
of \nh dense cores that have an associated SCUBA source should be complete
and unbiased.  The Palomar plate targets, however, were explicitly 
drawn from areas with no SCUBA detections.  
We anticipate that our total list of dense cores detectable in
\nh is mostly (but not fully) complete, and may have slight biases
in certain regions.
We therefore provide
{\it two} estimates of the number of cores observed per extinction region
-- the first being the number of \nh cores associated with a SCUBA source,
and the second being the total number of \nh cores in our
IRAM survey.  The latter should provide a closer estimate of the total
number, but different regions may have different fractions of cores
missed.  Table~\ref{obs_effic} provides
both of these numbers for the eight extinction regions analyzed.  The
table shows that the number of cores varies between a few to nearly thirty,
with only a small difference between the two estimates. 
Regions where star formation appears to have initiated more
recently have less than ten cores per region.

The observed total mass of the cores within each extinction region 
provides a further challenge in calculation, and provides a
second motivation for separating dense cores associated with SCUBA
sources from the total population.  The mass of the dense cores
cannot be determined from our \nh observations.  The SCUBA observations,
however, allow the mass to be determined (making
assumptions about the dust opacity, temperature, etc) but the
Palomar cores do not.  We use
our estimates of SCUBA masses from our 3\arcsec\ map analysis
(included in Table 6 of \citetalias{Kirk07}) and assume that the \nh cores
without an associated SCUBA source each have
the median mass of the SCUBA sources (0.8~M$_{\odot}$).
Since these \nh cores were not detected with
SCUBA, one might argue they are likely to have smaller masses than
those with associated SCUBA sources, thus the reader is urged to
treat the total core mass estimates for every extinction region as
approximate.  The fraction of mass in dense cores within 
every extinction region is calculated using
the mass of the extinction regions found in \citetalias{Kirk06}.
Table~\ref{obs_effic} shows the fraction of mass 
(core formation efficiency) found using both 
the \nh SCUBA cores only and all \nh cores.  
The fraction of mass in the dense cores tends to be on the order of a few 
percent using either measure, somewhat lower than in \citet{Lada03},
however, that measure includes already-formed stars.

In the simulations, both the number of dense cores and the fraction of mass 
contained within the dense cores (core formation efficiency) 
are straightforward to calculate.  
Table~\ref{sims_effic} shows these values for the various simulations.
As can be seen from the table, the simulations tend to form a small
number of cores ($<$10), with the fraction of mass contained within
them being of order a few percent.  Since the simulations are stopped
at a relatively early time-step (when a column density of ten times the
mean has been reached), it is possible that further accretion and /
or fragmentation could significantly alter these results.



\section{DISCUSSION -- INTERACTION OF THE MAGNETIC FIELD AND TURBULENCE}
\label{s_disc}
We take a step back from the simulations here to discuss some
general trends and behaviors which arise from the combination of
input turbulence and magnetic field.

The purely isothermal simulation ($M=0, \mu_{0}=10)$ reveals the
scales expected from gravitational instability (eqn 7), since the only other
physical parameter, the external pressure, provides merely a small
perturbation.  The simulation thus produces mostly round,
dense cores each with approximately a critcal mass, forming out
of a regular velocity field and undergoing gravitational collapse
after about a sound crossing time of the critical length,
or $\sim 12t_0$.

When turbulence is added to the isothermal scenario, there is the
addition of a non-regular velocity field, with the largest
kinematic modes found on the largest spatial scales.  These
supersonic velocity fields lead to shocks which create significant
density enhancements -- the ram pressure is $\sim \rho_0~M^2~c_s^2$,
which produces compressions of $\sim M^2$ in the post-shocked, isothermal
gas.  The strongest density enhancements originate from the largest
turbulent modes, forming long filaments in less than a turbulent crossing time
across the region.  For highly turbulent simulations, the initial conditions
of the highly supersonic velocity field can produce column density
enhancements strong enough to stop the simulations extremely quickly.
The dense objects formed in these simulations tend to be more elongated
than in the isothermal case due to non-uniform compressions, and the
sizes of the cores formed are smaller since the Jeans length
decreases with the higher density.

When magnetic fields are added without turbulence, the relevant 
size scale for collapse varies significantly as a function of the 
initial mass to flux ratio $\mu_0$ \citepalias{CB06}. The formed 
objects are mildly elongated \citepalias[see][for a detailed discussion 
of the shapes]{BCW09}, 
with a smooth velocity field. Strong magnetic fields, 
however, delay the formation of dense structures 
and gravitational collapse for an ambipolar-diffusion time, 
during which the neutrals slip past the load-bearing 
ions. For the conditions assumed here, the time to runaway 
collapse of the first core is about a factor of ten longer than in the 
isothermal nonmagnetic case.

When magnetic fields and turbulence are both present, the
situation becomes much more complicated.  For 
strong magnetic fields the magnetic pressure is sufficient to
effectively prevent strong compressions even in turbulent
velocity fields and evolution
occurs only slowly via ambipolar diffusion. 
(Note that since we are analyzing thin sheet models with
the magnetic field initially perpendicular to the sheet, compressions
cannot act along the direction of the magnetic field lines.)
In these cases, the
final outcome will be similar to the case where only the magnetic
field was present. For somewhat weaker magnetic fields, the turbulence
is able to create significant compressions but the ram pressure
is balanced  by a post-shock enhanced magnetic field pressure in
which the density increases by only $\sim M$ \citep{StahlerPalla} rather
than by a post-shock thermal pressure at an enhanced density.
In these compressed zones,
the magnetic field enhancement effectively overwhelms gravity and
collapse cannot occur until ambipolar diffusion removes the magnetic
support. Given the higher density and shorter length scales involved,
the ambipolar diffusion is nevertheless significantly shortened.
Finally, when the magnetic fields are insignificant, even in the
turbulently compressed regions, the simulation proceeds as in
the purely turbulent case.  Back of the envelope calculations
show that the simulations we analyze fall in the regime of turbulent
compression leading to significant magnetic field pressure; none have 
sufficiently weak fields to fall into the purely turbulent regime.

Will our results and conclusions differ dramatically if we consider 
three-dimensional simulations within the same physical parameter 
space?  This is not likely, as demonstrated by the recent results 
of \citet{Kudoh08}, who performed three-dimensional simulations of 
strongly and moderately magnetized clouds with initial turbulent 
fluctuations, including the effect of ambipolar diffusion. 
Their results are in general agreement with the thin-sheet calculations 
of \citet{Li04} and \citetalias{BCDW09}.  A large parameter study in 
three-dimensions, however, remains prohibitively expensive 
computationally.  The special feature of our current work is that we 
are able to sample a large range of parameter space. 
Nevertheless, we do anticipate performing a similar study with 
three-dimensional simulations in the future, as broad parameter surveys 
become available. 

Despite the above point, we note that different conclusions may yet be 
possible if we consider the case of continually driven turbulence or 
global (non-periodic) cloud models, in either the thin-sheet or 
three-dimensional cases. 


\section{CONCLUSIONS}
\label{s_conc}
Although the simulations we analyze have a thin sheet geometry, their
relative simplicity provide the advantage of taking a relatively 
short amount of time to run, and hence easily allow a parameter study.
While we do not expect the quantities measured here to be identical
with a more complex, three-dimensional study, we do expect the trends
we observe to be generally valid in the three-dimensional case as well.

From our analysis of the simulations we find the following:
\begin{itemize}
\item {\it The large-scale (regional) low density gas velocity 
	dispersions in the 
	simulations tend to be lower than found in our observations.}
	Simulations with strong magnetic fields tend to cause
	input turbulence to significantly damp by the time
	the simulation is halted for analysis (column overdensity of 
	$\sim10$).
\item {\it The velocity dispersions of the simulated dense cores tend to be low,
	in agreement with observations.}  The low density material
	along the same line of sight (LOS) tends to have larger 
	velocity dispersion, as is also observed.
\item {\it The motion of the simulated core within its local environment
	(difference in core and LOS low density material centroid velocities)
	is large in simulations with high turbulence.}  This
	contradicts our observations that show cores have little
	motion with respect to their local environment, even in
	the clustered environment of Perseus.
\end{itemize}

It therefore appears that reconciling turbulence on large scales and 
quiescence on small scales (both within the core and between the
core and envelope material) requires additional forces to soften
the small-scale dynamics while allowing turbulent motions
to remain on the large scales.  Magnetic fields are a promising
avenue to provide the small-scale damping of motions, however,
in the simulations we analyzed, the large-scale turbulent motions
also decayed, hence a mechanism would be
required to renew large-scale turbulence. 
In the simulations we analyzed, the input turbulence is neither 
long-lived nor refreshed through the global evolution of the cloud 
or the formation of the first protostars, etc.
The inclusion of further turbulent input after the start of the
simulation thus might allow for supersonic motions to be observed
on the large scale while preserving small scale quiescence.

The nature of the simulations themselves are also important 
to keep in mind -- the simulations are done in the thin-sheet 
approximation and have periodic boundaries in the horizontal 
directions. This has some limitations, for example, magnetic field 
lines and turbulent 
vorticity can become tangled in three-dimensions in a manner
that they cannot in thin-sheet models. The periodic boundaries 
can influence the coupling of turbulence on different scales, 
and also impose a rather arbitrary largest scale of turbulence 
in the simulation. We believe that the problem of maintaining very 
turbulent motions on large scales, while keeping 
cores relatively quiescent, may not be addressible 
in any periodic box model, whether thin-sheet or three 
dimensional, and may require a global approach to cloud modelling.

This study illustrates the power of utilizing kinematic `observables'
in simulations to discriminate between simulations with different
initial conditions and to compare with observations, rather than
limiting the analysis to quantities such as the mass function
and star formation efficiency.  This is an era where large (degree)
-scale surveys are becoming more feasible (the COMPLETE Survey
data used in part here
will be eclipsed by more than an order of magnitude by the
next generation of surveys, including the JCMT Gould's Belt
Legacy Surey \citep{Wardthomp07b} and the {\it Herschel} 
Gould's Belt Legacy Survey \citep{Andre05}).
Soon observers will be able to provide information on a statistically
significant set of dense cores and their environment over the extent
of many molecular clouds, which will provide much-needed information
to constrain simulations of star formation.  Simulators should start
to prepare to take advantage of this vast increase in information
when it arrives.

\section{ACKNOWLEDGEMENTS}
We thank the anonymous referee for the thorough review which
improved our paper.  HK would also like to acknowledge valuable 
discussions with people at the CfA, in particular Alyssa Goodman,
Charles Lada, and Phil Myers, and Wolf Dapp (UWO) for a careful
reading of our manuscript.

For the duration of this research, HK was supported by a Natural 
Sciences and Engineering Research
Council of Canada (NSERC) CGS Award and a National Research Council of 
Canada (NRC) GSSSP Award.  DJ and SB are supported by NSERC Discovery
grants. 

\appendix
\section{EFFECT OF SCALINGS}
\label{s_care}

The simulations we analyze are scale-free, which leads to the reasonable
question of how dependent the results we find above are on the 
values of the scale factors (temperature, external pressure and mean
density) we adopted in order to compare the simulations to observations.  
In this section, we address this question and demonstrate that a 
change in the scale factors has a minimal
impact on the `observables' derived and hence the results of our analysis.  
We start with a discussion of our motivation in choosing the values
of the scale factors used in the above analysis, which gives a sense
of the range that could have been applied.  Next we show how 
modifications to the scale factor would effect our results.

\subsection{Physical Motivation For Scalings}
Converting the simulations into observable units requires one to
assume a temperature, external pressure, and mean density.
(Comparison to our observations requires a further assumption
of the distance to the Perseus molecular cloud to convert the
size observed from angular to physical.)

The temperature is constrained to a fairly small
range by observations of star-forming regions, and hence does not
tend to play a significant role in comparisons.  Our previous
observational results assumed that the
temperature is 15~K everywhere.  Recent ammonia observations have shown
that dense cores in Perseus have a mean temperature of $\sim$12~K with
a spread of only a few Kelvin \citep{Rosolowsky08}; as discussed in
\citetalias{Kirk07}, assuming a temperature of 15~K rather than 12~K
does not have a large effect on our observational results.
For ease of comparison with
our previous work, we assume this temperature for the simulations
as well.  

The mean density, on the other hand, can be considered
over a wider range of values, since star formation regions
have a hierarchical structure in size and density. 
The dense cores we observed in our IRAM \nh survey require densities 
of greater than or equal to $\sim 10^{5}$~cm$^{-3}$, in order to be 
observable.  Thus the simulation must reach similar peak densities in
order to be `observable'.  Since the simulations are stopped when the 
maximum column density is  
ten times the mean column density, the corresponding mean 
(three-dimensional) density
cannot be too low for \nh to be detectable.
The mean density also cannot be too large, as this would
lead to a large fraction of the material in the simulation being
at relatively high densities, while observations show that this
is not the case (\S\ref{s_form}).
Also note that a higher assumed mean density implies a smaller size for the
region simulated (eqn 26 of CB06).  

While the above physical arguments place some boundaries on the
adopted scaling parameters, it is clear that there is an allowable
range.  A different choice in the adopted scaling parameters would
lead to a different set of appropriate scale-free parameters 
we adopted to `observe' the simulations (e.g., the beamsize in 
pixels).  In order to determine the magnitude of the effect on
our results that this would cause, we re-analyze the simulations 
using different scale-free parameters.

%

\subsection{Beamsize}
\label{s_care_beam}
We first re-analyze the simulations using a larger beamsize 
(in pixels) with everything else identical to our standard method.
Assuming a mean density $\sim$5 times higher than our nominal
value and keeping the other constants fixed, for example, would
shrink the physical size of each pixel and thus the beamsize 
would change to 7 pixels for \nh (and 3 pixels for \cons).
The observable quantities which
we are interested in are the velocity dispersion of the core and
LOS LDG material and the relative motion of the core and LOS LDG material.
It would be difficult to compare these quantities on a core-to-core
basis, since different cores could be identified in each (a larger
beamsize increases the minimum separation required between cores).
Instead, we use the mean and standard deviation of these quantities
to represent the full span of values found for each simulation.
(We use the absolute value for the relative motion of the core
and LOS LDG material.)

\begin{figure}[htb]
\plottwo{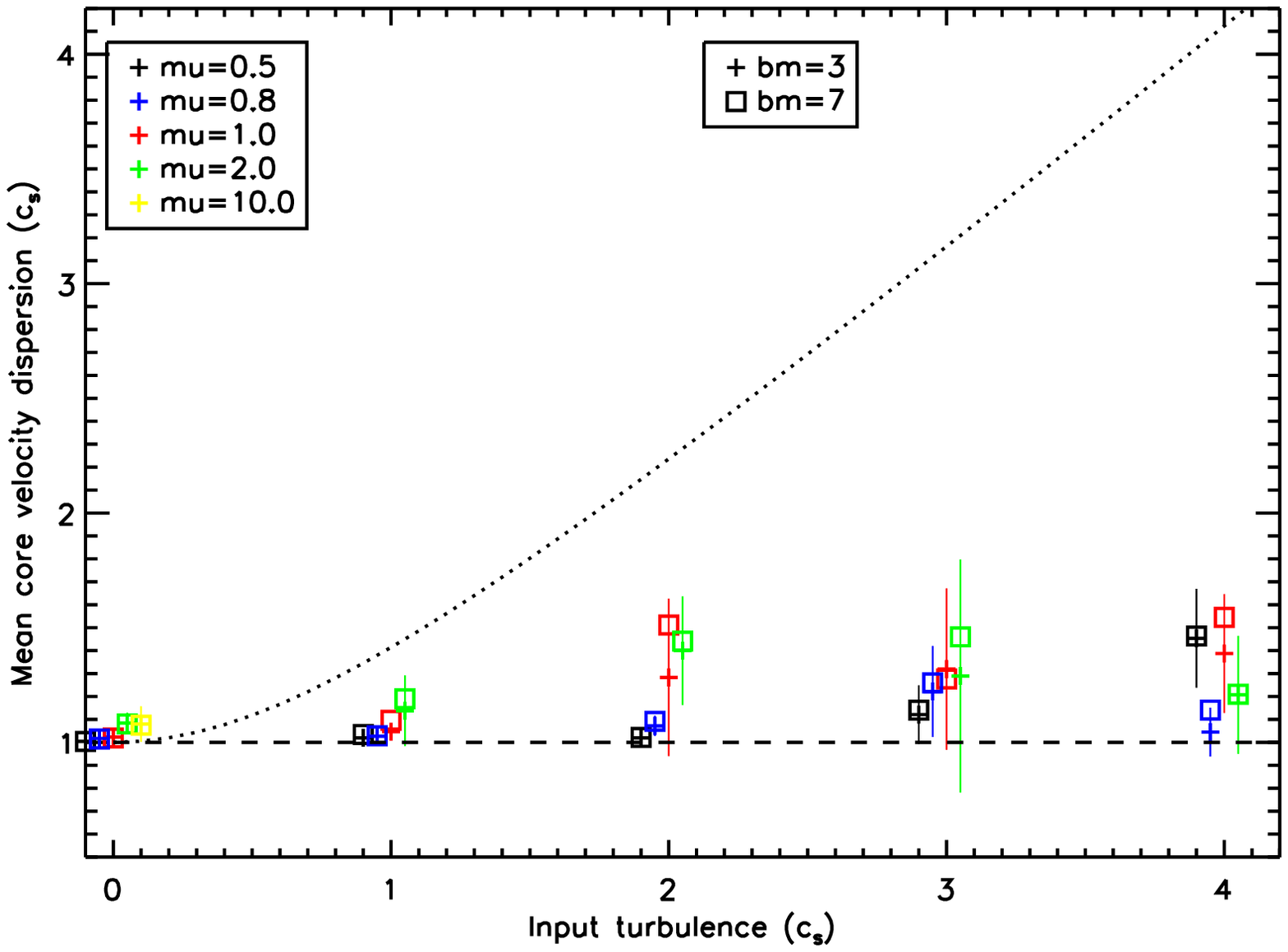}
	{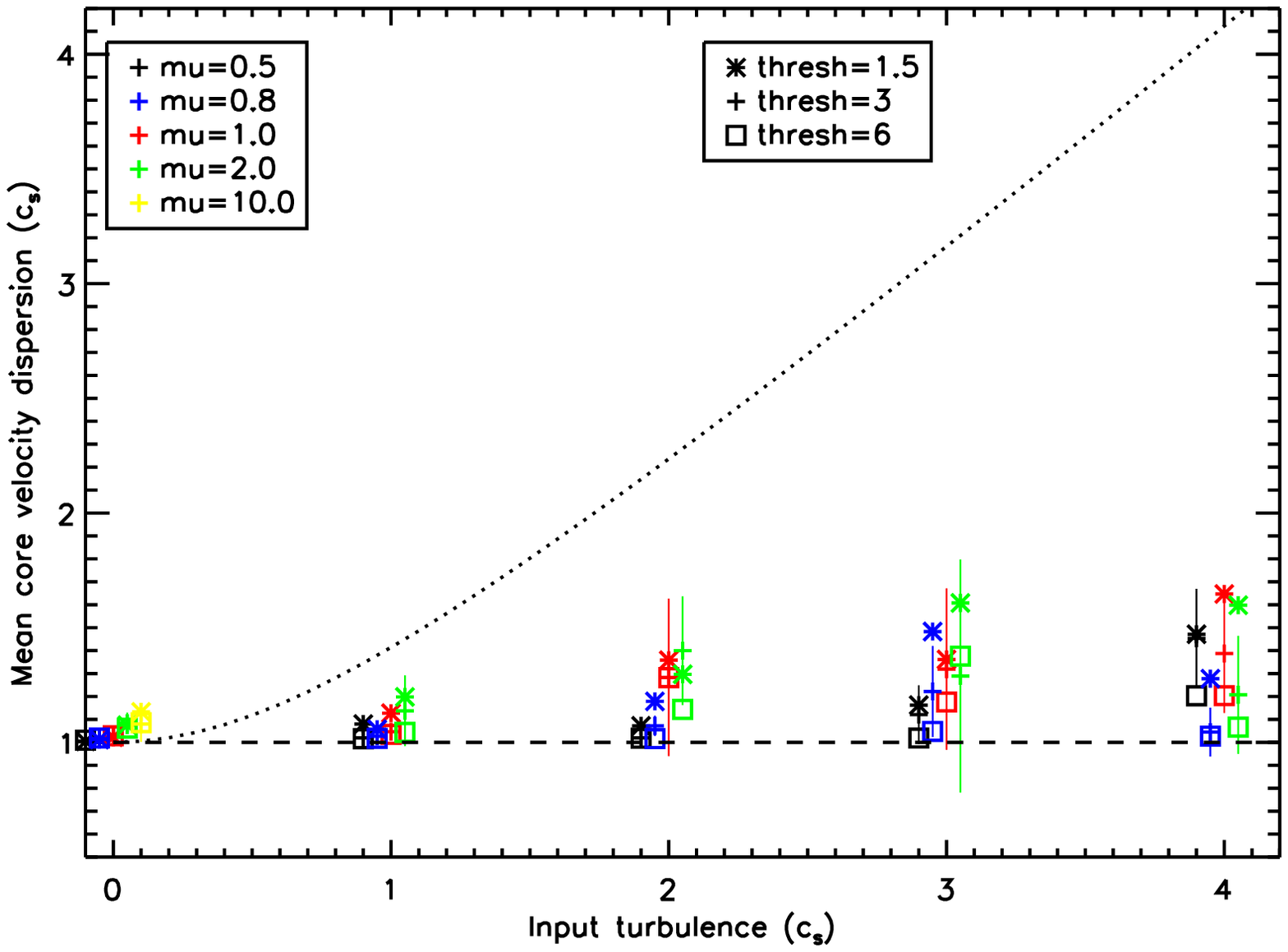}
  \caption{The variation in core velocity dispersion for cores identified
	with a differing beamsize (left) and minimum core column density
	threshold (right) versus the input turbulence.  The plus 
	signs and vertical lines indicate 
	the mean and standard deviation of the velocity dispersion found 
	for each simulation using our nominal values.  The squares and
	asterisks
	indicate the mean found for the simulations with a differing beamsize 
	(left) or core threshold (right).  Note the data points have been
	slightly offset from the input turbulence to allow better visibility.}
\label{var_dVcore} 
\end{figure}

\begin{figure}[htb]
\plottwo{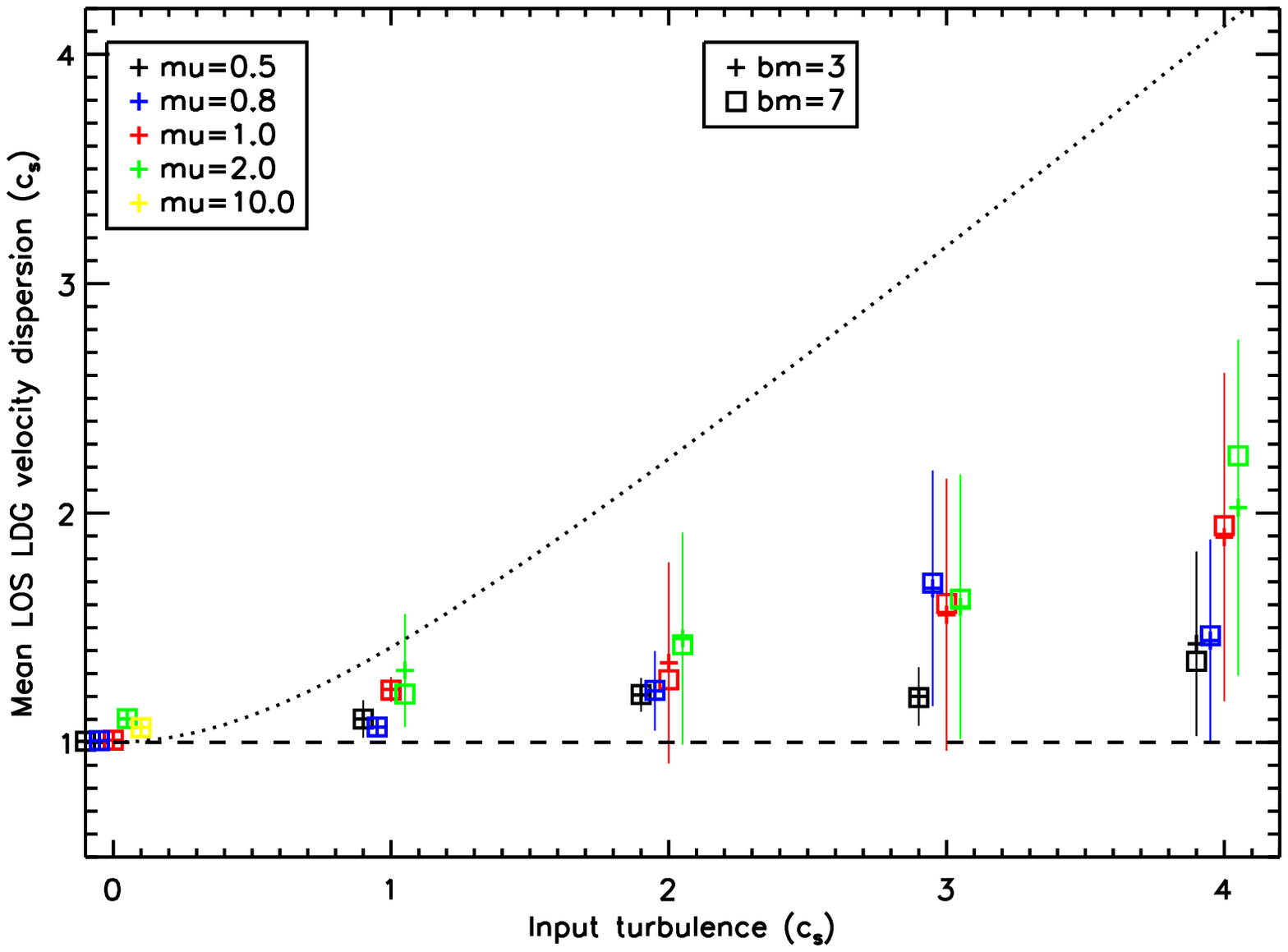}
	{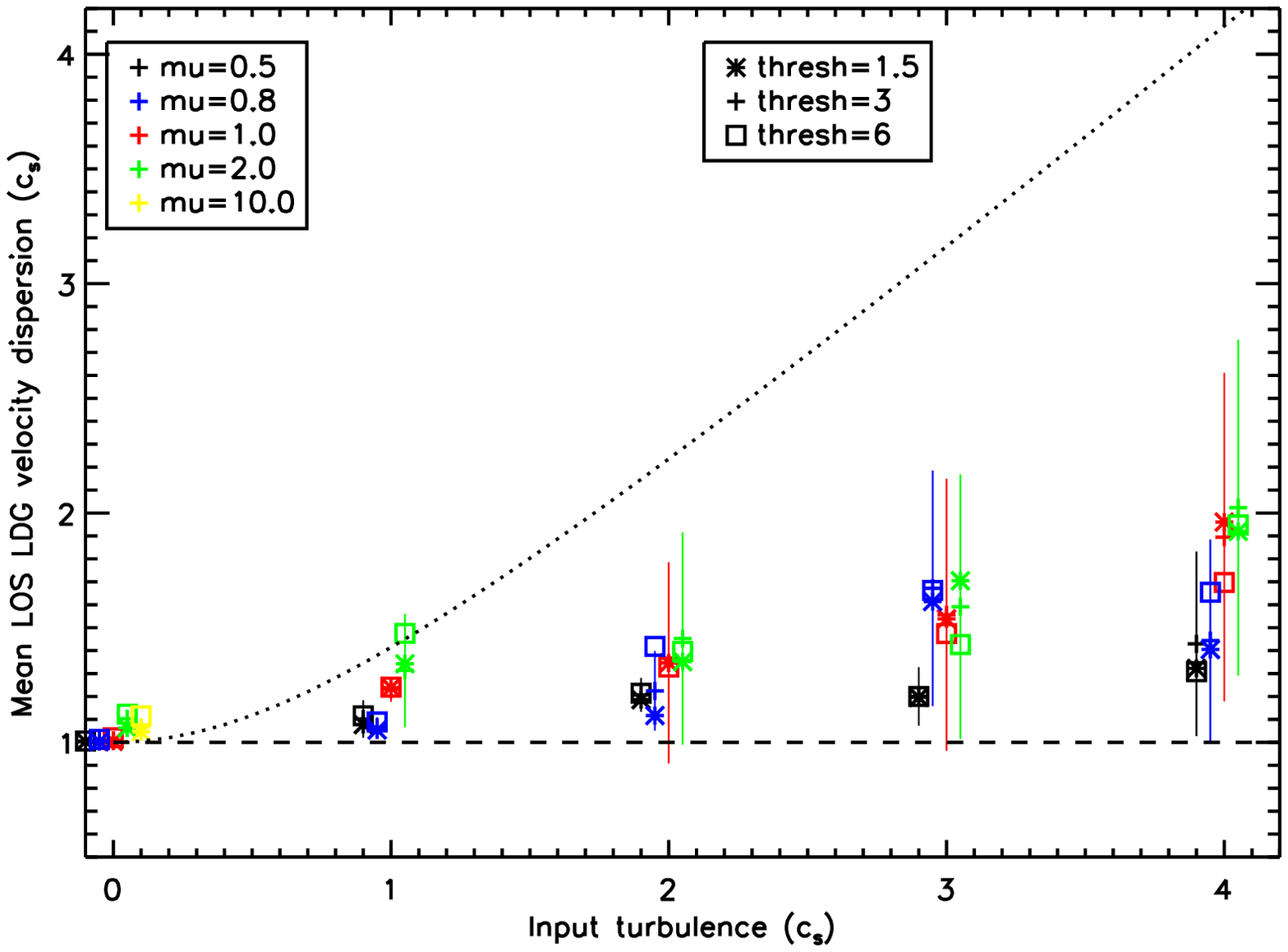}
  \caption{The variation in LOS LDG velocity dispersion for cores identified
	with a differing beamsize (left) and minimum core colum density
	threshold (right) versus the input turbulence.  See 
	Figure~\ref{var_dVcore} for the plotting conventions used.}
  \label{var_dVLOS}
\end{figure}

\begin{figure}[htb]
\plottwo{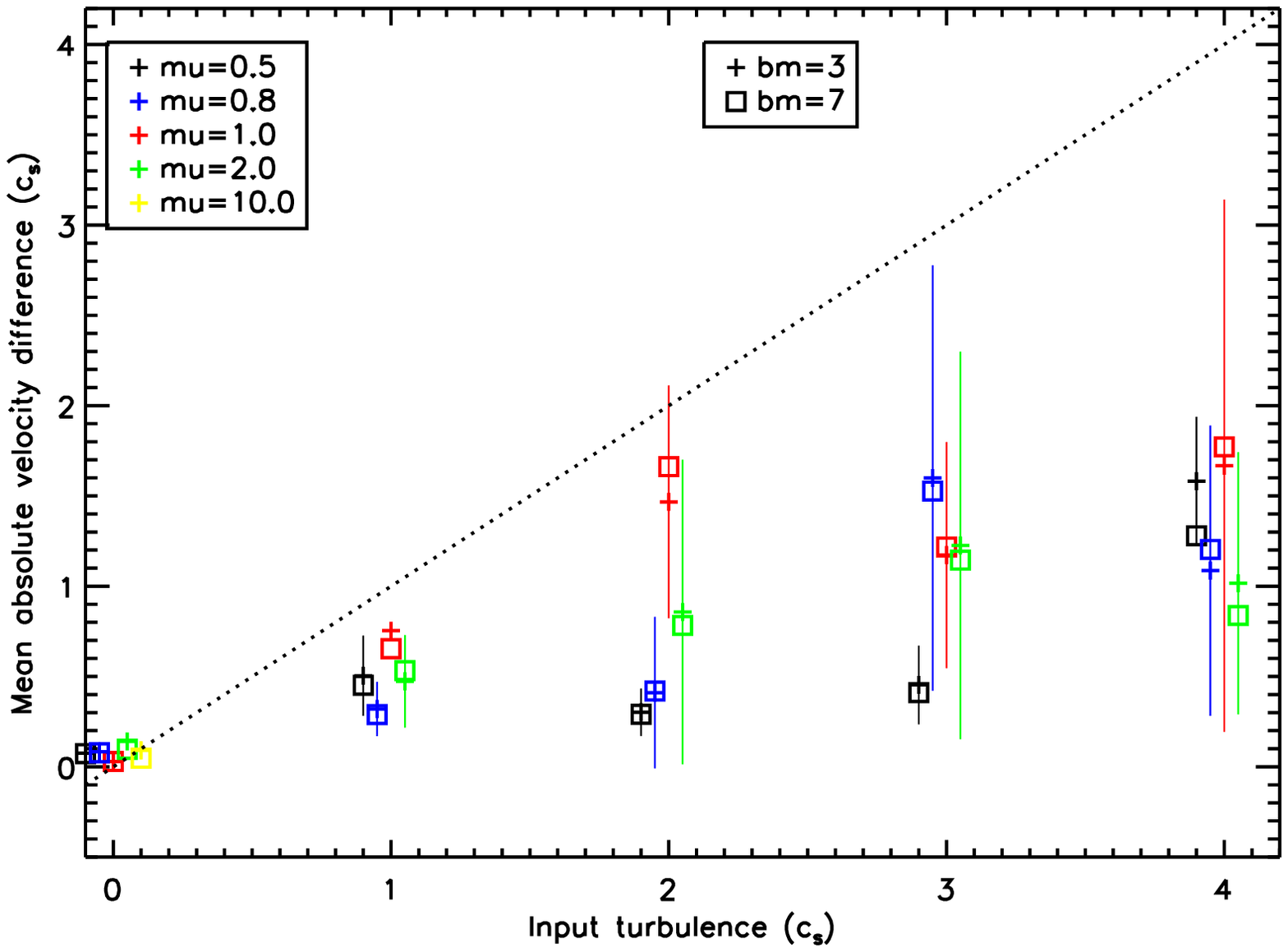}
	{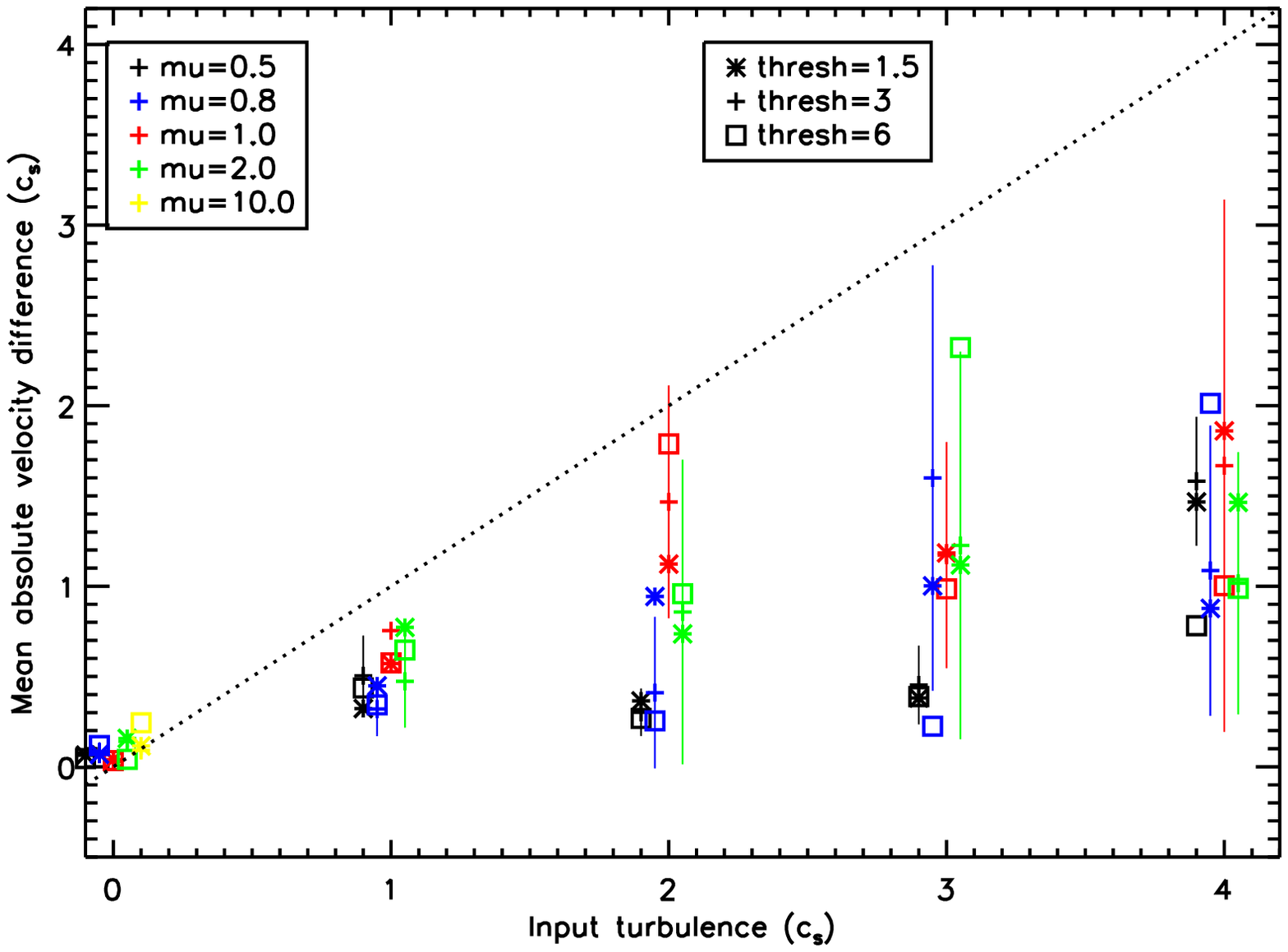}
  \caption{The variation in core to LOS LDG velocity difference for cores
	identified with differing beamsize (left) and minimum core
	column density threshold (right) versus the input turbulence.
	See Figure~\ref{var_dVcore} for the plotting conventions used.}
  \label{var_core_to_LOS}
\end{figure}

Figures~\ref{var_dVcore}, \ref{var_dVLOS}, and \ref{var_core_to_LOS} 
(left hand panels) show the variation of `observables' (the core 
velocity dispersion, LOS LDG velocity dispersion, and core to LOS LDG velocity 
difference) versus
the velocity dispersion of the region as a whole for 
the nominal beamsize (plus signs)
and the larger beamsize (squares).  In most cases, the
larger beamsize has relatively little effect on the `observables'.

The highly turbulent simulations tend to display filamentary structure
in the 2D column density, which translates to clustered peaks in the
1D projected column density distribution which we use to identify cores.
A larger beamsize tends to slightly decrease the number of cores 
identified, particularly in the highly turbulent simulations.
The filamentary structure also leads to a larger amount of higher 
(column) density material in the core's vicinity; hence the
core velocity dispersion in particular, and the LOS LDG velocity
dispersion to a lesser extent tend to increase with an increasing 
beamsize, while the core-to-LOS LDG motion is less affected.

The fraction of mass in dense cores and number of cores detected
(not plotted) show little to no variation with the varying beamsize,
with differences in values no more than a few percent of the mass
and one or two in number of cores.

\subsection{Core Threshold}
\label{s_care_core}
We next re-analyze the data using a higher and lower minimum core column density
threshold (relative to the mean).  A lower chosen physical mean density, for
example, would require a higher minimum core column density relative to the
mean in order to keep the same absolute value.  The simulations have a 
relatively small dynamic range (less than two orders of magnitude in
column density), so we test a minimum core column density threshold
of twice and half our nominal value (i.e., 6 and 1.5 times the mean). 

Figures~\ref{var_dVcore}, \ref{var_dVLOS}, and \ref{var_core_to_LOS}
(right hand panels) show the effect of varying the core identification 
threshold using the same plotting scheme as the left hand panels
described above.  
As is found for the beamsize, the majority of
`observables' do not vary significantly with a different core
identification threshold.  As is also found for the variation in
beamsize, the fraction of mass in the cores varies at most
by a few percent and the number of cores varies by at most a few.

The greatest effect of a higher core identification threshold is
that fewer cores are identified.  As mentioned above, the dynamic
range of the simulations is not large (the simulations are stopped
when the peak column density exceeds roughly 10 times the mean),
hence even the factor of two increase in the minimum core column
density can significantly reduce the number of cores identified,
in many cases to only one or two cores per simulation.  The
very small number statistics drawn from this appears to be the
largest contributor to the variations seen between the two
thresholds.

At the lower core identification threshold, more cores are identified,
although usually not substantially more than the nominal case.  Here,
the main differences originate from a larger fraction of material 
being included as part of the `core' and a smaller fraction as the `LOS LDG'.  
The amount of material excluded from the LOS LDG is a small fraction
of the total, hence the LOS LDG velocity dispersions tend to decrease by
only a small amount.  This material makes up a larger fraction of the
total core mass, hence the increase in core velocity dispersions
tend to be somewhat larger.
There is no obvious trend in the relative motions determined.

\subsection{LOS LDG Upper Threshold}
\label{s_care_los}
The other physical property which plays an implicit rather than explicit
role is a chemical one -- the density range in which
\nh and \co are sensitive.  The minimum density at which \nh is 
sensitive sets the value for the core identification threshold and
folds into the core identification threshold discussed above.  The
maximum density before \co becomes significantly depleted sets
the threshold for which material is considered to contribute to our
LOS LDG measures.  We nominally consider these two values to be
identical, so that all of the material in the simulation is 
traced either by \nh or \cons, but this might not be the case.
Here, we consider cases where \co is sensitive to half and twice
the minimum density for \nh emission.

The left hand panels of 
Figures~\ref{var_envthresh1} and \ref{var_envthresh2} show 
the LOS LDG velocity dispersion (Figure~\ref{var_envthresh1}) 
and core to LOS LDG velocity difference (Figure~\ref{var_envthresh2}) 
versus the velocity dispersion of the region as a whole for
three different maximum \co threshold values.
(Note the core velocity
dispersion remains unchanged, hence is not shown.)  The change in
LOS LDG column density threshold has little effect on the LOS LDG velocity 
dispersion (Figure~\ref{var_envthresh1}), as one would expect 
given that most of the mass
is found at the lower column densities which are included in all cases.
There is a slightly larger change in the core-to-LOS LDG motion since
the higher column density material tends to have a similar velocity
to that of core material.  Although the width of the spectral
feature is little affected by the inclusion of higher column
density material in the LOS LDG spectrum, the position of the peak
tends to be pushed towards velocities nearer to that of the core.
This is clearly illustrated in 
Figure~\ref{var_envthresh2} which shows the lower LOS LDG column
density threshold results in larger relative core-to-LOS LDG motions
while a higher threshold leads to smaller relative motions.
While the change in the core-to-LOS LDG motion is larger than the change
in the LOS LDG velocity dispersion, both changes are relatively small.
This result could have been anticipated from the brief discussion
in Section~4.3 and Figure~\ref{appendix_1}, where it was demonstrated
that the LOS LDG spectra look very similar to the spectrum that
would result from contributions due to {\it all} of the material along 
the line of sight, since only a small fraction of the material
has high column density.

\subsection{LOS LDG Lower Threshold}
Like any molecular tracer, \co also has a minimum density to which
it is sensitive.  In our analysis above, we 
assume that the minimum density reached in the simulation is
sufficiently high that \co will trace all of of the low density
material, but this might not be the case.  Here, we consider cases
where \co is sensitive to only material above one third and
one times the mean column density while keeping the maximum
column density the same as our standard analysis (three times the mean
column density).

\begin{figure}[htb]
\plottwo{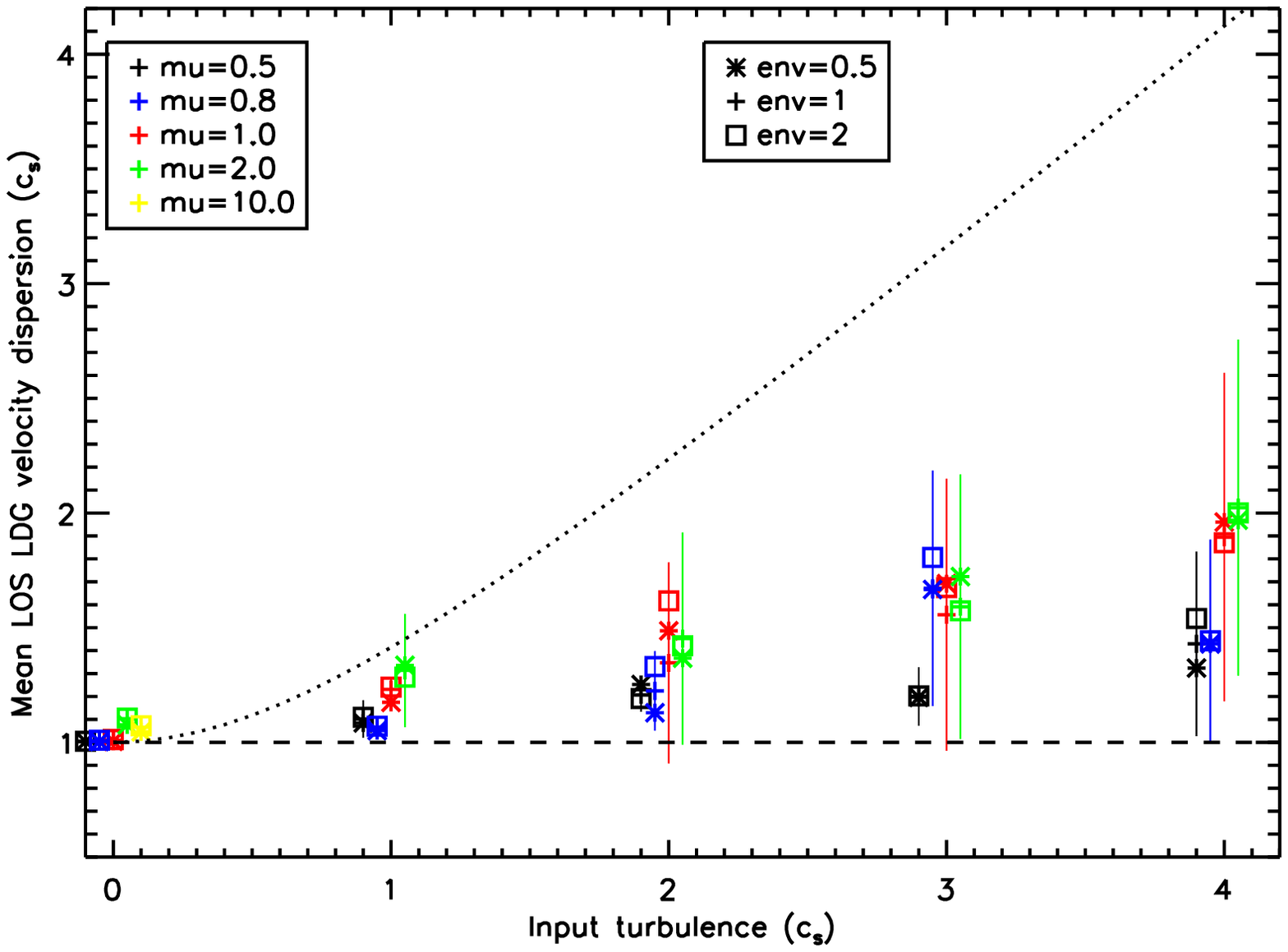}
	{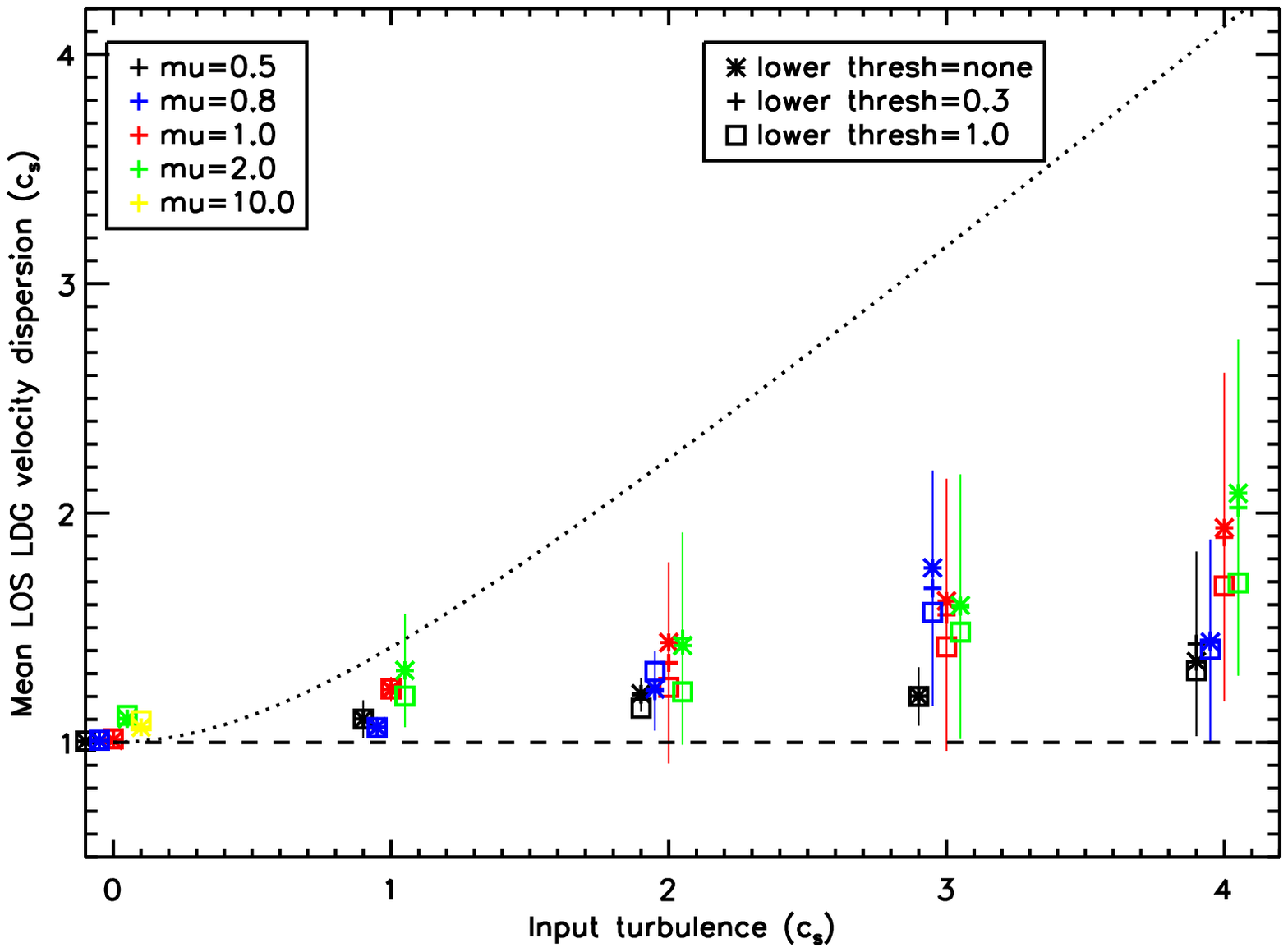}
  \caption{The variation in LOS LDG velocity dispersion for
	a differing LDG column density range.  The left hand panel
	shows the effect of changing the maximum LDG column density 
	threshold while the right hand panel shows the effect of 
	changing the minimum LDG column density threshold. 
	See Figure~\ref{var_dVcore} for the plotting conventions
	used.
	} 
  \label{var_envthresh1}
\end{figure}

\begin{figure}[htb]
\plottwo{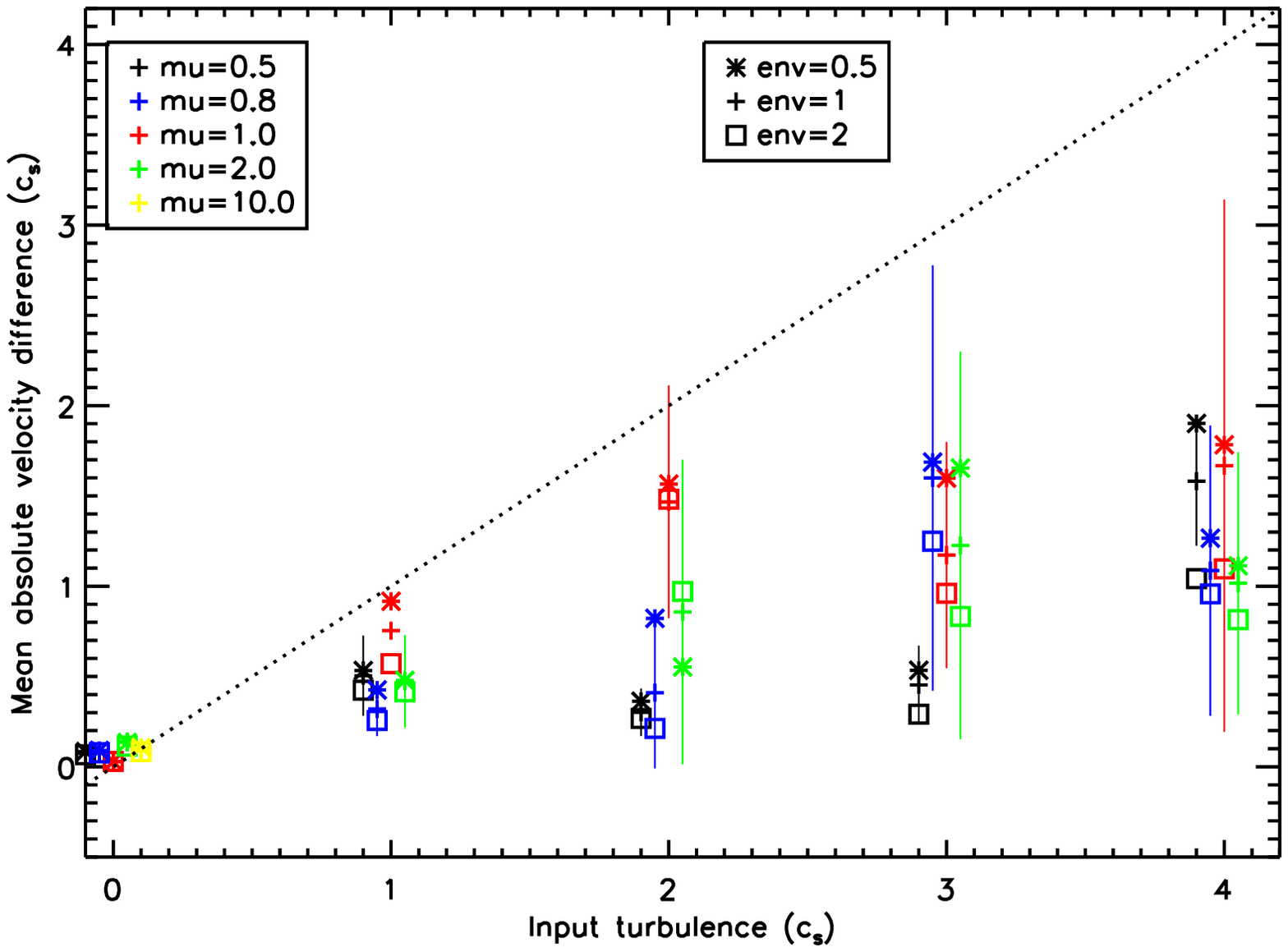}
	{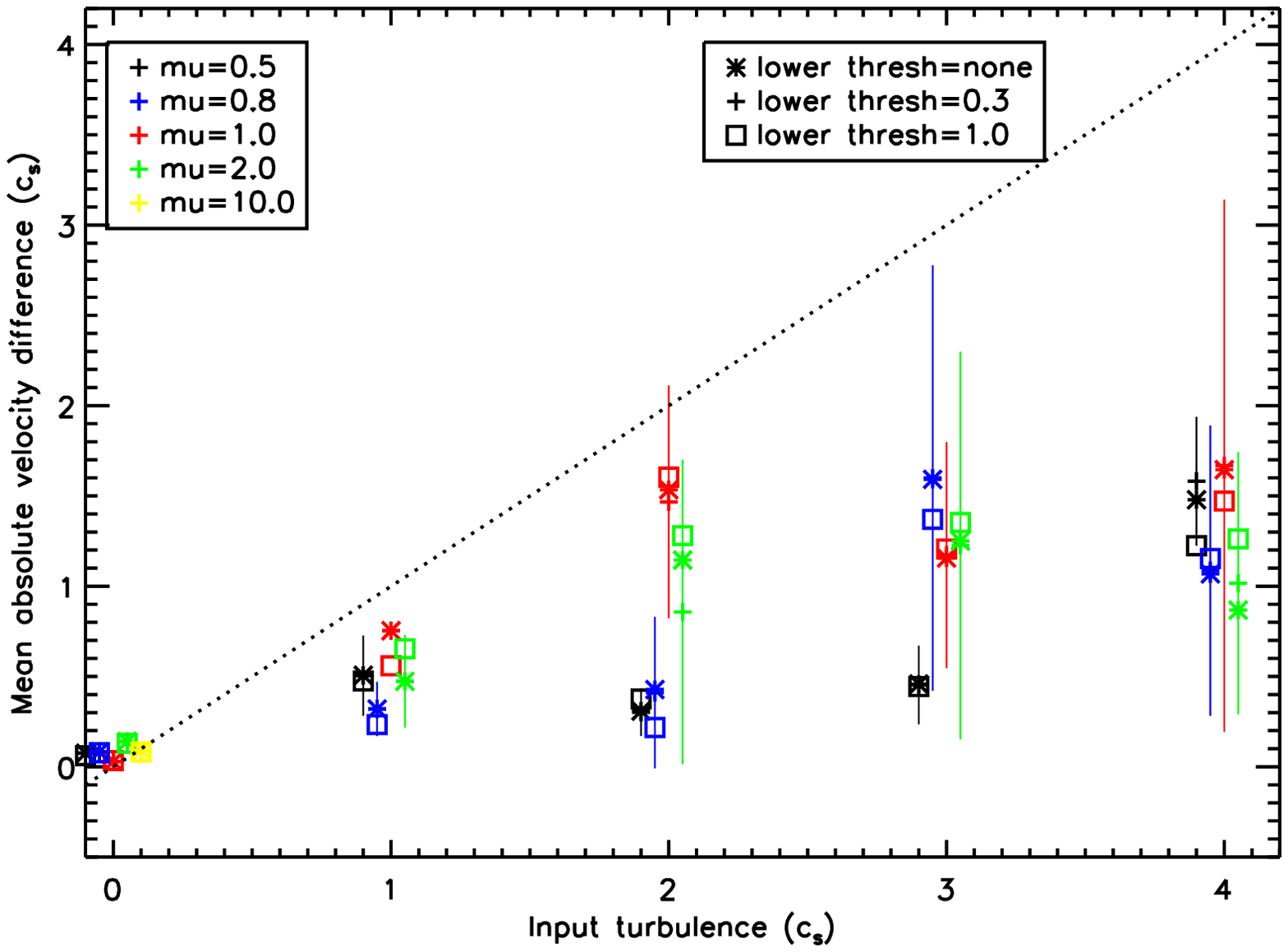}
  \caption{The variation in absolute core to LOS LDG velocity difference 
        for a differing LDG column density range.  The left hand panel
        shows the effect of changing the maximum LDG column density
        threshold while the right hand panel shows the effect of
        changing the minimum LDG column density threshold.
	See Figure~\ref{var_dVcore} for the plotting conventions used.
	} 
  \label{var_envthresh2}
\end{figure}

The right hand panels of Figures~\ref{var_envthresh1} and 
\ref{var_envthresh2} show
the LOS LDG velocity dispersion (Figure~\ref{var_envthresh1}) 
and core to LOS LDG velocity difference (Figure~\ref{var_envthresh2}) 
versus the velocity dispersion of the region as a whole for
three different maximum \co threshold values.  (Note the core velocity
dispersion remains unchanged, hence is not shown.)
The change in the minimum LOS LDG column density has
little effect on the LOS LDG velocity dispersion, except in 
a few of the highly turbulent simulations where the LOS LDG velocity
dispersion becomes somewhat lower when only material between
one and three times the mean column density is considered.  
Similarly, the change in the core to LOS LDG motion calculated
using different minimum LDG column density thresholds is
small, with higher thresholds tending to slightly reduce the velocity
difference.
This illustrates the fact that most of the mass in the simulations 
is at moderately low column densities and that even using a greatly
reduced range of column densities for the LOS LDG calculations
has very little effect on our results.

{\hk
\section{RESOLUTION}
As discussed in \S2.1, the simulations we analyze here have 128 by 128 cells,
in order for a large parameter study to be performed.
To test whether resolution influenced the results of our
calculations, we have also analyzed a subset of four times higher
linear resolution models (512 by 512 cells).  The results show that the
original simulations are sufficient for our purposes.
Here, we demonstrate that our results are not affected by the
resolution adopted.  We analyzed additional simulations with 
a Mach number of 4 and both weak ($\mu_0 = 2$) and strong
($\mu_0 = 0.5$) magnetic fields.  The high Mach number simulations
are the most likely to be affected by resolution since stronger
turbulence has a greater ability to compress material to smaller
scales.

We computed all of the `observable' properties of the two additional
simulations and found them to be consistent with their lower-resolution
simulation counterparts.  Tables~\ref{tab_highres1} and \ref{tab_highres2}
show the mean and standard deviation of the `observables' analyzed
in the paper at both the original and high resolution.  As can be 
seen from these tables, there is no obvious change in any of the
observables due to the higher resolution used.  Note that identical
results for the higher resolution simulations are not expected
because the initial random distribution of turbulent velocities is
different for each simulation run.  Initial conditions, and in
particular how these influence the time at which the simulation
reaches a factor of 10 in column overdensity, have a much larger
effect on the results than the change in resolution.
}

\include{tab1}
\include{tab2}
\include{tab3}
\include{tab4}
\include{tab5}
\include{tab6}
\include{tab7}

\clearpage


\clearpage



\clearpage

\end{document}

%% file: tab1.tex
\begin{deluxetable}{ccccccc}
\tabletypesize{\small}
\tablewidth{0pt}
\tablecolumns{7}
\tablecaption{Properties of Extinction Regions.\label{ext_regs}}
\tablehead{
\colhead{Number \tablenotemark{a}} &
\colhead{Mass\tablenotemark{a}} &
\colhead{Radius\tablenotemark{a}} &
\colhead{$\sigma V_{grav}$\tablenotemark{b}} &
\colhead{$\sigma V_{CO}$\tablenotemark{c}} &
\colhead{Coverage\tablenotemark{c}} &
\colhead{Notes \tablenotemark{d}}
\\
\colhead{} &
\colhead{(M$_{\odot}$)} &
\colhead{(\arcmin)} &
\colhead{(km~s$^{-1}$)} &
\colhead{(km~s$^{-1}$)} &
\colhead{(\%)} &
\colhead{ }
}
\startdata
  1& 859.6& 12.9& 0.89& 0.62& 100& B5\\
  2&1938.9& 18.7& 1.11& 0.91& 100& IC348\\
  3& 780.6& 12.3& 0.87& 1.03&  96& SW of IC348\\
  4& 560.5& 11.2& 0.77& 0.94&  56& E of B1\\
  5& 441.1&  9.7& 0.73& 0.87& 100& B1\\
  6& 257.6&  7.6& 0.63& 0.87&  52& SW of B1\\
  7& 973.3& 14.8& 0.88& 1.22& 100& NGC1333\\
  8& 246.2&  7.6& 0.62& 0.83& 100& S of NGC1333\\
  9& 240.1&  7.5& 0.62& 0.87&  40& L1455\\
 10& 173.7&  6.4& 0.56& 0.79& 100& L1448\\
 11& 107.4&  5.2& 0.50& 0.76&  84& S of L1448\\
\enddata
\tablenotetext{a}{Extinction region number, mass and size 
	from \citetalias{Kirk06}}
\tablenotetext{b}{Velocity dispersion required for the region to be
	in virial equilibrium (measured using a Gaussian sigma), 
	as calculated in \citetalias{Kirk07}.}
\tablenotetext{c}{Velocity dispersion (using a Gaussian sigma) 
	and fractional coverage of the 
	extinction region observed in \thirco COMPLETE data.  Those regions
	with less than 80\% coverage are not included in the analysis of this 
	paper.  See text for details.}
\tablenotetext{d}{Descriptive location of extinction region.}
\end{deluxetable}

%% file: tab2.tex
\begin{deluxetable}{cccccc}
\tabletypesize{\normalsize}
\tablewidth{0pt}
\tablecolumns{6}
\tablecaption{Observed Core Formation Statistics\label{obs_effic}}
\tablehead{
\colhead{Ext \# \tablenotemark{a} }&
\colhead{$\sigma V_{tot} / c_{S}$\tablenotemark{b} }&
\colhead{$N_{C,SCUBA}$\tablenotemark{c} } &
\colhead{$N_{C,tot}$\tablenotemark{d} } &
\colhead{CFE$_{C,SCUBA}$\tablenotemark{c} } &
\colhead{CFE$_{C,tot}$\tablenotemark{d} }
}
\startdata
  1  &  2.9  &    2  &    8  &  0.1  &  0.7 \\
  2  &  4.1  &   11  &   19  &  0.7  &  1.0 \\
  3  &  4.6  &    0  &    2  &  0.0  &  0.2 \\
  5  &  3.9  &    7  &   17  &  3.1  &  5.0 \\
  7  &  5.4  &   26  &   27  &  5.0  &  5.1 \\
  8  &  3.7  &    0  &    0  &  0.0  &  0.0 \\
 10  &  3.5  &    5  &    7  &  8.0  &  9.0 \\
 11  &  3.4  &    0  &    2  &  0.0  &  1.6 \\
\enddata
\tablenotetext{a}{Extinction region number as in Table~\ref{ext_regs}.
	Only regions where \thirco data coverage exceeds 80\% are shown.}
\tablenotetext{b}{Total velocity dispersion relative to the thermal
	value calculated from \thirco data with the thermal component
	corrected to that expected for the mean gas.}
\tablenotetext{c}{Number and percentage of the region's mass in 
	SCUBA-associated \nh dense cores.}
\tablenotetext{d}{Number of cores and core formation efficiency (CFE)
	as a percentage.  See \S\ref{s_form} for details.}
\end{deluxetable}

%% file: tab3.tex
\begin{deluxetable}{cccccc}
\tabletypesize{\small}
\tablewidth{0pt}
\tablecolumns{6}
\tablecaption{Simulation Timescales.\label{tab_time}}
\tablehead{
\colhead{$M$ \tablenotemark{a}} &
\colhead{} &
\colhead{} &
\colhead{Time ($t_{0}$) \tablenotemark{b}} &
\colhead{} &
\colhead{} 
\\
\colhead{($c_{s}$)} &
\colhead{$\mu_{0} = 0.5$\tablenotemark{c}} &
\colhead{$\mu_{0} = 0.8$\tablenotemark{c}} &
\colhead{$\mu_{0} = 1.0$\tablenotemark{c}} &
\colhead{$\mu_{0} = 2.0$\tablenotemark{c}} &
\colhead{$\mu_{0} = 10$\tablenotemark{c}} 
}
\startdata
0&  204 &  167 &  121  &  23  &  12 \\
1&  49.6&  31.6&    4.3&   2.5&  -- \\
2&  28.7&  10.8&    2.5&   1.1&  -- \\
3&  36.0&   0.9&    0.8&   0.8&  -- \\
4&   2.1&   1.0&    0.7&   0.5&  -- \\
\enddata
\tablenotetext{a}{Input turbulence (in sound speed units).}
\tablenotetext{b}{Time for a point in the simulation to reach
	a column density of roughly ten times the mean in units
	of $t_0$ (eqn 5).  Note that the sound crossing time
	of the length scale of maximum growth is $4\pi$ (see
	\S4.1 for more details).}
\tablenotetext{c}{Input mass to magnetic flux ratio.}
\end{deluxetable}

%% file: tab4.tex
\begin{deluxetable}{ccccccccccc}
\tabletypesize{\scriptsize}
\tablewidth{0pt}
\tablecolumns{11}
\tablecaption{Simulation Dynamic Observables\label{sims_dyn}}
\tablehead{
\colhead{$M$\tablenotemark{a}} &
\colhead{$\mu_{0}$\tablenotemark{a}} &
\colhead{Projection\tablenotemark{a}} &
\multicolumn{2}{c}{$\sigma V_{LOS LDG,all}$\tablenotemark{b}} &
\multicolumn{2}{c}{$\sigma V_{LOS LDG,cores}$\tablenotemark{c}} &
\multicolumn{2}{c}{$\sigma V_{core}$\tablenotemark{c}} &
\multicolumn{2}{c}{Core to LOS LDG\tablenotemark{c}} \\
\colhead{} &
\colhead{} &
\colhead{} &
\colhead{mean} &
\colhead{stddev} &
\colhead{mean} &
\colhead{stddev} &
\colhead{mean} &
\colhead{stddev} &
\colhead{mean} &
\colhead{stddev} 
}
\startdata
 0 & 0.5 & X &  1.00 &  0.00 &  1.00 &  0.00 &  1.00 &  0.00 &  0.07 &  0.05 \\
 0 & 0.5 & Y &  1.00 &  0.00 &  1.00 &  0.00 &  1.00 &  0.01 &  0.08 &  0.06 \\
 1 & 0.5 & X &  1.10 &  0.10 &  1.12 &  0.13 &  1.03 &  0.03 &  0.40 &  0.34 \\
 1 & 0.5 & Y &  1.07 &  0.05 &  1.09 &  0.04 &  1.01 &  0.00 &  0.58 &  0.05 \\
 2 & 0.5 & X &  1.12 &  0.06 &  1.16 &  0.01 &  1.02 &  0.00 &  0.23 &  0.01 \\
 2 & 0.5 & Y &  1.17 &  0.11 &  1.29 &    -- &  1.02 &   --  &  0.45 &   --  \\
 3 & 0.5 & X &  1.24 &  0.19 &  1.25 &  0.16 &  1.17 &  0.15 &  0.38 &  0.05 \\
 3 & 0.5 & Y &  1.13 &  0.07 &  1.13 &  0.02 &  1.05 &  0.03 &  0.55 &  0.34 \\
 4 & 0.5 & X &  2.22 &  0.43 &  1.37 &  0.32 &  1.34 &  0.13 &  1.68 &  0.22 \\
 4 & 0.5 & Y &  1.95 &  0.83 &  1.54 &  0.55 &  1.68 &  0.17 &  1.39 &  0.61 \\
 0 & 1.0 & X &  1.00 &  0.00 &  1.01 &  0.00 &  1.02 &  0.03 &  0.02 &  0.01 \\
 0 & 1.0 & Y &  1.00 &  0.00 &  1.01 &  0.01 &  1.03 &  0.03 &  0.04 &  0.04 \\
 1 & 1.0 & X &  1.08 &  0.07 &  1.19 &   --  &  1.07 &   --  &  0.74 &   --  \\
 1 & 1.0 & Y &  1.16 &  0.11 &  1.27 &   --  &  1.03 &   --  &  0.76 &   --  \\
 2 & 1.0 & X &  1.85 &  0.54 &  1.17 &  0.31 &  1.23 &  0.33 &  1.48 &  0.13 \\
 2 & 1.0 & Y &  1.73 &  0.49 &  1.49 &  0.50 &  1.34 &  0.42 &  1.46 &  1.01 \\
 3 & 1.0 & X &  2.59 &  0.97 &  1.77 &  0.76 &  1.16 &  0.16 &  1.06 &  0.69 \\
 3 & 1.0 & Y &  2.18 &  0.37 &  1.31 &  0.14 &  1.52 &  0.45 &  1.31 &  0.60 \\
 4 & 1.0 & X &  3.28 &  0.97 &  2.09 &  0.90 &  1.48 &  0.26 &  1.35 &  1.02 \\
 4 & 1.0 & Y &  3.00 &  0.81 &  1.70 &  0.44 &  1.31 &  0.25 &  1.95 &  1.80 \\
 0 & 2.0 & X &  1.06 &  0.04 &  1.10 &   --  &  1.05 &   --  &  0.17 &   --  \\
 0 & 2.0 & Y &  1.04 &  0.03 &  1.10 &  0.00 &  1.10 &  0.05 &  0.13 &  0.00 \\
 1 & 2.0 & X &  1.55 &  0.22 &  1.35 &  0.34 &  1.23 &  0.18 &  0.36 &  0.23 \\
 1 & 2.0 & Y &  1.34 &  0.09 &  1.27 &  0.08 &  1.05 &  0.04 &  0.58 &  0.26 \\
 2 & 2.0 & X &  1.92 &  0.73 &  1.61 &  0.54 &  1.32 &  0.25 &  0.63 &  0.99 \\
 2 & 2.0 & Y &  1.86 &  0.21 &  1.22 &  0.18 &  1.56 &  0.15 &  1.32 &  0.04 \\
 3 & 2.0 & X &  3.17 &  0.52 &  1.64 &  0.70 &  1.14 &  0.15 &  1.07 &  1.03 \\
 3 & 2.0 & Y &  2.41 &  0.65 &  1.50 &  0.22 &  1.60 &  0.84 &  1.54 &  1.24 \\
 4 & 2.0 & X &  3.02 &  0.70 &  2.14 &  1.02 &  1.52 &  0.14 &  1.46 &  0.90 \\
 4 & 2.0 & Y &  3.17 &  0.76 &  1.96 &  0.57 &  1.08 &  0.17 &  0.84 &  0.61 \\
 0 &10.0 & X &  1.04 &  0.02 &  1.08 &  0.03 &  1.10 &  0.09 &  0.09 &  0.04 \\
 0 &10.0 & Y &  1.03 &  0.03 &  1.05 &  0.03 &  1.06 &  0.07 &  0.09 &  0.06 \\
 0 & 0.8 & X &  1.00 &  0.00 &  1.02 &   --  &  1.04 &   --  &  0.11 &   --  \\
 0 & 0.8 & Y &  1.00 &  0.00 &  1.00 &  0.00 &  1.01 &  0.01 &  0.07 &  0.06 \\
 1 & 0.8 & X &  1.03 &  0.01 &  1.05 &  0.03 &  1.02 &  0.01 &  0.29 &  0.21 \\
 1 & 0.8 & Y &  1.05 &  0.05 &  1.08 &  0.05 &  1.03 &  0.03 &  0.34 &  0.15 \\
 2 & 0.8 & X &  1.37 &  0.24 &  1.16 &  0.20 &  1.09 &  0.07 &  0.54 &  0.73 \\
 2 & 0.8 & Y &  1.20 &  0.13 &  1.29 &  0.14 &  1.06 &  0.03 &  0.32 &  0.23 \\
 3 & 0.8 & X &  2.75 &  0.45 &  1.64 &  0.35 &  1.11 &  0.11 &  1.87 &  1.25 \\
 3 & 0.8 & Y &  2.25 &  0.66 &  1.72 &  0.75 &  1.45 &  0.08 &  1.06 &  1.00 \\
 4 & 0.8 & X &  2.75 &  0.44 &  1.48 &  0.52 &  1.05 &  0.09 &  0.89 &  0.99 \\
 4 & 0.8 & Y &  2.22 &  0.49 &  1.41 &  0.38 &  1.04 &  0.13 &  1.24 &  0.64 \\
\enddata
\tablenotetext{a}{Initial Mach number, mass to magnetic flux ratio, and
	projection for each simulation.  See \S\ref{s_sim} for more detail.}
\tablenotetext{b}{The LOS LDG velocity dispersion measured along every line 
	of sight using the FWQM and converted into the equivalent
	Gaussian sigma measure.  See \S4.4 for more details.}
\tablenotetext{c}{The mean and standard deviation of the LOS LDG velocity 
 dispersion, core velocity dispersion, and absolute core to LOS LDG velocity
 difference for every core 
 identified.  (No standard deviation is listed where only one measurement
 exists).}
\end{deluxetable}

%% file: tab5.tex
\begin{deluxetable}{cccccc}
\tabletypesize{\scriptsize}
\tablewidth{0pt}
\tablecolumns{6}
\tablecaption{Simulation Core Formation Statistics\label{sims_effic}}
\tablehead{
\colhead{$M$\tablenotemark{a}} &
\colhead{$\sigma V_{box}$\tablenotemark{b}} &
\colhead{$\mu_{0}$\tablenotemark{a}} &
\colhead{Projection\tablenotemark{a}} &
\colhead{$N_{C}$\tablenotemark{c}} &
\colhead{$CFE_{C}$\tablenotemark{c}} 
}
\startdata
  0  &  1.0  &  0.5  &   X  &   3  &   1.1 \\
  0  &  1.0  &  0.5  &   Y  &   0  &   0.0 \\
  1  &  1.2  &  0.5  &   X  &   3  &   1.9 \\
  1  &  1.1  &  0.5  &   Y  &   1  &   0.0 \\
  2  &  1.7  &  0.5  &   X  &   2  &   0.9 \\
  2  &  1.3  &  0.5  &   Y  &   1  &   0.2 \\
  3  &  2.1  &  0.5  &   X  &   4  &   4.0 \\
  3  &  1.3  &  0.5  &   Y  &   2  &   0.9 \\
  4  &  2.7  &  0.5  &   X  &   4  &   2.8 \\
  4  &  3.7  &  0.5  &   Y  &   2  &   1.4 \\
  0  &  1.0  &  1.0  &   X  &   2  &   0.9 \\
  0  &  1.0  &  1.0  &   Y  &   0  &   0.0 \\
  1  &  1.3  &  1.0  &   X  &   1  &   2.8 \\
  1  &  1.2  &  1.0  &   Y  &   0  &   0.0 \\
  2  &  1.9  &  1.0  &   X  &   2  &   1.6 \\
  2  &  1.8  &  1.0  &   Y  &   1  &   0.1 \\
  3  &  3.2  &  1.0  &   X  &   4  &   2.8 \\
  3  &  3.3  &  1.0  &   Y  &   2  &   1.6 \\
  4  &  3.5  &  1.0  &   X  &   5  &   4.9 \\
  4  &  3.6  &  1.0  &   Y  &   4  &   2.7 \\
  0  &  1.1  &  2.0  &   X  &   1  &   0.6 \\
  0  &  1.0  &  2.0  &   Y  &   1  &   0.3 \\
  1  &  1.6  &  2.0  &   X  &   4  &   2.3 \\
  1  &  1.5  &  2.0  &   Y  &   1  &   0.4 \\
  2  &  2.1  &  2.0  &   X  &   4  &   3.0 \\
  2  &  2.5  &  2.0  &   Y  &   1  &   0.4 \\
  3  &  3.6  &  2.0  &   X  &   6  &   3.9 \\
  3  &  3.3  &  2.0  &   Y  &   2  &   0.4 \\
  4  &  4.3  &  2.0  &   X  &   4  &   1.3 \\
  4  &  4.1  &  2.0  &   Y  &   3  &   0.5 \\
  0  &  1.0  & 10.0  &   X  &   5  &   1.6 \\
  0  &  1.0  & 10.0  &   Y  &   1  &   0.3 \\
  0  &  1.0  &  0.8  &   X  &   1  &   0.6 \\
  0  &  1.0  &  0.8  &   Y  &   1  &   0.1 \\
  1  &  1.1  &  0.8  &   X  &   2  &   1.2 \\
  1  &  1.3  &  0.8  &   Y  &   0  &   0.0 \\
  2  &  1.7  &  0.8  &   X  &   2  &   1.8 \\
  2  &  1.4  &  0.8  &   Y  &   0  &   0.0 \\
  3  &  2.9  &  0.8  &   X  &   4  &   2.0 \\
  3  &  3.2  &  0.8  &   Y  &   1  &   0.1 \\
  4  &  4.4  &  0.8  &   X  &   5  &   2.0 \\
  4  &  4.0  &  0.8  &   Y  &   4  &   0.6 \\
\enddata
\tablenotetext{a}{For simulation details, see \S\ref{s_sim}.}
\tablenotetext{b}{Velocity dispersion measured for the region in units
	of the sound speed; see
        \S4.3 for further details.}
\tablenotetext{c}{Number of cores and core formation efficiency (CFE) as
	a percentage.}
\end{deluxetable}

%% file: tab6.tex
\begin{deluxetable}{ccccc}
\tabletypesize{\small}
\tablewidth{0pt}
\tablecolumns{5}
\tablecaption{Comparison Between $\mu_0=2.0$, Mach 4 Simulations
	\label{tab_highres1}}
\tablehead{
\colhead{ } &
\multicolumn{2}{c}{N=128} &
\multicolumn{2}{c}{N=512} \\
\colhead{Quantity \tablenotemark{a} } &
\colhead{Mean} & \colhead{Std Dev} &
\colhead{Mean} & \colhead{Std Dev} 
}
\startdata
Region Velocity Dispersion & 4.24 & 0.14 & 4.21 & 0.04 \\
All LOS Vel Disp X (matching \thirco observations) &
	2.90 & 0.94 & 2.66 & 0.62 \\
All LOS Velocity Dispersion Y Projection & 3.28 & 1.16 & 2.64 & 0.65 \\
LOS Vel Disp where Cores (matching \co observations) &
	2.02 & 0.73 & 1.89 & 0.67 \\
Core Velocity Dispersion & 1.21 & 0.26 & 1.26 & 0.41 \\
Absolute Core-LOS Vel Difference & 1.02 & 0.73 & 2.04 & 1.19 \\
Absolute Core-LOS Vel Diff / LOS Vel Disp & 0.49 & 0.34 & 0.92 & 0.45 \\
Absolute Core Velocities & 2.23 & 2.07 & 3.54 & 2.27 \\
Number of Cores & 3.5 & 0.7 & 6.5 & 2.1 \\
Percentage of Mass in Cores & 0.9 & 0.5 & 0.5 & 0.6 \\
\enddata

\tablenotetext{a}{All velocities in units of the sound speed.}

\end{deluxetable}

%% file: tab7.tex
\begin{deluxetable}{ccccc}
\tabletypesize{\small}
\tablewidth{0pt}
\tablecolumns{5}
\tablecaption{Comparison Between $\mu_0=0.5$, Mach 4 Simulations 
	\label{tab_highres2}}
\tablehead{
\colhead{ } &
\multicolumn{2}{c}{N=128} &
\multicolumn{2}{c}{N=512} \\
\colhead{Quantity \tablenotemark{a} } &
\colhead{Mean} & \colhead{Std Dev} &
\colhead{Mean} & \colhead{Std Dev} 
}
\startdata
Region Velocity Dispersion & 3.17 & 0.71 & 3.90 & 0.73 \\
All LOS Vel Disp X (matching \thirco observations) &
	2.27 & 0.57 & 2.85 & 1.11 \\
All LOS Velocity Dispersion Y Projection & 1.96 & 0.79 & 2.16 & 0.92 \\
LOS Vel Disp where Cores (matching \co observations) &
	1.43 & 0.40 & 1.53 & 0.33 \\
Core Velocity Dispersion & 1.45 & 0.22 & 1.28 & 0.21 \\
Absolute Core-LOS Vel Difference & 1.58 & 0.36 & 1.90 & 1.32 \\
Absolute Core-LOS Vel Diff / LOS Vel Disp & 1.09 & 0.33 & 0.91 & 0.27 \\
Absolute Core Velocities & 1.91 & 1.24 & 2.01 & 1.53 \\
Number of Cores & 2.5 & 1.4 & 3.0 & 0.0 \\
Percentage of Mass in Cores & 2.1 & 1.0 & 1.4 & 1.0 \\
\enddata

\tablenotetext{a}{All velocities in units of the sound speed.}

\end{deluxetable}

%% file: ms.bbl
\begin{thebibliography}{}
\bibitem[Andr\'e, Ward-Thompson, \& Barsony(1993)]{Andre93} Andr\'e, P.,
	Ward-Thompson, D., \& Barsony, M. 1993, ApJ, 406, 122
\bibitem[Andr\'e \& Saraceno(2005)]{Andre05} Andr\'e, P. \& Saraceno, P.
	2005, Proceedings of the dusty and molecular universe: a prelude
	to Herschel and ALMA, A. Wilson (eds), ESA Publications, Noordwijk,
	Netherlands, p179-184
\bibitem[Ayliffe et al.(2007)]{Ayliffe07} Ayliffe, B.~A., Langdon, J.~C., 
	Cohl, H.~S., \& Bate, M.R. 2007, MNRAS, 374, 1198
\bibitem[Ballesteros-Paredes et al.(2003)]{BP03} Ballesteros-Paredes, J.
	Klessen, R.~S. \& V\'azquez-Semadeni, E. 2003, ApJ, 592, 188
\bibitem[Basu \& Ciolek(2004)]{BC04} Basu, S. \& Ciolek, G.~E. 2004, ApJ,
	607, 39L
\bibitem[Basu, Ciolek \& Wurster(2009)]{BCW09} Basu, S., Ciolek, G.~E.,
	\& Wurster, J. 2009, NewA, 14, 221
\bibitem[Basu, Ciolek, Dapp \& Wurster(2009)]{BCDW09} Basu, S., 
	Ciolek, G.~E., Dapp, W.~B. \& Wurster, J. NewA, 14, 483
\bibitem[Benson \& Myers(1989)]{Benson89} Benson, P.~J. \& Myers, P.~C.
	1989, ApJS, 71, 89
\bibitem[Bonnell, Larson, \& Zinnecker(2007)]{Bonnell07} Bonnell, I.~A.,
	Larson, R.~B., Zinnecker, H. 2007, Protostars and Planets V,
	B. Reipurth, D. Jewitt, \& K. Keil (eds), University of Arizona
	Press, Tuscon, p149-164
\bibitem[Burkert \& Hartmann(2004)]{Burkert04} Burkert, A. \& Hartmann, L.
	2004, ApJ, 616, 288
\bibitem[${\rm \breve{C}}$ernis(1993)]{Cernis93} ${\rm \breve{C}}$ernis, K. 
	1993, BaltA, 2, 214
\bibitem[Ciolek \& Basu(2006)]{CB06} Ciolek, G.~E. \& Basu, S. 2006, ApJ, 
	652, 442
\bibitem[Di~Francesco et al.(2007)]{Difran07} Di~Francesco, J., 
	Evans, N.~J.~II, Caselli, P., Myers, P.~C., Shirley, Y.,
	Aikawa, Y., Tafalla, M. 2007, Protostars and Planets V,
	B. Reipurth, D. Jewitt, \& K. Keil (eds), University of Arizona
	Press, Tuscon, p17-32
\bibitem[Elmegreen (2007)]{Elmegreen07} Elmegreen, B.~G. 2007, ApJ, 668, 1064
\bibitem[Gammie \& Ostriker(1996)]{Gammie96} Gammie, C.~F. \& Ostriker, E.~C.
	1996, ApJ, 466, 814
\bibitem[Hatchell et al.(2005)]{Hatchell05} Hatchell, J., Richer, J.~S.,
	Fuller, G.~A., Qualtrough, C.~J., Ladd, E.~F., \& Chandler, C.~J.
	2005, A\&A, 440, 151
\bibitem[Henriksen et al.(1997)]{Henriksen97} Henriksen, R., Andr\'e, P., \&
	Bontemps, S. 1997, A\&A, 323, 549
\bibitem[Jijina et al.(1999)]{Jijina99} Jijina, J., Myers, P.~C., Adams, F.~C.
	1999, ApJS, 125, 161
\bibitem[Johnstone, Di Francesco \& Kirk(2004)]{Johnstone04} Johnstone, D.,
	Di Francesco, J., \& Kirk, H. 2004, ApJ, 611, 45L
\bibitem[J\o rgensen et al.(2008)]{Jorgensen08} J\o rgensen, J.~K.,
	Johnstone, D., Kirk, H., Myers, P.~C., Allen, L., Shirley, Y.
	2008, ApJ, 683, 822
\bibitem[J\o rgensen et al.(2007)]{Jorgensen07} 
	J\o rgensen, J.~K., Johstone, D., Kirk, H., \& Myers, P.~C. 2007,
	ApJ, 656, 293
\bibitem[Kirk, Johnstone \& Di~Francesco(2006)]{Kirk06} Kirk, H.,
	Johnstone, D. \& Di~Francesco, J. 2006, ApJ, 646, 1009
\bibitem[Kirk, Johnstone \& Tafalla(2007)]{Kirk07} Kirk, H.,
	Johnstone, D. \& Tafalla, M. 2007, ApJ,  668, 1042
\bibitem[Klessen et al.(2005)]{Klessen05} Klessen, R.~S., 
	Ballesteros-Paredes, J., V\'azquez-Semadeni, E., Duran-Roj\'as, C.
	2005, ApJ, 620, 786
\bibitem[Kudoh \& Basu(2008)]{Kudoh08} Kudoh, T., Basu, S. 2008, ApJ, 679, L97
\bibitem[Lada(1987)]{Lada87} Lada, C.~J. 1987, Proceedings of the
	Symposium, Tokyo, Japan, Dordrecht, D. Reidel Publishing Co.,
	p1-17
\bibitem[Lada \& Lada(2003)]{Lada03} Lada, C.~J. \& Lada, E.~A.
	2003, ARA\&A, 41, 57
\bibitem[Larson(1981)]{Larson81} Larson, R.~B. 1981, MNRAS, 194, 809
\bibitem[Li \& Nakamura(2004)]{Li04} Li, Z.-Y., \& Nakamura, F. 2004, ApJ,
	609, L83
\bibitem[Lombardi \& Alves(2001)]{Lombardi01} Lombardi, M. \& Alves, J.
	2001, A\&A, 337, 1023
\bibitem[MacLow \& Klessen(2004)]{MacLow04} MacLow, M-M. \& Klessen, R.
	2004, RvMP, 76, 125
\bibitem[Offner et al.(2008)]{Offner08} Offner, S.~S.~R., Klein, R.~I.,
	\& McKee, C.~F. 2008, AJ, 136, 404 
\bibitem[Onishi et al.(1998)]{Onishi98} Onishi, T., Mizuno, A.,
        Kawamura, H.~O., \& Fukui, Y.\ 1998, ApJ, 502, 296
\bibitem[Rosolowsky et al.(2008)]{Rosolowsky08} Rosolowsky, E.~W., 
	Pineda, J.~E., Foster, J.~B., Borkin, M.~A., Kauffmann, J., 
	Caselli, P., Myers, P.~C., Goodman, A.~A. 2008, ApJS, 175, 509
\bibitem[Ridge et al.(2006)]{Ridge06} Ridge, N.~A., Di~Francesco, J., Kirk, H.,
	Li, D., Goodman, A.~A., Arce, H.~G., Borkin, M.~A., Caselli, P.,
	Foster, J.~B., Heyer, M.~H., Johnstone, D., Kosslyn, D.~A.,
	Lombardi, M., Pineda, J.~E., Schnee, S.~L., \& Tafalla, M. 
	2006, AJ, 131, 2921
\bibitem[Shu et al.(1994)]{Shu94} Shu, F., Najita, J.,Ostriker, E., Wilkin, F.,
	Ruden, S., \& Lizano, S. 1994, ApJ, 429, 781
\bibitem[Stahler \& Palla(2004)]{StahlerPalla} Stahler, S.~W. \&
	Palla, F. 2004, The Formation of Stars (Wiley-Vch)
\bibitem[Tafalla et al.(2002)]{Tafalla02} Tafalla, M., Myers, P.~C., 
	Caselli, P., Walmsley, C.~M., Comito, C. 2002, ApJ, 569, 815
\bibitem[Terebey et al.(1984)]{Terebey84} Tereby, S., Shu, F.~H., \& Cassen, P.
	1984, ApJ, 286, 529
\bibitem[Ungerechts et al.(1997)]{Ungerechts97}  Ungerechts, H., 
	Bergin, E.~A., Goldsmith, P.~F., Irvine, W.~M., Schloerb, F.~P.
	\& Snell, R.~L. 1997, ApJ, 482, 245
\bibitem[Walsh et al.(2004)]{Walsh04} Walsh, A.~J., Myers, P.~C. \&
	Burton, M.~G. 2004, ApJ, 614, 194
\bibitem[Walsh et al.(2007)]{Walsh07} Walsh, A.~J., Myers, P.~C., 
	Di Francesco, J., Mohanty, S., Bourke, T.~L., Gutermuth, R.,
	\& Wilner, D. 2007, ApJ, 655, 958
\bibitem[Ward-Thompson et al.(2007)]{Wardthomp07} Ward-Thompson, D.,
	Andr\'e, P., Crutcher, R., Johnstone, D., Onishi, T., \& Wilson, C.
	2007, Protostars and Planets V, B. Reipurth, D. Jewitt, 
	\& K. Keil (eds), University of Arizona Press, Tuscon, p33-46
\bibitem[Ward-Thompson et al.(2007b)]{Wardthomp07b} Ward-Thompson, D., 
	et al. 2007, PASP, 119, 855
\end{thebibliography}
